\documentclass[%
 amsmath,amssymb,
 aip,
reprint,
]{revtex4-1}

\usepackage[english]{babel}  %Gets rid of annoying babel error messages

\usepackage{listings}       %source code blocks
\usepackage[colorlinks=true,linkcolor=blue]{hyperref} %add hyperlinks to tables and citations and others, probably superlfuous
\usepackage{graphicx} %black draft images
\usepackage{epstopdf}       %compile eps files to pdf for inclusion in document
\usepackage[utf8]{inputenc} %input encoding
\usepackage[T1]{fontenc}
\usepackage{bm} %bold math
\usepackage{mathptmx}
\usepackage{multirow}
\usepackage{siunitx}         %Otherwise getting angsttrom symbol in math mode is annyoing

\begin{document}

\title[]{Global Optimization of Copper Clusters at the ZnO(10\=10) Surface Using a DFT-based Neural Network Potential and Genetic Algorithms}
\author{Mart\'in Leandro Paleico}
\affiliation{Universit\"{a}t G\"{o}ttingen, Institut f\"{u}r Physikalische Chemie, Theoretische Chemie, Tammannstra\ss{}e 6, 37077 G\"{o}ttingen, Germany}
\email{martin.paleico@uni-goettingen.de}
\author{J\"org Behler}
\email{joerg.behler@uni-goettingen.de}
\affiliation{Universit\"{a}t G\"{o}ttingen, Institut f\"{u}r Physikalische Chemie, Theoretische Chemie, Tammannstra\ss{}e 6, 37077 G\"{o}ttingen, Germany}
\affiliation{International Center for Advanced Studies of Energy Conversion (ICASEC), Universit\"at G\"ottingen, Tammannstra\ss{}e 6, 37077 G\"ottingen, Germany}
\date{\today}

\begin{abstract}
The determination of the most stable structures of metal clusters supported at solid surfaces by computer simulations represents a formidable challenge due to the complexity of the potential-energy surface. Here we combine a high-dimensional neural network potential, which allows to predict the energies and forces of a large number of structures with first-principles accuracy, with a global optimization scheme employing genetic algorithms. This very efficient setup is used to identify the global minima and low-energy local minima for a series of copper clusters containing between four and ten atoms adsorbed at the ZnO(10$\bar{1}$0) surface. A series of structures with common structural features resembling the Cu(111) and Cu(110) surfaces at the metal-oxide interface has been identified, and the geometries of the emerging clusters are characterized in detail. We demonstrate that the frequently employed approximation of a frozen substrate surface in global optimization can result in missing the most relevant structures.
\end{abstract}

\maketitle
\bigskip

%\MLP{Temporary table of contents for easier navigation}
%\tableofcontents
%\newpage

\section{Introduction}

\textbf{NOTE: The following article has been submitted to The Journal of Chemical Physics. After it is published, it will be found at \url{https://aip.scitation.org/journal/jcp}.}

Supported clusters, with the cluster and the support being either metals or oxides, are an active area of research~\cite{gates_supported_1995, santra_oxide-supported_2002, wang_zinc_2004, heiz_fundamental_2004, yu_review_2012} in surface science and catalysis. Accordingly, they have been studied extensively both with theoretical~\cite{cabria_theoretical_2010, robles_oxidation_2010, tang_theoretical_2012} and experimental~\cite{heiz_chemical_1997, haas_nucleation_2000, yamaguchi_catalytic_2010, peters_size-dependent_2013} methods. A prominent example for such catalysts that is used in the industrial synthesis of methanol~\cite{bowker_mechanism_1988, saito_development_1996, sa_catalysts_2010}, consists of ZnO and Cu nanoparticles. Given the commercial importance of methanol as a solvent and as a reagent, this system has received a lot of attention in recent years, and many theoretical~\cite{french_growth_2008, cheng_cu_2014, artrith_neural_2013, hellstrom_structure_2016, mora-fonz_development_2017} as well as experimental~\cite{dulub_stm_2002, koplitz_stm_2003, kroll_small_2007, behrens_active_2012, kuld_quantifying_2016, kattel_active_2017} studies are available with the aim of unraveling the structural details of this system. 

Computer simulations can provide valuable information, complementary to experimental data, but theoretical studies of supported clusters are hampered by several challenges. The first one concerns the system size, as supported clusters can be very large containing hundreds or even thousands of atoms, and large supercells are needed for realistic structural models of the support to prevent spurious interactions between adsorbed clusters and other finite-size effects. This rules out the use of demanding electronic structure methods like density-functional theory (DFT) for all but the most simple systems, as most simulation techniques like molecular dynamics (MD) or Monte Carlo (MC) require the computation of thousands to millions of configurations. 

Simple and thus computationally cheap empirical potentials would allow to study a large number of structures, but reliable atomistic potentials for supported clusters, which provide a direct functional relation between the atomic positions and the potential energy, are difficult to construct due to the complex and diverse bonding and interactions present in these systems, and consequently they are available only for a few systems~\cite{zhou_modified_2004,P2884}. Thus the second challenge is the lack of reliable potentials, which are able to cover the wide variety of atomic environments and bonding patterns present in the nanoparticle and the support.

A possible solution to this dilemma is offered by machine learning (ML) potentials~\cite{P2559,behler_perspective_2016,P4938}, which can provide very accurate potential energy surfaces (PES) based on reference data sets for representative configurations obtained in electronic structure calculations. Once constructed, ML potentials allow to compute the energies and forces of large systems in a very efficient way enabling the study of realistic structural models, while the first-principles accuracy of the underlying reference method is essentially maintained. Starting with the seminal work of Doren and coworkers in 1995~\cite{P0316}, the technology of ML potentials has made rapid progress in recent years, and many algorithms are available, like neural network-based approaches~\cite{P0421,P1388,P5366,P5577,jiang_permutation_2013,P4945,P4419}, Gaussian approximation potentials~\cite{bartok_gaussian_2010}, moment tensor potentials~\cite{shapeev_moment_2016}, spectral neighbor analysis potentials~\cite{thompson_spectral_2015}, and many others. Here, we utilize high-dimensional neural network potentials (NNPs) proposed by Behler and Parrinello in 2007~\cite{behler_generalized_2007,behler_constructing_2015, behler_first_2017} to study small copper clusters supported on the (10$\bar{1}$0) surface of ZnO. High-dimensional NNPs of this type have been used successfully in the past for a series of related systems like copper~\cite{P3114}, brass nanoparticles~\cite{weinreich_properties_2020}, zinc oxide~\cite{artrith_high-dimensional_2011}, water at copper~\cite{P4886} and zinc oxide surfaces~\cite{P4988}, and also for the copper-zinc oxide system~\cite{artrith_neural_2013}.

Specifically, here we use NNPs to perform genetic algorithm (GA)~\cite{michalewicz_genetic_1991, deaven_molecular_1995} based global optimization (GO) searches of pure copper clusters containing between four and ten atoms supported on the low index zinc oxide (10$\bar{1}$0) surface. Other recent papers \cite{kolsbjerg_neural-network-enhanced_2018, jennings_genetic_2019} have shown the usefulness of such a combined NNP+GA approach. Combining both algorithms allows for generating a new potential for a target system ``on-the-fly'', while at the same time requiring a minimal amount of electronic structure calculations. The generated NNP is then used to run GA searches, and good candidates from this GA search are used to further refine the potential. This procedure could, in the future, easily be adapted to run automated global optimization searches on a wide range of materials.

Previous GO searches of this or similar systems utilized either parametrized classical force fields~\cite{mora-fonz_development_2017}, or relied on \textit{ab-initio} calculations, which requires them to pre-optimize the clusters in an isolated configuration in vacuum~\cite{wan_single_2018} before depositing them on the surface. The NNP allows us to easily include the influence of the support on the optimized clusters in the GO search, without needing to resort to isolated clusters. Additionally, it allows us to run much longer GA searches which provides for better sampling. 

The paper is organized as follows: Section~\ref{sec:methods} summarizes the key concepts of high-dimensional NNPs and GA searches. In Section~\ref{sec:compdets} we describe our computational setup for the reference DFT calculations, the procedure for generating the NNP data set, the parameters for the NNP construction and the details of the GA search. In Section~\ref{sec:results} we present our results addressing the structural properties of the clean ZnO support (\ref{sec:znosupport}), the global minimum and low-energy local minima from Cu$_4$ to Cu$_{10}$ (\ref{sec:garesults}), the interface structures (\ref{sec:interface}), the geometric properties of the optimized clusters (\ref{sec:clusterprops}) and the role of a possibly frozen support surface (\ref{sec:frozensurface}). Finally, we draw our conclusions in Section~\ref{sec:conclusion}.

%\MLP{If time: Add Section shortly demionstrating the effectiveness of different mutation moves of the GA, and its consistency in finding the putative GM.}

%Additionally, in section \ref{sec:bah} we introduce a new method for quickly analyzing atomic environments, which aids in the construction and improvement of NNP datasets.

%%%%%%%%%%%%%%%%%%%%%%%%%%%%%%%%%%%%%%%%%%%%%%%%%%%%%%%%%%%%%%%%%%%%%%%%%%%%
%%%%%%%%%%%%%%%%%%%%%%%%%%%%%%%%%%%%%%%%%%%%%%%%%%%%%%%%%%%%%%%%%%%%%%%%%%%%
\section{Methods} \label{sec:methods}
%%%%%%%%%%%%%%%%%%%%%%%%%%%%%%%%%%%%%%%%%%%%%%%%%%%%%%%%%%%%%%%%%%%%%%%%%%%%
%%%%%%%%%%%%%%%%%%%%%%%%%%%%%%%%%%%%%%%%%%%%%%%%%%%%%%%%%%%%%%%%%%%%%%%%%%%%

%%%%%%%%%%%%%%%%%%%%%%%%%%%%%%%%%%%%%%%%%%%%%%%%%%%%%%%%%%%%%%%%%%%%%%%%%%%%
%%%%%%%%%%%%%%%%%%%%%%%%%%%%%%%%%%%%%%%%%%%%%%%%%%%%%%%%%%%%%%%%%%%%%%%%%%%%
\subsection{Neural Network Potential} \label{sec:methods_nnp}
%%%%%%%%%%%%%%%%%%%%%%%%%%%%%%%%%%%%%%%%%%%%%%%%%%%%%%%%%%%%%%%%%%%%%%%%%%%%
%%%%%%%%%%%%%%%%%%%%%%%%%%%%%%%%%%%%%%%%%%%%%%%%%%%%%%%%%%%%%%%%%%%%%%%%%%%%

Neural networks (NN) are one of the main methods in use in the realm of machine learning. They are capable of reproducing any well-behaved function \cite{hornik_approximation_1991}, such as the potential energy surface (PES) of a system of atoms, if a set of points where the function has been evaluated is available. In the present work we utilize fully connected, high-dimensional, feed-forward neural networks introduced by Behler and Parrinello~\cite{behler_generalized_2007, behler_first_2017}, which are constructed to reproduce reference data sets containing the energies and forces obtained from \textit{ab-initio} calculations for a diverse set of structures. 

In the Behler-Parrinello approach, the system for which the energy and forces are to be predicted is broken down into atomic environments, each of which consists of a central atom and all other atoms within a certain cutoff radius. Each of these subsystems is processed by a separate feed-forward NN, unique to each element but shared across atoms of the same element. Each of these elemental NNs possesses the same set of weight parameters, and predicts an atomic energy contribution as a function of the chemical environment. These atomic energies are then added to provide the total potential energy of the system. Analytic gradients to determine the atomic forces are readily available.

In the input layers of the atomic NNs the information about the atomic environments is provided using vectors of symmetry functions~\cite{P2882}. These serve as structural fingerprints with the required translational, rotational and permutational invariance. 
In a training process the weight parameters and biases connecting the different nodes of the network are adjusted until the output matches the available \textit{ab-initio} reference data. For this, it is necessary to sample the PES of the system, by performing single point calculations at different positions of this PES. Once the NNP has been trained, it is able to predict the energies and forces at a fraction of the computational costs of the underlying electronic structure method, and can be applied to much larger systems than are accessible by DFT. For all details about the method, the training process and the validation strategies for high-dimensional NNPs the interested reader is referred to a series of recent reviews of the method~\cite{P4106,behler_constructing_2015,behler_first_2017}

%%%%%%%%%%%%%%%%%%%%%%%%%%%%%%%%%%%%%%%%%%%%%%%%%%%%%%%%%%%%%%%%%%%%%%%%%%%%
%%%%%%%%%%%%%%%%%%%%%%%%%%%%%%%%%%%%%%%%%%%%%%%%%%%%%%%%%%%%%%%%%%%%%%%%%%%%
\subsection{Genetic Algorithm} \label{sec:methods_ga}
%%%%%%%%%%%%%%%%%%%%%%%%%%%%%%%%%%%%%%%%%%%%%%%%%%%%%%%%%%%%%%%%%%%%%%%%%%%%
%%%%%%%%%%%%%%%%%%%%%%%%%%%%%%%%%%%%%%%%%%%%%%%%%%%%%%%%%%%%%%%%%%%%%%%%%%%%

Genetic algorithms~\cite{michalewicz_genetic_1991, deaven_molecular_1995} are a class of global optimization algorithms. They rely on codifying the properties of the system in form of ``genes'' of a given configuration, assigning  a fitness criterion, which in the case of global structural optimization often is the potential energy, but other target criteria are possible~\cite{jensen_designing_2014}, and then selecting and crossing over these genes with the purpose of generating new, fitter candidates. Along the way, variability is introduced into the candidate pool in the form of mutations that directly affect this genome. The final goal of this GO algorithm is to quickly generate and evaluate candidates, exploring the configuration space until the fittest candidate corresponding to the global minimum is found.

GA can be utilized for optimizing different kinds of systems~\cite{bozkurt_genetic_2018}, and the procedure is rather standardized for structural optimization of atomic systems~\cite{vilhelmsen_systematic_2012, vilhelmsen_identification_2014, huang_structural_2019, buendia_study_2017, heydariyan_new_2018}. The genes in this case consist of the coordinates of the atoms of interest, and elemental composition for multi-elemental systems. Crossovers between candidates are achieved by cutting the system along random planes~\cite{deaven_molecular_1995} and recombining the generated parts while maintaining the system's size and stoichiometry. After crossover events, the new configurations are minimized, in an approach similar to basin hopping Monte Carlo~\cite{wales_global_1997} (BHMC), and it is acknowledged that both algorithms in fact sample the same transformed PES configuration space~\cite{wales_global_1997}. 

Strictly speaking, a genetic algorithm requires that the properties of the system are encoded into strings that can then be mutated and combined, such as for example listing the elements at given positions of a defined crystal structure~\cite{jensen_designing_2014}. Modifying directly the coordinates of the system as we do in this work is more akin to modifying the phenotype rather than the genotype of the configurations. This kind of algorithms are more accurately described as ``evolutionary'' algorithms. We nonetheless refer to it as a genetic algorithm throughout the text as this is the more commonly in-use designation, and evolutionary algorithms are a subset of genetic algorithms.

The fitness is usually evaluated through the potential energy since the goal is to find the lowest energy configuration. Mutations are applied as direct, large-scale modifications to the coordinates of the candidates, such as mirrorings through a plane, i.e., discarding one half of the cluster and mirroring the other half, which leads to new configurations if the original cluster was not symmetrical through that plane; or rotations, such as partial rotations of a section of a cluster, or rotations of a supported cluster over a support. This results in a rapid exploration of the PES by generating and evaluating thousands of structurally different candidates in quick succession.

Some GAs make use of a generational \textit{ansatz}, where the algorithm proceeds in steps by fixing the fittest population, generating new candidates, evaluating them, and proceeding to recalculate and update the fittest population. This \textit{ansatz} is ideal for \textit{ab-initio} based searches, since there is a long wait time for evaluating the new population that can be partially reduced by parallelization within a generation. This has been shown not to be necessary for a successful GA search~\cite{vilhelmsen_genetic_2014}, and our NNP based calculations are extremely fast compared to similarly sized \textit{ab-initio} relaxations. For this reason we utilize a continuous approach, where the fittest population is updated every time a relaxation is finished, and new candidates are continually pulled from this population.

Mutations are the key for efficiently exploring the PES. Similar to Monte Carlo simulations, poor mutations will advance the system through an inefficient route, and only rarely new good structures will be generated. Mutations should thus result in big changes to the candidate, but at the same time generate ``reasonable'' structures, which is aided by the final geometry relaxation every candidate goes through. In our case we have made use of the following mutations, which are applied only to the atoms belonging to the cluster, while the atoms of the support only change positions in the subsequent local geometry optimizations:

%\MLP{If time, add figure for mutations: , as shown in Fig.} %JB I don't think we need this, would be nice but we are under time pressure and it is not essential

\begin{enumerate}
    \item Rattle mutation: Inspired by basin hopping Monte Carlo~\cite{wales_global_1997}, all atoms in the cluster are displaced at random up to a maximum amount. The displacement is slightly biased in the positive $z$ direction to avoid atoms penetrating into the support. This move helps the cluster to expand across the interface with the support and also to grow in height.
    \item Twist mutation: The whole cluster is twisted by a random angle around a vector that passes through its center of geometry (COG) and is perpendicular to the slab surface. We note that the COM is used here as an element-independent generalization instead of the also frequently used center of mass (COM) of the cluster, since the COM does not coincide with the COG for clusters with more than one element. The cluster is also lifted a small random amount in the $z$ direction to avoid repulsive overlaps with the topmost surface atoms. This move helps the cluster to align itself with the atomic patterns of the support.
    \item Angular mutation: Several atoms of the cluster are picked based on their distance to the COG with the furthest away atoms being picked first, and displaced to positions on a hemisphere centered on the COG and whose radius is the average distance of all atoms in the cluster to the COG. This move has been inspired by Rondina \textit{et al.} \cite{rondina_revised_2013}, which describe other possible interesting moves that can be adapted from BHMC. This move helps the cluster grow out in a 3D mode, avoiding flat clusters.
    \item Mirror and shift mutations: A random plane passing through the COG of the cluster is selected, and the cluster is mirrored through this plane. To avoid frequent overlaps at the mirror plane, the mirror image is shifted a small amount away from the plane at random, which increases the acceptance ratio of the move. This move helps to generate more symmetric configurations, which tend to have a lower energy.
    \item Molecular dynamics mutation: A short MD simulation in the $NVT$ ensemble (constant number of particles, volume and temperature) is performed including cluster and support atoms at high enough temperatures ensuring that the cluster atoms are mobile, which can even be above the melting temperature of the cluster, and then the system is cooled down slowly to avoid completely amorphous configurations. This move is inspired by simulated annealing~\cite{kirkpatrick_optimization_1983} and minima hopping~\cite{goedecker_minima_2004} optimization procedures, and it is well-known that MD is good at finding transition paths over low energy barriers~\cite{goedecker_minima_2004}. The cluster is thus allowed to naturally jump over potential energy barriers as provided by the available kinetic energy. This takes advantage of the shape of the PES, instead of being completely random such as the rattle mutation. This move is more expensive to run, since it requires a couple of thousand of MD steps, and it would be extremely expensive to implement in an \textit{ab-initio} based search, but it is straightforward to perform with the NNP approach. Even then, this move requires more computing time than the other simple mutation plus local relaxation moves. For this reason, this move is given a lower occurrence probability.
\end{enumerate}

%\begin{figure}
%    \centering
%    \includegraphics{example-image-a}
%    \caption{Example pictures of mutations}
%    \label{fig:mutations-examples}
%\end{figure}

Interatomic distances, within the cluster as well as between cluster and support, are checked after every type of move to avoid too repulsive atomic overlaps, and if this check fails the mutation is repeated a number of times until successful, or discarded if too many failures accumulate. This is most relevant for \textit{ab-initio} based GA searches, where long relaxations due to closely overlapping atoms would cost a lot of computation time. But, this is still important in our case, since unphysical atomic positions can give rise to configurations that are outside the training set of the NNP, leading to less reliable energy and force predictions due to extrapolation.

We find it necessary to remark that although the moves might bias the final cluster shape in certain directions, they only provide an avenue of change, since in the end the energy of the cluster as evaluated through a particular force field dictates whether the structures are accepted or not. This means that it is hard to over-bias the cluster, but on the other hand it is possible to not offer enough mutations, resulting in the cluster exploring only a part of the configuration space.

A key component to a successful GA run is maintaining structural diversity in the breeding population. Otherwise, this population can become overrun by a small number of very fit candidates, which leads to always breeding the same children configurations and the GA search stalling. This structural diversity can be achieved in a number of ways, such as penalizing candidates that have already been chosen for breeding~\cite{vilhelmsen_genetic_2014}. In general, a way of quantifying structural similarity is needed. In this case we use an interatomic distance comparator~\cite{vilhelmsen_systematic_2012}, which is insensitive to rotations and translations of the compared atoms. A first filter is applied comparing the energies of the structures in question. Afterwards, the comparator value is calculated employing the interatomic distances. This comparator is quick to compute and easy to calculate, which makes it ideal for this application. In our experience it is effective at keeping the breeding population distinct.

The size of the actively breeding population is kept fixed at a predefined maximum number of candidates, which correspond to the fittest candidates that are also structurally different. The fitness criterion is calculated as~\cite{vilhelmsen_structure_2012}:
\begin{equation}
    f_i = 0.5 * (1 - \textrm{tanh}(2 \cdot (E_\textrm{max}-E_i) / (E_\textrm{max} - E_\textrm{min}) - 1))
\end{equation}
where $f_i$ is the fitness of a particular candidate, $E_i$, $E_\textrm{min}$ and $E_\textrm{max}$ correspond to the energies of the candidate and the minimum as well as maximum energies in the whole population. This creates a fitness distribution that falls sharply as we get away form the optimal candidate and has a constant low value for the worst candidates.

Candidates are then extracted at random from this population utilizing a roulette wheel algorithm~\cite{johnston_evolving_2003}, where the following criterion is evaluated,
\begin{equation}
    f_i > f_{\textrm{max}} \cdot \textrm{ran}(0.0, 1.0) \quad ,
\end{equation}
to decide if a candidate is kept, where $f_i$ is the fitness of the randomly chosen candidate, $f_{\textrm{max}}$ is the fitness of the fittest candidate, and $\textrm{ran}$ generates a random number between 0.0 and 1.0. With this equation, fitter candidates will tend to get chosen more often since their ratio of $f_i/f_{\textrm{max}}$ will be closer to 1.0, but all candidates in the active population have a non-zero chance of being picked.

Other GO methods have been considered for this system. Grid-based~\cite{rahm_beyond_2017} optimization methods are not ideal for our purpose, since the resulting atomic grid of the clusters is not known beforehand, and as seen in the Results sections, is can be strongly modified by the support. Instead, methods with flexible grids could be tested~\cite{paleico_flexible_2020}. BHMC~\cite{wales_global_1997} is in our experience too slow, since structures are still connected through a Monte Carlo-like chain. Instead, GA offers the advantage of evaluating many structures in parallel. Of course, a number of BHMC simulations could be run in a trivial parallelization approach, but GA benefits from biasing future candidates with information from previous candidates, thanks to the crossover operations.

%%%%%%%%%%%%%%%%%%%%%%%%%%%%%%%%%%%%%%%%%%%%%%%%%%%%%%%%%%%%%%%%%%%%%%%%%%%%
%%%%%%%%%%%%%%%%%%%%%%%%%%%%%%%%%%%%%%%%%%%%%%%%%%%%%%%%%%%%%%%%%%%%%%%%%%%%
\section{Computational Details} \label{sec:compdets}
%%%%%%%%%%%%%%%%%%%%%%%%%%%%%%%%%%%%%%%%%%%%%%%%%%%%%%%%%%%%%%%%%%%%%%%%%%%%
%%%%%%%%%%%%%%%%%%%%%%%%%%%%%%%%%%%%%%%%%%%%%%%%%%%%%%%%%%%%%%%%%%%%%%%%%%%%

%%%%%%%%%%%%%%%%%%%%%%%%%%%%%%%%%%%%%%%%%%%%%%%%%%%%%%%%%%%%%%%%%%%%%%%%%%%%
%%%%%%%%%%%%%%%%%%%%%%%%%%%%%%%%%%%%%%%%%%%%%%%%%%%%%%%%%%%%%%%%%%%%%%%%%%%%
\subsection{Density Functional Theory Calculations} \label{sec:compdets_dft}
%%%%%%%%%%%%%%%%%%%%%%%%%%%%%%%%%%%%%%%%%%%%%%%%%%%%%%%%%%%%%%%%%%%%%%%%%%%%
%%%%%%%%%%%%%%%%%%%%%%%%%%%%%%%%%%%%%%%%%%%%%%%%%%%%%%%%%%%%%%%%%%%%%%%%%%%%

The reference method we have used to generate the training set for parameterizing the NNP is density functional theory (DFT)~\cite{hohenberg_inhomogeneous_1964, kohn_self-consistent_1965}, as implemented in the Vienna Ab initio Simulation Package (VASP)~\cite{kresse_efficient_1996, kresse_ultrasoft_1999} version 5.4.4. This software performs periodic DFT calculations with plane wave basis functions treating the core electrons with PAW pseudopotentials~\cite{blochl_projector_1994, kresse_norm-conserving_1994}. We have utilized the pseudopotentials recommended by the VASP manual for each element. We have chosen the generalized gradient approximation (GGA) functional suggested by Perdew, Burke, and Ernzerhof (PBE)~\cite{perdew_generalized_1996} to describe electronic exchange and correlation.

We have run convergence tests for all relevant settings of the DFT calculations, until energy and force values converged below the aimed NNP root mean squared errors (RMSE), i.e., 1~meV/atom for total energies and 100 meV/Bohr for the forces. For this purpose, we utilized a plane wave cutoff of 500~eV, and a k-point grid of $12\times 12\times 12$ for the 3.61~\r{A} cubic unit cell of fcc Cu containing 4 atoms. For larger supercells the number of k-point has been scaled accordingly to ensure a similar level of convergence. For slab systems, at least 13~\r{A} vacuum have been used in the $z$ direction perpendicular to the surface to minimize interactions with the periodic images of the slab. We applied a Gaussian smearing at the Fermi level, with a 0.1 eV $\sigma$ factor. An example input file for VASP with a complete list of the settings is given in the supporting information.

Although the NNP can be trained to DFT reference data including dispersion corrections~\cite{kondati_natarajan_self-diffusion_2017, quaranta_structure_2019, P4556}, which are often problematic in GGA functionals, we have not done this in the present work. The reason is that in particular the dispersion-corrected PBE functional often yields overbinding, which is present to some extent also for the plain PBE functional, and too small lattice constants are obtained with commonly used correction schemes \cite{P3112,P3296,bucko_extending_2014}, which is in agreement with previous reports of overbinding~\cite{antony_density_2006} for certain exchange-correlation functionals. As we consider a good prediction of the bulk properties of copper and zinc oxide as important, we have chosen the uncorrected PBE functional providing the most accurate description. It should be noted that, as discussed below, the energy differences between local minima can be very small and the subtle energetic ordering can change depending on if and which dispersion correction scheme is used.

%%%%%%%%%%%%%%%%%%%%%%%%%%%%%%%%%%%%%%%%%%%%%%%%%%%%%%%%%%%%%%%%%%%%%%%%%%%%
%%%%%%%%%%%%%%%%%%%%%%%%%%%%%%%%%%%%%%%%%%%%%%%%%%%%%%%%%%%%%%%%%%%%%%%%%%%%
\subsection{Generation and Composition of the Reference Data Set} \label{sec:genpotential}
%%%%%%%%%%%%%%%%%%%%%%%%%%%%%%%%%%%%%%%%%%%%%%%%%%%%%%%%%%%%%%%%%%%%%%%%%%%%
%%%%%%%%%%%%%%%%%%%%%%%%%%%%%%%%%%%%%%%%%%%%%%%%%%%%%%%%%%%%%%%%%%%%%%%%%%%%

\begin{figure*}
    \centering
    \includegraphics[width=\linewidth]{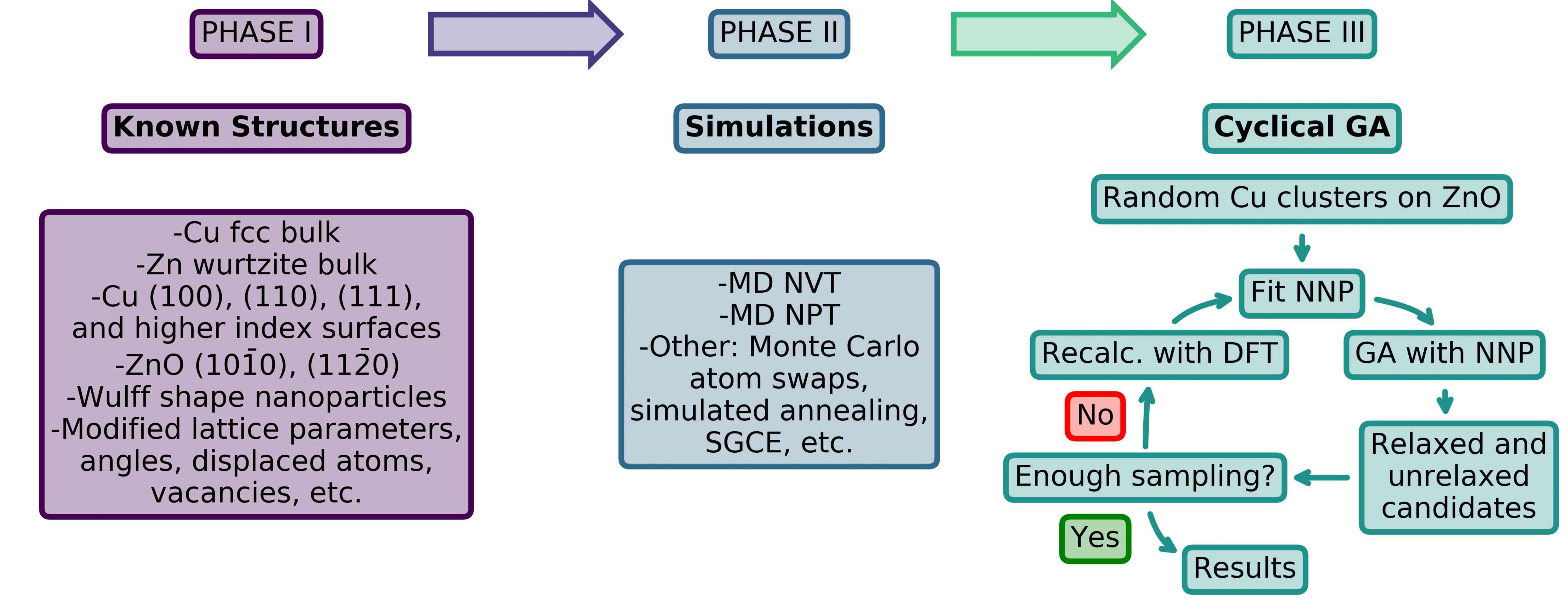}
    \caption{Phases followed in the construction of the NNP dataset, as described in more detail in the main text. Phase I corresponds to generating known structures with small modifications, in phase II we perform simulations with these structures with the usual simulation methods such as MD and MC, and in phase III we proceed to perform a cyclical GA search for new structures.}
    \label{fig:ga_phases}
\end{figure*}

The key to generating a reliable NNP is to produce a reference data set covering all atomic environments that are relevant for the chosen type of simulation. Since carrying out these electronic structure calculations is the most time-consuming step of the NNP construction, this data set should be kept as small as possible. This avoids wasting computational time on calculating \textit{ab-initio} data on configurations that a given type of simulation will not visit, or the reverse, missing training data for configurations that emerge in a simulation. For example, Monte Carlo simulations are more likely to visit high energy configurations corresponding to less-favorable relative atomic positions when compared to molecular dynamics simulations, that generate trajectories guided by the interatomic forces.

It is thus important to tailor the structure generation method to the simulations being performed, and if possible to integrate both processes~\cite{P3114}. Several aspects need to be considered. Structures need to be generated covering a wide range of the PES, but not at random since this samples irrelevant high-energy regions. Structures also need to fit within the available computational resources, i.e., they should contain only up to about 200-300 atoms per structure in case of DFT, while also avoiding to compute identical or very similar environments repeatedly. 

The protocol we have followed for the generation of the NNP in the present work can be summarized and categorized into the following phases, also presented in fig.\ref{fig:ga_phases}:

In phase I, structures are generated from known configurations, i.e., known crystal structures and surface models, for which either experimental data or theoretical information is available. In our case, this corresponds to Cu and ZnO bulk structures in their fcc and wurtzite crystal structures, respectively; low Miller index surfaces of both materials, and Wulff-shape nanoparticles. Additionally, the NNP includes some additional data for brass structures~\cite{weinreich_properties_2020}. These base structures are then modified in a well-defined way. For example, the experimental lattice parameters and interatomic distances of the bulk structures are varied in a small range, to accommodate for possible DFT deviations from the experimental values, and to provide the NNP with more information around the equilibrium conditions. Additionally, thermal distortions are emulated by randomly displacing the atoms around their equilibrium positions, which is also important to break the symmetry of the atomic configurations. For slab structures, different numbers of layers are included. This results in a NNP that is applicable to a narrow range of conditions centered around the base structures, and is useful for finding simple structural and energetical parameters~\cite{artrith_neural_2013}, such as lattice constants, bulk moduli, and surface energies.

In phase II, $NVT$ and $NPT$ ensemble simulations are utilized. With these simulations the system is naturally allowed to explore the available configurations at given temperatures and pressures. Included in this is melting and subsequently cooling the systems to introduce realistic liquid and amorphous environments into the data set, which would be difficult to achieve with the procedural methods of phase I. Structures are sampled at regular intervals, e.g. every couple of picoseconds, or by utilizing an ensemble of NNPs~\cite{P3114,behler_constructing_2015} to scan specifically for new configurations.  After phase II, the NNP is usually able to perform long $NVT$ and $NPT$ MD simulations without visiting unsampled regions of the PES. In the present case we require both high energy configurations, generated e.g. by GA mutations and crossings, which are not well sampled in a canonical ensemble, and close to equilibrium low energy structures. Other simulation methods  that allow the NNP to explore reasonable sections of configuration space are also useful, such as Monte Carlo atom swaps~\cite{weinreich_properties_2020} that exchange elements, simulated annealing~\cite{kirkpatrick_optimization_1983},  and semi-grand canonical ensemble simulations~\cite{kofke_monte_1988}.

Phase III aims to include high-energy structures and consists of two parts. First, random copper clusters are generated on different ZnO surfaces. Structures are checked for short interatomic distances and rejected if atoms are too close together to avoid unphysically large, spurious repulsive forces, which can lead to problems both in the DFT calculations and in the fitting procedure. We set this threshold at 20\% below the nearest neighbor distance for the respective material, which is sufficient to cover the usual deviations from this distance observed in clusters and provides the required information about the relevant repulsive parts of the PES. 
This enables the NNP to perform the second part, where iterative GA searches are carried out. In this iterative approach, GA searches are run based on a given NNP for different cluster sizes, which is very fast. Both minimized and unrelaxed structures resulting from a mutation or crossover operation are then extracted and evaluated, again utilizing an ensemble of NNPs~\cite{P3114,behler_constructing_2015}. With this procedure we obtained about 200-500 new structures per GA run. These structures then have been recalculated by DFT, a set of new NNPs has been generated for the extended data set, and this process has been repeated iteratively several times until convergence has been reached, which is defined here by two NNPs trained on the same data set agreeing on the predictions for the obtained clusters within the accepted energy and force tolerance. 

Two problems need to be solved in this hierarchical approach. The first is to avoid recalculating structures that are already well represented in the data set. The use of ensembles of NNPs can assist the selection, since structures that are already included will exhibit small errors in all NNPs trained on this data set, while unknown structures will have very different errors for different NNPs. A complementary method based on the quick classification of SF sets and atomic environment descriptions has been employed here as well%~\cite{paleico_bah_2020}
, and also other algorithms like CUR are in principle available to select the most distinct training structures~\cite{P5398}.
Of course, repeated environments cannot be avoided completely, since for example the presence of a slab support will often lead to redundant atomic environments, in particular in the bottom-most layers of the support, which hardly change while the cluster and the top layers are optimized. 

The second problem to solve is extracting smaller structures from the very large systems used in the simulations, once atomic environments that need to be added to the DFT data set have been identified. This is a problem particular to the subset of structures derived from supported clusters, since other structures that we have used in training the NNP (bulk solids, slabs, etc.) are already intrinsically periodic and designed to fit within the available \textit{ab-initio} calculations.

To avoid periodic images affecting the GA search, which can lead to self-interacting periodic configurations such as wires and sheets of copper atoms on the surface, the GA simulations are performed on large slab supports with up to a thousand atoms. These structures are too large for routine DFT calculations, and even if they would be feasible, they would offer very little new information with respect to the required computational cost since most support atoms are too far away from the cluster to notably interact. 
To reduce the system size for the DFT calculations, we need to extract the relevant atoms from the simulations~\cite{eckhoff_molecular_2019, weinreich_properties_2020} while preserving the local atomic environment of these atoms. In the particular case of supported clusters we also need to maintain the periodicity of the extracted subsystems, which is required by the VASP code, which is not able to perform non-periodic calculations, and, more importantly, for obtaining realistic structural models without artificially merging ZnO surfaces at the edges of the periodic boundary condition. 

\begin{figure*}
    \centering
    \includegraphics[width=\linewidth]{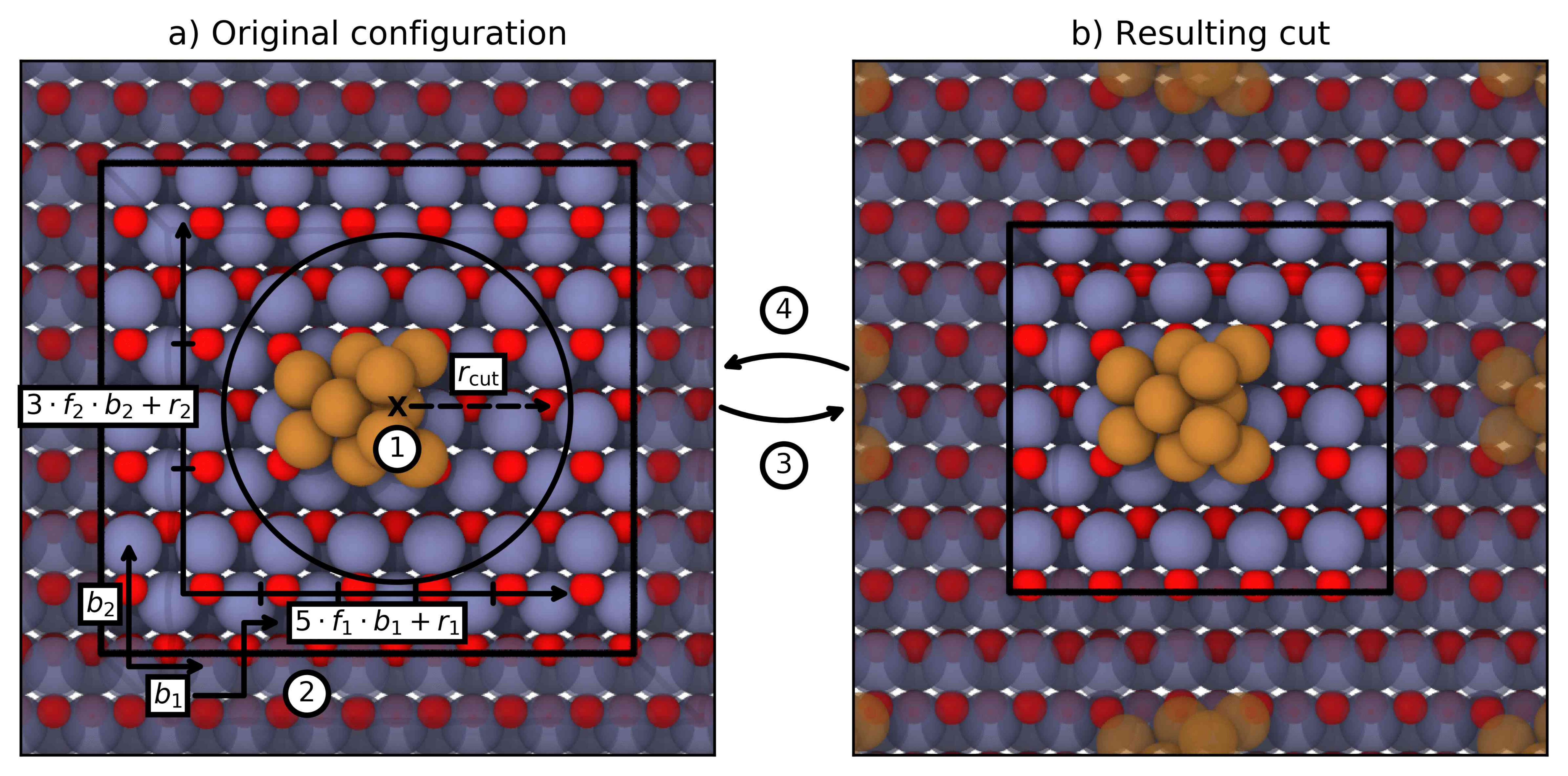}
    \caption{Periodic cut algorithm. a) Initial configuration. b) Resulting structure. 1) The atom closest to (or farthest from) the COG is chosen. 2) A superlattice of the original lattice vectors $b_i$ is created that completely contains the cutoff radius ($r_{\rm cut}$) around the chosen atom. Ticks on the superlattice show multiples of the original lattice vectors. 3) The created structure is checked for consistent stoichiometry, and reasonable interatomic distances at the periodic boundary. 4) If rejected, modify lattice vectors slightly by a factor $f_i$ ($f_1 = 0.97, f_2 = 1.03$) and try again. If the factor is already too large/small, pick the next atom according to the distance from the COG and repeat the procedure.}
    \label{fig:cut-cube}
\end{figure*}

To achieve this, the key realization is that rectangular cuts that preserve the periodicity of the underlying surface lattice will also tile periodically correctly. This assumption holds, of course, if the atoms in the slab have not been substantially distorted from their equilibrium positions. To account for possible small distortions due to cluster-support interactions, our cutting algorithm checks that i) the number of copper atoms in the cell and in particular the zinc-to-oxygen ratio has been preserved and ii) in the cut-out subsystem, atomic distances across the periodic boundary condition are reasonable. This sometimes requires centering the cutting surfaces on different atoms of the system until a cut fulfilling requirements i) and ii) has been found. Additionally, slightly adapting the size of the cutting box from the the size dictated by the ZnO lattice constants by 1-3\% might be required. Taking into account these considerations, it is possible to obtain periodic subsystems with minimal mismatch at the cell boundaries that are suitable for routine DFT calculations. This procedure is shown in fig.~\ref{fig:cut-cube}.

%JB this figure may not be needed
%\begin{figure}
%    \centering
%    \includegraphics[width=\linewidth]{example-image-a}
%    \caption{Showing different environmental decompositions for a big cluster}
%    \label{fig:env-decomposition}
%\end{figure}

%%%%%%%%%%%%%%%%%%%%%%%%%%%%%%%%%%%%%%%%%%%%%%%%%%%%%%%%%%%%%%%%%%%%%%%%%%%%
%%%%%%%%%%%%%%%%%%%%%%%%%%%%%%%%%%%%%%%%%%%%%%%%%%%%%%%%%%%%%%%%%%%%%%%%%%%%
\subsection{Construction of the Neural Network Potential} \label{sec:compdets_nnp}
%%%%%%%%%%%%%%%%%%%%%%%%%%%%%%%%%%%%%%%%%%%%%%%%%%%%%%%%%%%%%%%%%%%%%%%%%%%%
%%%%%%%%%%%%%%%%%%%%%%%%%%%%%%%%%%%%%%%%%%%%%%%%%%%%%%%%%%%%%%%%%%%%%%%%%%%%

The final data set obtained from the procedures described in the previous section consists of 73,136 structures covering about 5 million atomic environments. Therefore, in total 73,136 energies and about 15 million force components are available for training the NNP. The data set contains a wide range of configurations, including bulk, slab, cluster, and amorphous structures for the subsystems Cu, ZnO, and CuZn as well as supported cluster structures for the full ternary (Cu,Zn,O) system.

We have utilized our in-house program RuNNer~\cite{behler_constructing_2015,behler_first_2017} for the construction of the NNP, which is freely available under the GPL3 license.  To choose the best architecture, we have optimized different NN architectures using the hyperbolic tangent activation function in the hidden layers and a linear function for the output layer. For simplicity we have chosen to use the same architecture for each element in the system. The parameters of the symmetry functions used to describe the atomic environments, which have been derived from previous work on ZnO and Cu~\cite{artrith_neural_2013} with some functions removed without significant loss of accuracy, are given in the supporting information. The resulting errors are shown in Table~\ref{tab:nnpfits}.   Based on these errors we have selected the architecture with 2 hidden layers and 15 nodes in each layer for the subsequent simulations. This architecture yields a 2.5 meV/atom root mean square error for the training energies and 59.7 meV/Bohr for the training forces, with very similar values for the testing data not included in the fitting process indicating the absence of significant overfitting. As further evidence of the quality of the potential, we present a predicted vs. DFT energy plot in our supporting information for the structures in our dataset.

\begin{table*}
    \centering
    \begin{tabular}{|c|c|c|c|c|}
    \hline
    \multirow{2}{*}{Architecture} & \multicolumn{2}{|c|}{Training Set RMSE} & \multicolumn{2}{|c|}{Test Set RMSE} \\
    \cline{2-5}
     & $E$ (meV/atom) & $F$ (meV/Bohr) & $E$ (meV/atom) & $F$ (meV/Bohr)\\
    \hline
    
    \textbf{15-15}      & \textbf{2.5} &  \textbf{59.7} & \textbf{3.1} &  \textbf{59.6} \\
    20-20      & 2.7 &  70.5 & 3.5 &  69.4 \\
    15-15-15   & 7.4 & 254.3 & 9.5 & 257.3 \\
    20-20-20   & 2.6 &  73.5 & 3.4 &  74.4 \\
    \hline
    \end{tabular}
    \caption{NNPs obtained for different architectures (neurons per hidden layer are given) and root mean square errors (RMSE) for the predicted energies $E$ and forces $F$. In bold, the chosen architecture for the GA simulations is highlighted.}
    \label{tab:nnpfits}
\end{table*}

%%%%%%%%%%%%%%%%%%%%%%%%%%%%%%%%%%%%%%%%%%%%%%%%%%%%%%%%%%%%%%%%%%%%%%%%%%%%
%%%%%%%%%%%%%%%%%%%%%%%%%%%%%%%%%%%%%%%%%%%%%%%%%%%%%%%%%%%%%%%%%%%%%%%%%%%%
\subsection{Genetic Algorithm} \label{sec:compdets_ga}
%%%%%%%%%%%%%%%%%%%%%%%%%%%%%%%%%%%%%%%%%%%%%%%%%%%%%%%%%%%%%%%%%%%%%%%%%%%%
%%%%%%%%%%%%%%%%%%%%%%%%%%%%%%%%%%%%%%%%%%%%%%%%%%%%%%%%%%%%%%%%%%%%%%%%%%%%

Utilizing a NNP for the GA search allows us to perform the simulations in large cells, with 448 support atoms, with even larger cells also being possible. Each cell was generated to have a lateral diameter of at least 20~\r{A} in each direction of the ZnO(10$\bar{1}$0) surface to provide enough separation of the clusters from their periodic images, resulting in a final size of 23.03 $\times$ 21.19 \r{A}$^2$, or a 6 $\times$ 4 supercell.  Each slab was generated with enough layers to provide a thickness of at least 7~\r{A}, which avoids atoms on top of the layer from ``seeing'' past the atoms on the bottom-most layer due to the NNP symmetry function cutoff. The vacuum added in the $z$ direction (at least 13~\r{A}) ensures that the top-most atoms of the cluster do not interact with the bottom-most atoms of the slab through the PBC.

For each cluster size, the algorithm is seeded by generating and optimizing 15 clusters starting from random geometries. These clusters are initialized in an area in the center of the supercell to avoid the randomly placed atoms being spread too far apart, which leads to poor initial structures. Random initial structures too far from reasonable configurations might lead to extrapolations of the NNP since the underlying data set samples mostly the energetically favored compact clusters, while very high energy configurations, e.g. isolated atoms may not be covered in all situations with a very high precision, which is acceptable as they are physically less relevant. Indeed, after this initial random sampling of the PES the GA search quickly abandons these initial random structures due to their high energies.

The supporting clean slab is pre-optimized at the beginning of the GA search, and then every new candidate cluster is relaxed together with the supporting interface. We note that this is sometimes avoided in DFT-based GA simulations because of the added costs, but as we will show in Sec.~\ref{sec:frozensurface}, this can lead to different results. For the minimization we use the LBGFS algorithm (limited memory BFGS~\cite{liu_limited_1989}), with a maximum force criterion of 0.005 eV/\r{A} and a limit of 1000 evaluation steps, as implemented in the ASE library~\cite{larsen_atomic_2017}. This optimizer was chosen mainly for being the fastest for our particular setup and rather robust. Care should be taken that the structures are properly optimized, as we found that otherwise false LM can appear in the final candidate pool that relax into other LM when stricter criteria are used, particularly for small clusters.

The emerging active breeding population is kept at a maximum of 50 structures during the simulations. It is ensured to be structurally distinct by utilizing the interatomic distance comparator~\cite{vilhelmsen_systematic_2012}. This comparator has three criteria that need to be set. We use a minimum total energy difference of 0.1 eV. If the energy difference between two structures is larger, they are assumed to be different. A maximum correlation distance of 0.7~\r{A}, and a maximum cumulative difference of 0.025~\r{A} are also utilized for the comparison of interatomic distances. For the last two values, the larger the number, the less ``strict'' the comparator is. If a too-strict comparator setting would be used, small differences in structures within the minimization convergence threshold might be detected as significant, resulting in many clusters corresponding to the same structure. In our experience, these values are suitable to keep the breeding population structurally distinct (that is, varied).

The mutation moves we employed have a number of parameters that can be adjusted. We have utilized the following set of values:

\begin{enumerate}
    \item Rattle mutation:
    \begin{itemize}
        \item 60\% of the atoms in the cluster are rattled, chosen at random.
        \item Maximum displacements of up to 3~\r{A} in the $x$ and $y$ directions are used. This value is chosen because the distance between sites on the ZnO surface is also in the order of 3~\r{A}, and we want the cluster atoms to be able to jump to a neighboring site. Displacements in the negative $z$ direction towards the surface are limited to 1~\r{A} to avoid cluster atoms burrowing into the support, while positive displacements remain the same as in the other directions. The probability is uniform in this range.
    \end{itemize}
    \item Twist mutation:
    \begin{itemize}
        \item A vertical offset of 0.5~\r{A} is added to the cluster after twisting it, to avoid cluster atoms overlapping directly with the support. This offset is removed by the geometry relaxation that takes place after the mutation.
    \end{itemize}
    \item Angular mutation:
    \begin{itemize}
        \item between 25 and 50\% of the atoms in the cluster are displaced by the move, starting with the atoms farthest away from the COG.
        \item Atoms are positioned at random on a hemisphere centered on the COG of the cluster, with a radius that is the average distance of all the atoms in the cluster to the COG. This radius is then modified by a random amount between -0.5 and 3.0 ~\r{A}.
        \item The mutation can be biased to position atoms towards the top of the hemisphere and thus promote cluster growth in the $z$ direction by restricting the allowed spherical angles, but in our case we have not added this bias.
    \end{itemize}
    \item Mirror and shift mutations:
    \begin{itemize}
        \item The angle between the mirroring plane and the surface plane can be biased to favor mirrorings parallel or perpendicular to the surface, but we have left this unbiased in our simulations.
        \item Mirrored atoms are shifted to avoid overlaps close to the mirroring plane. This shift is a random value between 0.0 (that is, no shifting) and 1.5~\r{A}. Since there is usually already a small gap at the mirror plane, this amount is sufficient to avoid serious overlaps given the usual Cu-Cu nearest neighbor distance ($d_o$) of approximately 2.55~\r{A}.
    \end{itemize}
    \item Molecular dynamics mutation:
    \begin{itemize}
        \item The MD is run for 3000 steps with a 5~fs timestep, for a total of 15 ps.
        \item The temperature is set to 1000~K, which enables to melt small Cu clusters (the bulk copper melting temperature is around 1360~K, but due to the high surface to volume ratio clusters have lower melting temperatures~\cite{cui_phase_2017, weinreich_properties_2020}).
    \end{itemize}
\end{enumerate}

\begin{figure*}
    \centering
    \includegraphics[width=\linewidth]{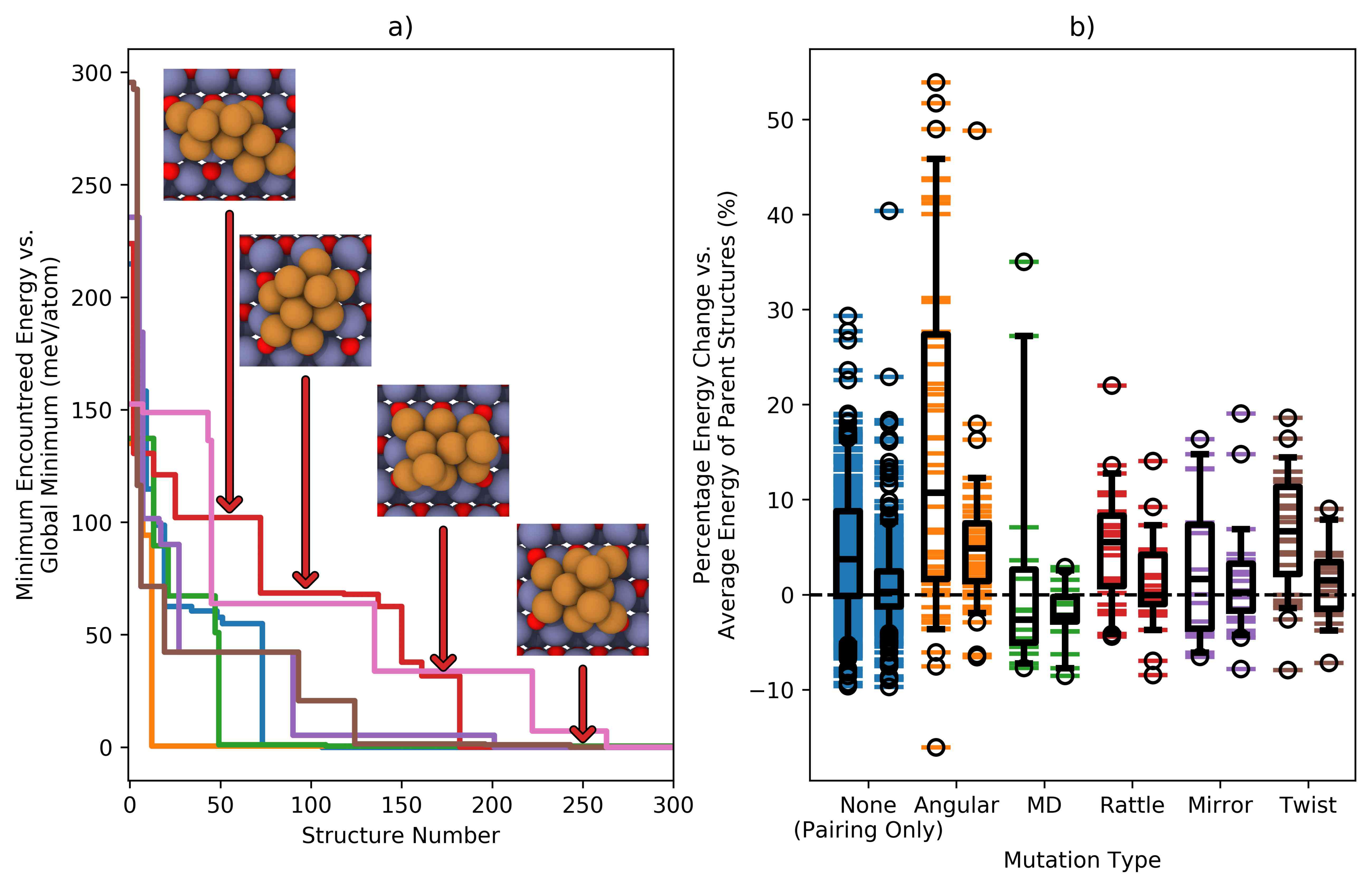}
    \caption{a) Energy progression of seven GA searches for the Cu$_{10}$ cluster. Lines indicate the energy of the lowest energy configuration found in a given GA search until that point, with respect to the putative GM. All runs find the minimum in less than 300 structures, and no lower energy configuration is observed for the following 2700 structures for a total of 3000 structures in each independent run. Arrows and inserts show selected cluster structures for the red run.  b) Percentage energy changes compared to the average energy of the parent structures (see Eq.~\ref{eq:percentage}), for different mutations for two GA searches on the Cu$_{10}$. ``None'' means that only a cut and splice pairing has been performed, with no added mutation. ``MD'' is the molecular dynamics move described in sec.~\ref{sec:methods_ga}. The lines inside the boxplots shows the median energy change, the boxplots extend from the lower to the upper quartile of the data, whiskers extend from the 5th to the 95th percentile of the data, circles show data points outside of the whisker range.}
    \label{fig:ga-progress}
\end{figure*}

Each search for each cluster size is repeated 4 times, until 3000 structures have been evaluated in each run, for a total of 12.000 structures. Clusters at each cluster size are then merged into a single dataset before continuing with the analysis performed in Sec.~\ref{sec:garesults}. The same structures are usually found in each run (see fig.~\ref{fig:ga-progress} a), described below), but the independent runs are kept as a consistency check, and we can expect this situation to be different for larger clusters where finding a given putative GM becomes more difficult~\cite{paleico_flexible_2020}.

The GA search described here has been implemented around the already available functions in the ASE (Atomic Simulation Environment) Python library~\cite{larsen_atomic_2017}, version 3.17.0, with some extra functions as required. In particular all of the mutations needed to be either modified (rattle and mirror) or implemented from scratch (the rest) following the library's templates. The GA library is based around a database module, which makes it easy to store and retrieve structures in an organized way and query the results of the GA search. Due to the large amount of structures that can be calculated with the NNP, the database needs to be purged regularly from about 500 stored structures onward keeping the fittest candidates, otherwise the updating process for the current population becomes excessively slow. Energy and force evaluations were provided by the n2p2 neural network library~\cite{singraber_library-based_2019} in combination with LAMMPS~\cite{plimpton_fast_1995}.

To illustrate the course of a typical simulation Figure~\ref{fig:ga-progress} shows the results of running four GA searches for the GM of Cu$_{10}$. In a) we can see how the GA search quickly finds the GM in fewer than 300 structures for all runs. The energy before this point decreases rapidly, and no lower energy structures are found in the following 2700 configurations generated in each run (for a total of 3000 structures per independent run). 

Figure~\ref{fig:ga-progress} b) shows how the different mutations (or no mutation, just pairing) affect the energy of the generated relaxed candidates, for two GA searches for Cu$_{10}$. The energy is expressed relative to the parents' average energy, as given by:

\begin{equation}
    P = 100*\Big(\frac{E_{\rm candidate}}{(E_{\rm parent~1}+E_{\rm parent~2})/2}-1\Big) \label{eq:percentage}
\end{equation}
where $E_{\rm candidate}, E_{\rm parent~1}$ and $E_{\rm parent~2}$ are the energies of the relaxed candidate and its two parents, respectively. If the relaxed candidate has a lower energy than the average of its parents, $P$ is negative, and we can say that the mutation was ``successful'' in progressing the GA search.

Just pairing with no mutation seems to be the most consistent operation, with energy changes clustering around the 0\% mark. Angular mutation moves generate the highest energy configurations, which is usually a consequence of fewer atoms being attached to the surface. Molecular dynamic runs are very effective at generating lower energy configurations, with their data set the only one with a median energy change below zero. They still can generate very high energy configurations due to the occasional formation of separated isolated atoms. Rattle, mirror and twist mutations exhibit a similar behavior between them, with the twist mutation obtaining the largest energy gain. 

%%%%%%%%%%%%%%%%%%%%%%%%%%%%%%%%%%%%%%%%%%%%%%%%%%%%%%%%%%%%%%%%%%%%%%%%%%%%
%%%%%%%%%%%%%%%%%%%%%%%%%%%%%%%%%%%%%%%%%%%%%%%%%%%%%%%%%%%%%%%%%%%%%%%%%%%%
\section{Results} \label{sec:results}
%%%%%%%%%%%%%%%%%%%%%%%%%%%%%%%%%%%%%%%%%%%%%%%%%%%%%%%%%%%%%%%%%%%%%%%%%%%%
%%%%%%%%%%%%%%%%%%%%%%%%%%%%%%%%%%%%%%%%%%%%%%%%%%%%%%%%%%%%%%%%%%%%%%%%%%%%

%%%%%%%%%%%%%%%%%%%%%%%%%%%%%%%%%%%%%%%%%%%%%%%%%%%%%%%%%%%%%%%%%%%%%%%%%%%%
%%%%%%%%%%%%%%%%%%%%%%%%%%%%%%%%%%%%%%%%%%%%%%%%%%%%%%%%%%%%%%%%%%%%%%%%%%%%
\subsection{Structure of the ZnO Support} \label{sec:znosupport}
%%%%%%%%%%%%%%%%%%%%%%%%%%%%%%%%%%%%%%%%%%%%%%%%%%%%%%%%%%%%%%%%%%%%%%%%%%%%
%%%%%%%%%%%%%%%%%%%%%%%%%%%%%%%%%%%%%%%%%%%%%%%%%%%%%%%%%%%%%%%%%%%%%%%%%%%%

\begin{figure*}
    \centering
    \includegraphics[width=0.95\linewidth]{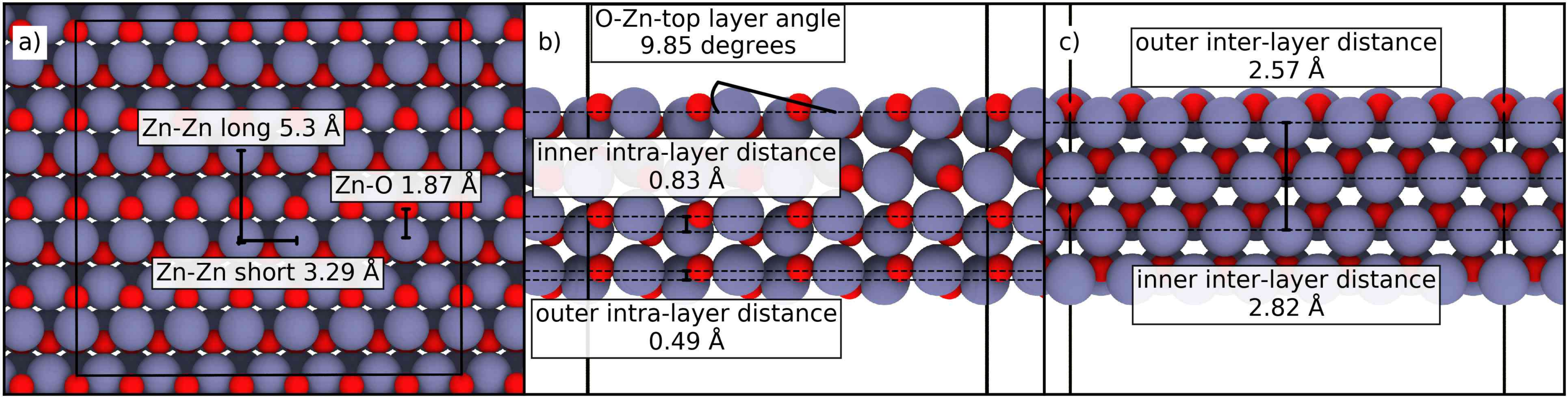}
    \caption{Structure of the ZnO(10$\bar{1}$0) support, indicating important distances and angles. a) Top view.  b) Side view normal to the [11$\bar{2}$0] direction. c) Side view normal to the [0001] direction.}
    \label{fig:zno-structure} 
\end{figure*}

The structure of the support is very important for the obtained optimized clusters. For this reason, we have carefully checked the description of the bulk material as well as of the relaxed geometry of the ZnO surface provided by the NNP. The optimized lattice parameters of the bulk ZnO wurtzite structure we obtained are presented in Table~\ref{tab:zno}. They are basically indistinguishable from the reference DFT results, and are only 2\% above the experimental values\cite{lide_david_r_handbook_2009}, which can be ascribed to the employed exchange correlation functional. 

\begin{table}[]
    \centering
    \begin{tabular}{|c|c|c|c|}
        \hline
         Property & NNP & DFT & Exp.~\cite{lide_david_r_handbook_2009} \\
         \hline
        a (\r{A})       & 3.29 & 3.29 & 3.25 \\
        c (\r{A})       & 5.30 & 5.30 & 5.21\\
        u (frac. dist.) & 0.385 & 0.385 & 0.382\\
         \hline
    \end{tabular}
    \caption{Lattice parameters of the ZnO wurtzite structure as predicted with the NNP, the DFT reference method, and experimental values.}
    \label{tab:zno}
\end{table}

For the  ZnO(10$\bar{1}$0) surface that we use in this work, Figure \ref{fig:zno-structure} shows all the relevant angles and distances as obtained from a geometry relaxation of a 4 layer, $(4\times 7)$ ZnO(10$\bar{1}$0) slab. We observe the known geometry changes due to relaxation from the introduction of vacuum interfaces, with layer distances shortening in particular close to the surface, and tilting of the ZnO dimers to an angle of 9.85 degrees. These values are very similar to the ones obtained from DFT calculations, and to previous results reported in the literature~\cite{meyer_density-functional_2003}.

%%%%%%%%%%%%%%%%%%%%%%%%%%%%%%%%%%%%%%%%%%%%%%%%%%%%%%%%%%%%%%%%%%%%%%%%%%%%
%%%%%%%%%%%%%%%%%%%%%%%%%%%%%%%%%%%%%%%%%%%%%%%%%%%%%%%%%%%%%%%%%%%%%%%%%%%%
\subsection{Global Optimization Results} \label{sec:garesults}
%%%%%%%%%%%%%%%%%%%%%%%%%%%%%%%%%%%%%%%%%%%%%%%%%%%%%%%%%%%%%%%%%%%%%%%%%%%%
%%%%%%%%%%%%%%%%%%%%%%%%%%%%%%%%%%%%%%%%%%%%%%%%%%%%%%%%%%%%%%%%%%%%%%%%%%%%
\begin{figure*}
    \centering
    \includegraphics[width=\linewidth]{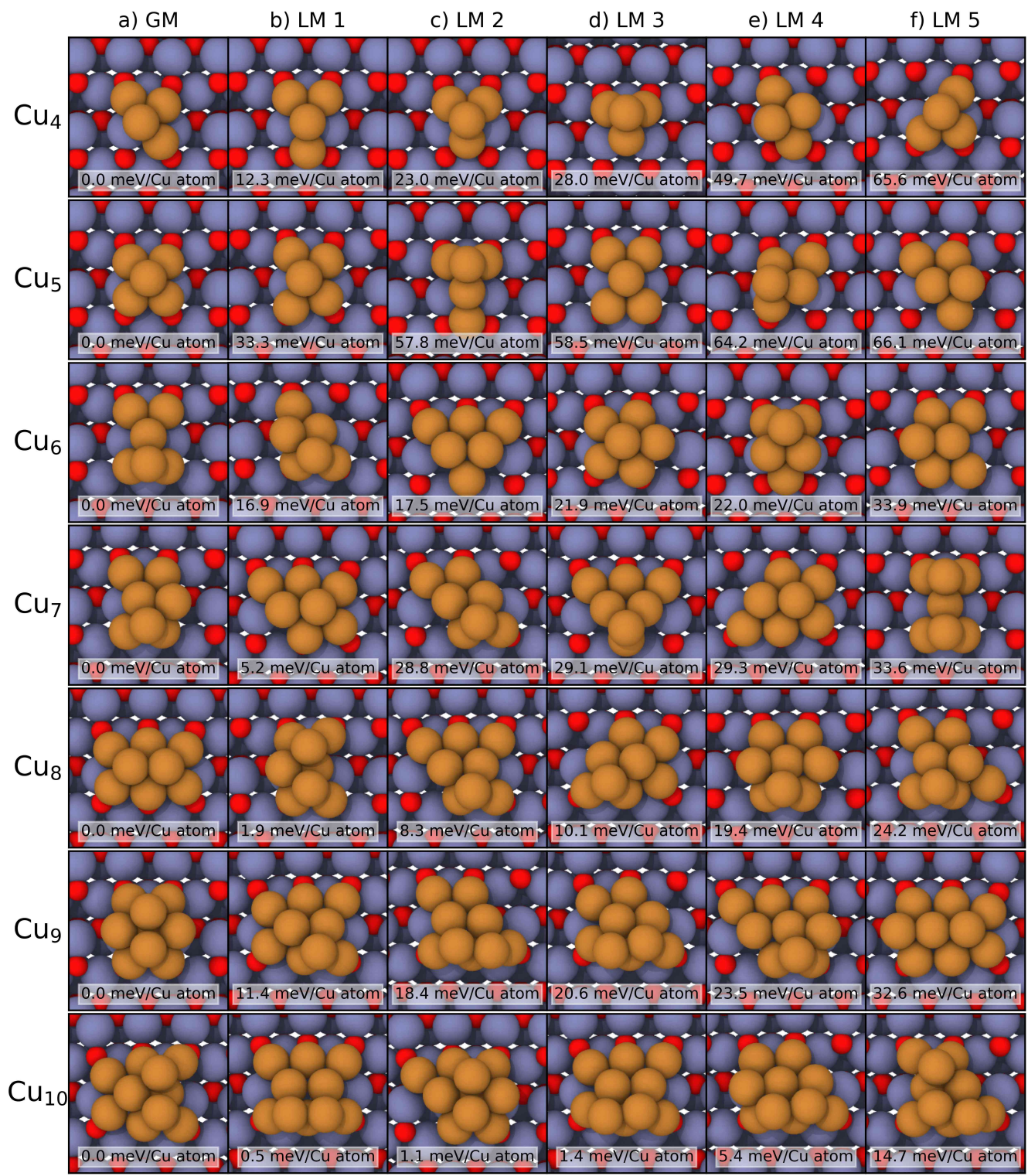}
    \caption{Global minimum (GM) and first five local minima (LM) for different copper cluster sizes from GA search on the ZnO(10$\bar{1}$0) surface. Each row corresponds to a cluster size with a number of atoms as indicated on the left, each column corresponds to the ranking within a size. For each structure the relative energy per copper atom with respect to the GM is given.}
    \label{fig:gaminima410}
\end{figure*}

\begin{figure*}
    \centering
    \includegraphics[width=\linewidth]{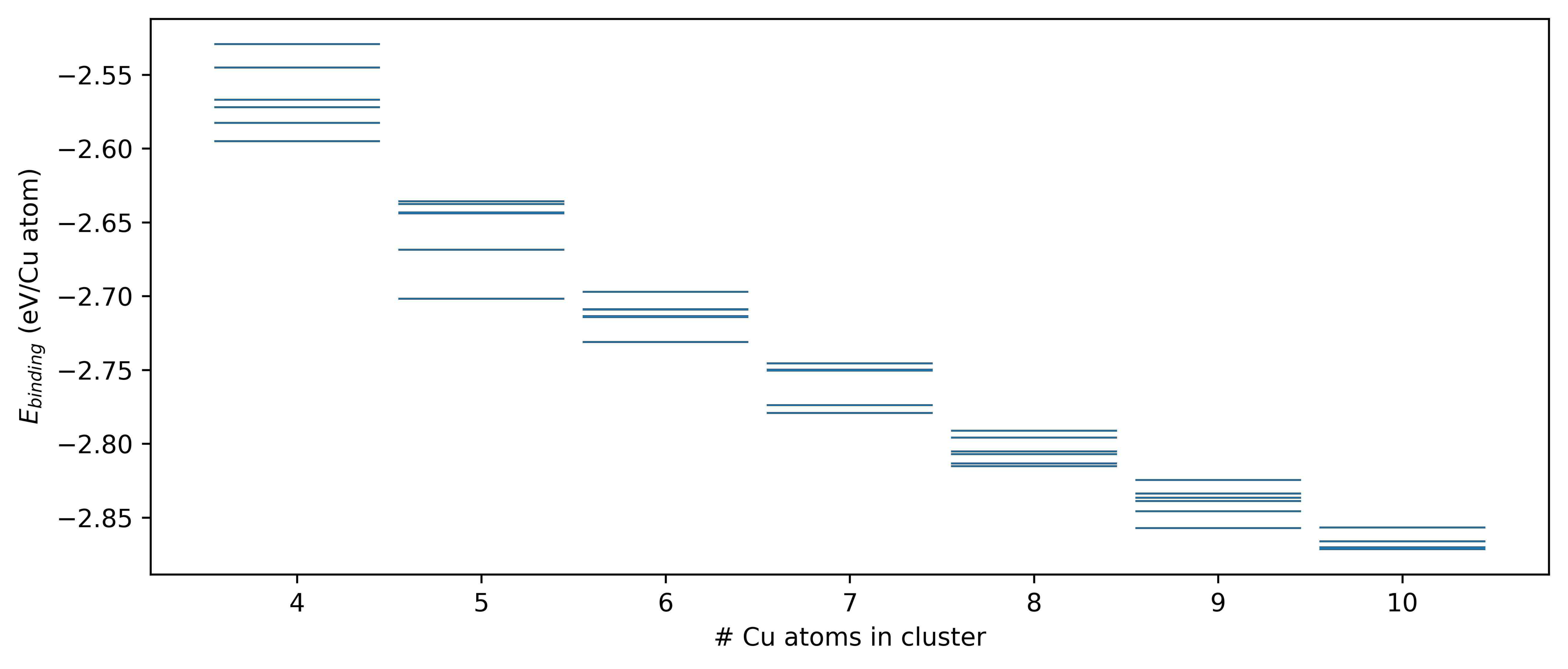}
    \caption{Binding energy per copper atom (see Eq.~\ref{eq:ebind}) for the GM and first 5 LM for the Cu$_{4-10}$ clusters.}
    \label{fig:ebinding}
\end{figure*}

In the next step, we have employed GA simulations to identify the GM and energetically favorable LM structures of a series of cluster sizes at the  ZnO(10$\bar{1}$0) surface.
Figure~\ref{fig:gaminima410} shows the results of running the GA algorithm for Cu clusters between 4 and 10 atoms. Four separate runs have been carried out per system yielding 3000 structures each, and then the results of the runs for each cluster size have been combined and analyzed. Before doing this, as shown in Fig.~\ref{fig:ga-progress} it was confirmed that different repeated runs tend to find the same GM although; since the process is stochastic, this is not necessarily always the case~\cite{vilhelmsen_genetic_2014}. Still, most of the minima have been found multiple times, within a single run and also across runs. 

We find a rich variety of structural patterns. First, we note that it is possible to detect a number of ``geometric families'' that are consistently present in the best ranked clusters. One important general trend is that Cu atoms tend to interact strongly with the oxygen atoms of the surface, preferring a direct adsorption, or the adsorption between the Zn and O of a ZnO dimer. Further, they can also be located between two surface oxygen atoms, like the Cu atom at the bottom of the image of LM 2 for Cu$_5$ in Fig. \ref{fig:gaminima410}.

For Cu$_4$, we identify the simplest structural patterns, and here we can derive our second important principle: Even for small sizes the clusters already adopt a 3D shape, and do not lie flat on the surface. The same is observed for all other cluster sizes, with only a few rather flat configurations being found among the lowest energy clusters although they do appear at higher energies. Additionally, if somewhat trivial, the Cu atoms do indeed form clusters and do not spread across the support into individual adsorption sites. This clustering is well known from experimental results even at low coverages~\cite{dulub_stm_2002}, and can be predicted from the relative binding and surface energies of the materials involved~\cite{henry_morphology_2005}. Some separated clusters are observed in simulations of the larger clusters as a consequence of certain mutation moves, but these always correspond to high energy configurations. Finally, we notice that neither here nor for any other size do the clusters form extended periodic structures in one or two dimensions, as has been proposed in other studies~\cite{mora-fonz_development_2017}. These kinds of structures were only ever observed in our simulations for very small periodic cells, which prevent the formation of non-interacting clusters due to rather high copper coverages.

Due to the small number of atoms in the Cu$_4$ cluster, it appears to form only two main structural patterns at low energies. The first pattern is a truncated square pyramid with only 3 atoms present on the support, which is found in the GM in Fig.~\ref{fig:gaminima410} and in LM 5. Notice that the GM and LM 5 are in fact very different with respect to the support, as the adsorption footprint on the ZnO(10$\bar{1}$0) surface is not symmetric because of the orientation of the ZnO dimers corresponding to the polar (0001) and (000$\bar{1}$) surfaces. In this direction, we have a short Zn-O distance within the dimers, and a long distance between different dimers. For both structures, the GM and LM 5, the three copper atoms attached to the surface are in the vicinity of oxygen atoms. However, in the GM two of them are in between two surface ZnO dimers and only one is found on top of a dimer, while in LM 5 the opposite is observed. This indicates a preferential direction, since rotating the cluster by 180$^{\circ}$ does not result in the same support environment. 

The second pattern for the Cu$_4$ clusters found in LM 1, LM 2 and LM 3 corresponds to small, more or less distorted tetrahedra. Perfectly stacked structures like LM 2 and 3 cannot easily be realized for bigger clusters, but are sometimes observed for specific cluster sizes at higher energies. Notice that also these structures exhibit the preferential directionality mentioned in the previous paragraph, and are elongated to some extent in the ZnO dimer direction. Due to their structural similarity, we cannot fully exclude that these structures are connected by only small energy barriers or even correspond to the same shallow minimum, which might also be dependent on the employed exchange correlation functional or subtleties of the NNP, and further investigations using e.g. DFT-based nudged elastic band calculations~\cite{P0951} might be used to finally clarify this question. However, such calculations would be very expensive to carry out with DFT in our simulation setup. 

Finally, we notice that all the Cu$_4$ clusters are seemingly contained within a $(1\times 1)$ surface cell. The clusters do not form flat 1D or 2D structures, but rather gain height in the $z$ direction while remaining confined within this box.

For Cu$_5$, many of the structures are natural extensions of the Cu$_4$ structures. In particular, the GM is the completed square pyramid with the missing atom added in one corner, while LM 1 is a distortion of this structure. LM 2 looks like a extended tetrahedron emerging from LM 1 of Cu$_4$, and shows a preferred orientation. In LM 3 the distortion of LM 1 extends to both sides of the pyramid base making it very likely that this distortion represents true local minima in LM 1 and LM 3. The local minima LM 4 and LM 5 resemble distorted structures emerging from the GM of Cu$_4$ by the attachment of an additional copper atom.

For the Cu$_6$ clusters we notice a number of interesting developments. The GM now corresponds to the structure of LM 3 of Cu$_5$ with an atom attached. We will call this pattern a saddle. For LM 3 we observe a first cluster with distorted five-fold symmetry, and a number of rather flat clusters occur that follow the corrugation of the surface in the cases of LM 2 and LM 5. More interestingly, LM 2 is the first cluster to extend beyond the $(1\times 1)$ surface cell, which will become the norm for larger clusters. This new extension is larger in the ``short'' direction of the surface, perpendicular to the direction of the Zn-O dimers. This is reasonable, since this distance is shorter and thus easier to ``bridge over'' as the cluster grows.

For Cu$_7$, many of the clusters appear to be directly related to those at Cu$_6$. The new GM is a direct evolution of the previous saddle-like GM of Cu$_6$, with another related structure at LM 5. LM 1 and 4 are related to the five-fold symmetry cluster at the LM 3 of Cu$_6$, and LM 3 is similar to the previously flat triangular cluster. Of these clusters, only 2 remain within the $(1 \times 1)$ cell, which are interestingly the lowest and the highest energy structure, with the rest increasing in footprint.

For Cu$_8$, the GM is a moderately distorted version of the cluster at the LM 1 of Cu$_7$, with an extra atom added in the lower left corner of the image, but it can also be thought of being derived from LM 4 thus unifying two rather stable sub-patterns. For  LM 1 and 5 we notice new modifications of the saddle, a new five-fold derived cluster emerges in LM 3, and a first cluster with a buckled hexagonal base pattern plus an attached atom at LM 4.

For Cu$_9$, the saddle-based pattern once again becomes the GM. Additionally, at this size all the shown clusters are for the first time outside the $(1\times 1)$ cell.

Finally, for Cu$_{10}$, by attaching an additional atom the previous LM 1 of Cu$_9$ becomes the GM, with the saddle configuration falling to LM 2. However, it must be noted that in particular for Cu$_{10}$ the energy differences are extremely small, and cannot be resolved with full confidence, neither by the NNP nor by the underlying DFT calculations.

These analyses of the structural and energetic trends leads us to a general conclusion: the shapes of these clusters are a compromise between the interaction between copper and the support, which would tend to favor spread out, flat clusters, and the cohesive energy of the copper clusters, which  tends to favor three-dimensional clusters. Additionally, the granular nature, corrugation and symmetry of the support, with preferred adsorption sites, and the strain that it induces on the clusters, complicates this interaction. This is a known effect that can be observed even with simple potentials~\cite{eckhoff_structure_2017}. In Fig. \ref{fig:ebinding} we plot the binding energy of the GM structures as a function of the cluster size. We define the binding energy as:
\begin{align}
E_{\rm binding}(N) &= E_{\rm cluster~+~slab} + \nonumber \\
                &- E_{\rm ZnO~slab~clean~relaxed} - N \cdot E_{\rm Cu~atom} \label{eq:ebind}
\end{align}
where $N$ is the number of Cu atoms in the cluster under consideration, $E_{\rm cluster + slab}$ is the energy of the whole system, including cluster and slab; $E_{\rm ZnO~slab~clean~relaxed}$ is the energy of the initial supporting slab, relaxed and without cluster atoms, and $E_{\rm Cu~atom}$ is the energy of an isolated Cu atom as obtained from DFT calculations, since the NNP has not been trained with this information. All other energies can be obtained from NNP calculations. 
%\MLP{For comparison, the cohesive energy for fcc Cu is 3.66 eV/atom, calculated both with DFT and with the NNP.} 
Notice that since the only term that changes between clusters is $E_{\rm cluster + slab}$, this plot also shows the relative energies between all the clusters. As expected, we observe a stronger binding (the binding energy becomes more negative) with increasing number of copper atoms. Additionally, the spread in energies within a cluster size becomes smaller as the clusters become larger.

%%%%%%%%%%%%%%%%%%%%%%%%%%%%%%%%%%%%%%%%%%%%%%%%%%%%%%%%%%%%%%%%%%%%%%%%%%%%
%%%%%%%%%%%%%%%%%%%%%%%%%%%%%%%%%%%%%%%%%%%%%%%%%%%%%%%%%%%%%%%%%%%%%%%%%%%%
\subsection{Interface Structure} \label{sec:interface}
%%%%%%%%%%%%%%%%%%%%%%%%%%%%%%%%%%%%%%%%%%%%%%%%%%%%%%%%%%%%%%%%%%%%%%%%%%%%
%%%%%%%%%%%%%%%%%%%%%%%%%%%%%%%%%%%%%%%%%%%%%%%%%%%%%%%%%%%%%%%%%%%%%%%%%%%%
\begin{figure*}
    \centering
    \includegraphics[width=\linewidth]{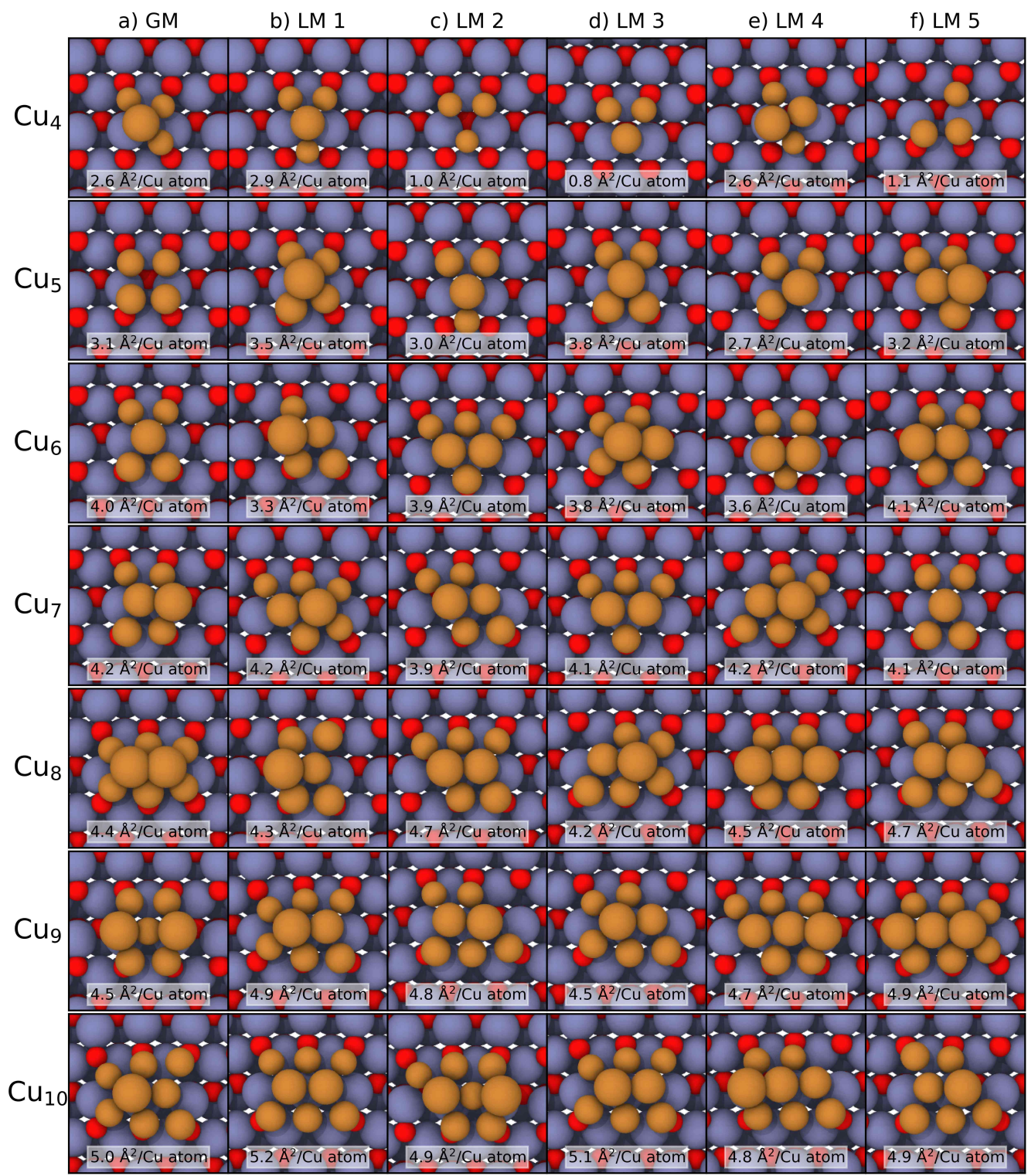}
    \caption{Interfacial Cu atoms of the optimized clusters in Fig.~\ref{fig:gaminima410}, with the number indicating the interface area per cluster atom as calculated with the convex hull algorithm. We have defined as interface atoms all those atoms belonging to the cluster that are within 1.6~\r{A} vertical distance of the lowermost atom of the cluster, enough to encompass up to one monolayer of Cu (110) plus a margin of error. The radius for Cu atoms has been scaled by distance to the surface (larger as the distance increases) for visualization purposes.}
    \label{fig:gaminima410interface}
\end{figure*}

%\begin{figure*}
%    \centering
%    \includegraphics[width=12cm]{figures/clusters-interface-marked.pdf}
%    \caption{View of the GM for Cu$_{10}$. a) Distance from the bridge position between two Zn atoms (approx. the end of a ZnO dimer) to the middle of the next dimer; this falls between the long distances of the Cu (111) and (110) surfaces.\JBC{I am not yet happy. We should not mention the term "end of a dimer" at all, because it is not existing, and also if a middle of a dimer exist, I would search for the middle of the bond, which you do not use here? Why don't you use the distance between one O and the middle of the ZnO bond, this is your calculation in the main text anyway} b, c, d) Distances between opposed pairs of Cu atoms. e, f) Projection of the distance between an edge and a central atom (defined in text). This value is 0.0 for a perfect (111) surface, and 0.5 for an unrelaxed (110) surface. g) Distance between two ZnO dimers on the surface, this distance is too long for any of the Cu surfaces under consideration due to the lattice mismatch.}
%    \label{fig:interface-cu10}
%\end{figure*}

An interesting question is what local structure the clusters adopt at the contact with the interface. We can expect that for very large clusters, far away from the influence of the support, the central atoms will organize in the geometry of fcc bulk copper. However, right at the interface, a number of opposing trends influence the structure. First, it is known that finite solids such as nanoparticles tend to adopt shapes dominated by the need to minimize the surface energy, as described by Wulff~\cite{wulff_zur_1901}. This, in principle, applies to large structures where surface and bulk energetic terms dominate, that are in equilibrium, with only one element, and more importantly, considering solid-vacuum interfaces, although attempts have been made to extend Wulff's theory to interfacing materials~\cite{henry_morphology_2005} and alloys~\cite{ringe_wulff_2011}. Since here we are addressing very small clusters, which in addition are strongly influenced by the ZnO interface, we have to take into consideration its effect. The degree of interaction between the cluster atoms and the support can lead to new preferred geometries. This can be due to low-energy interaction sites dictating a given geometry, and specific surface structures favoring these interactions. This can be interpreted in various forms, such as strong metal-support interaction\cite{tauster_strong_1987, pan_tuning_2017, figueiredo_understanding_2019}, or the theory of lattice mismatch~\cite{eckhoff_structure_2017, koda_coincidence_2016}. 

The Cu-ZnO system presents such an inherent lattice mismatch between the two materials. If we take one low index surface from each material, it is not possible to produce a coincident supercell without straining, i.e., expanding, contracting, or even reconstructing, at least one of the two subsystems. Such lattice mismatches can induce defects and strain in structures, which are known to enhance catalytic properties~\cite{mavrikakis_effect_1998, amakawa_how_2013, bissett_enhanced_2013}, and it is still present to some extent for finite structures such as clusters. It is partially reduced by losing the requirement for a periodic structure, which allows for lateral changes of the interatomic distances, but the atoms in the interface area are still guided by the underlying geometry of the support to adopt regular positions. 

We note that the clusters studied in this work are very small and thus regular surface facets corresponding to the low-Miller index surfaces in a Wulff-like picture cannot be expected, but still small structural features geometrically resembling these surfaces may emerge, which we will now investigate. The interfacial atoms of the minimized clusters of Fig.~\ref{fig:gaminima410} are shown in Fig.~\ref{fig:gaminima410interface}. For this purpose we define interfacial atoms as those within a 1.6~\r{A} vertical distance of the lowermost Cu atom of the respective cluster. This interval was chosen as it is thick enough to potentially contain a monolayer of all the low index Cu surfaces. The ideal (100) and (111) monolayers are flat, but the (110) monolayer has a thickness of about 1.275~\r{A}, i.e., one-half of the nearest neighbor distance, $d_0$. The distance criterion has been chosen about 0.3~\r{A} larger than this value to accommodate also distorted positions of Cu atoms on top of the support, with vertical distances falling in the range of 1.3 to 1.6~\r{A}. Additionally, those atoms further away from the substrate are displayed in a perspective view with larger radii, providing a sense of the 3D structure of these interface atoms.
Identifying surface features at the interface in such a heterogeneous system is not trivial, but we attempt an analysis by a comparison with the well-known low Miller index surfaces of copper, (111), (100) and (110), which are known to be the most stable ones~\cite{tyson_surface_1977} for the material in vacuum, in the given order~\cite{tran_surface_2016}. 

For Cu$_4$ and Cu$_5$ we observe a variety of patterns, including some triangular configurations. The GM of Cu$_5$ is reminiscent of the (100) surface, organized in approximately square units of Cu atoms. This surface, which is the second most stable out of the three low index cuts in vacuum, only appears for the smallest clusters.
In LM 3 of Cu$_5$ we observe the first example of a structural pattern that repeats for many of the larger clusters: five interfacial atoms, structured as a rectangle with an atom in the center that is slightly above the plane formed by the other atoms. Both, the buckled (110) and the planar (111) surfaces of fcc crystals are somewhat related. The (111) surface is usually described as a planar layer of hexagonally arranged atoms, but can also be reinterpreted as a rectangle plus a central atom-in-plane pattern. Its hexagons are completely symmetrical with equivalent bonds. The rectangle has two characteristic distances: a short one equal to $d_0$, the nearest neighbor distance of 2.55~\r{A}, and a long one equal to $\sqrt{3} a_0$, the lattice constant of bulk copper, yielding 4.42~\r{A}.
The (110) surface is usually interpreted as the rectangle plus central atom pattern, with the central atom belonging to a higher or deeper layer resulting in a buckled and rather open surface, but it can actually also be reinterpreted as asymmetrical tilted hexagons, i.e., they are longer in one dimension than the other one. The distances in the rectangle consist of:  a short one, which is again equal to $d_0$, 2.55~\r{A}, and a long one equal to $a_0$, 3.64~\r{A}.
This means that actually the (111) and (110) surfaces are in fact quite similar, with the key difference being whether the hexagonal pattern is flat or buckled and the related consequences for the interatomic distances. We can smoothly convert from one surface to the other by flattening the (110) configuration, which causes the other Cu atoms to move away and increases the length of the long side of the rectangle.

As can be seen in Fig.~\ref{fig:gaminima410interface}, at the interface we obtain a mixture of both behaviors, with clusters showing a structural intermediate of both surfaces. Some interfaces are flatter with larger distances in the long direction (closer to (111)), while in others we observe the opposite configuration (closer to (110)). In fact, this can vary within one cluster size or even one structure, such as the GM at Cu$_{10}$, which exhibits two parts with the two different extremes. In effect we observe structures that are in a continuum between the two extremes of Cu (110) and (111), which is a consequence of the cluster shape at the interface adapting to the ZnO surface geometry, while still subtle distortions resulting from the Cu-Cu interactions are present. Due to the small size of the clusters they are rather flexible in shape and none of the atoms has the full copper environment of the bare metal surfaces in vacuum. This structural behavior is expected to change with an increasing number of copper atoms, finally resulting in copper interface structures more similar to one of the ideal copper surfaces.

\subsection{Properties} \label{sec:clusterprops}
%%%%%%%%%%%%%%%%%%%%%%%%%%%%%%%%%%%%%%%%%%%%%%%%%%%%%%%%%%%%%%%%%%%%%%%%%%%%
%%%%%%%%%%%%%%%%%%%%%%%%%%%%%%%%%%%%%%%%%%%%%%%%%%%%%%%%%%%%%%%%%%%%%%%%%%%%

\begin{figure*}
    \centering
    \includegraphics[width=\linewidth]{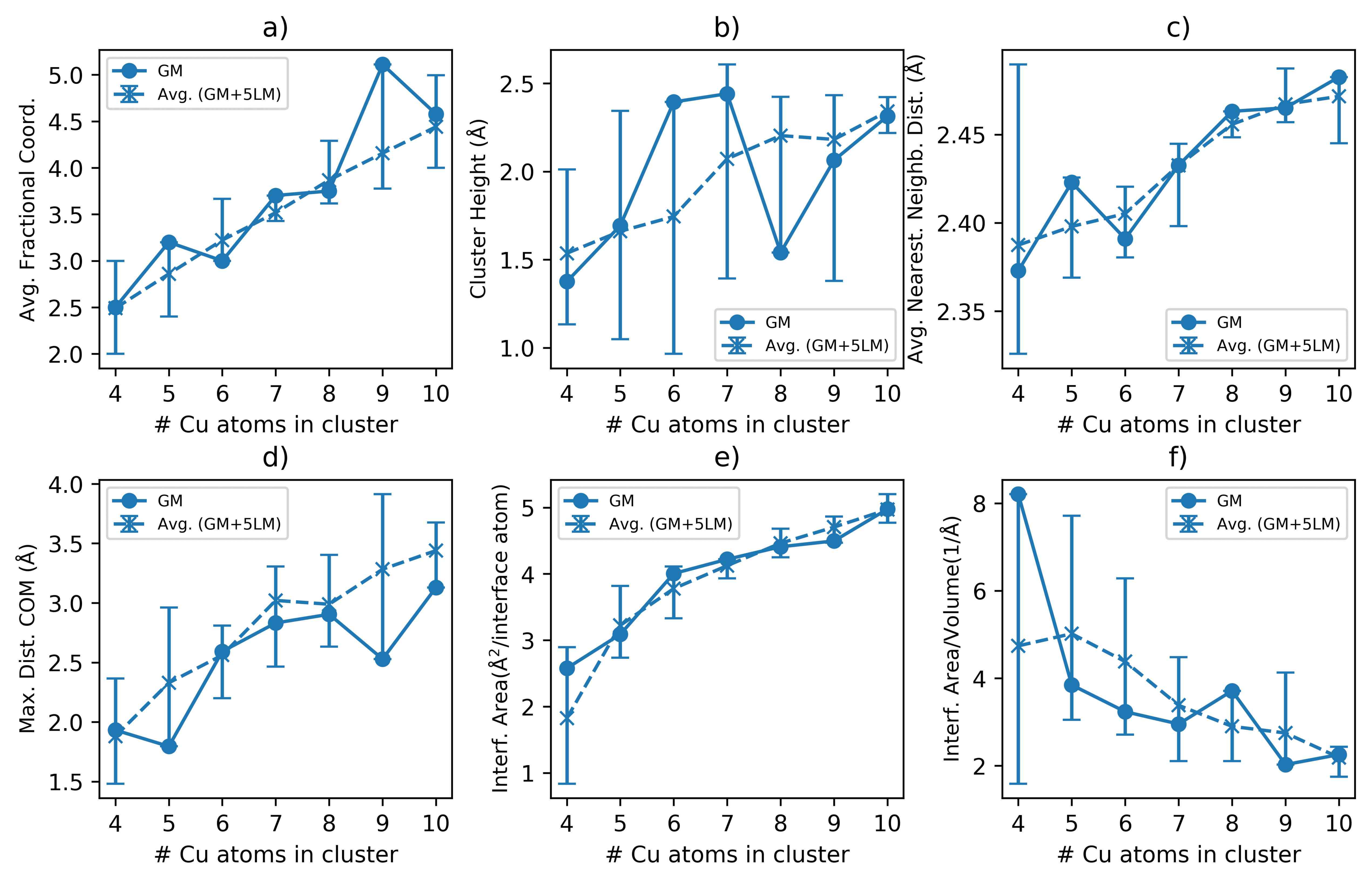}
    \caption{Various structural and geometrical properties of the clusters in Fig.~\ref{fig:gaminima410} versus cluster size. Each plot contains two curves: the full line shows the values of the given property for the GM, the dashed line shows the average for the GM plus the 5 LM, with vertical bars indicating the minimum and maximum values within the group.}
    \label{fig:cluster410properties}
\end{figure*}

\begin{figure*}
    \centering
    \includegraphics[width=\linewidth]{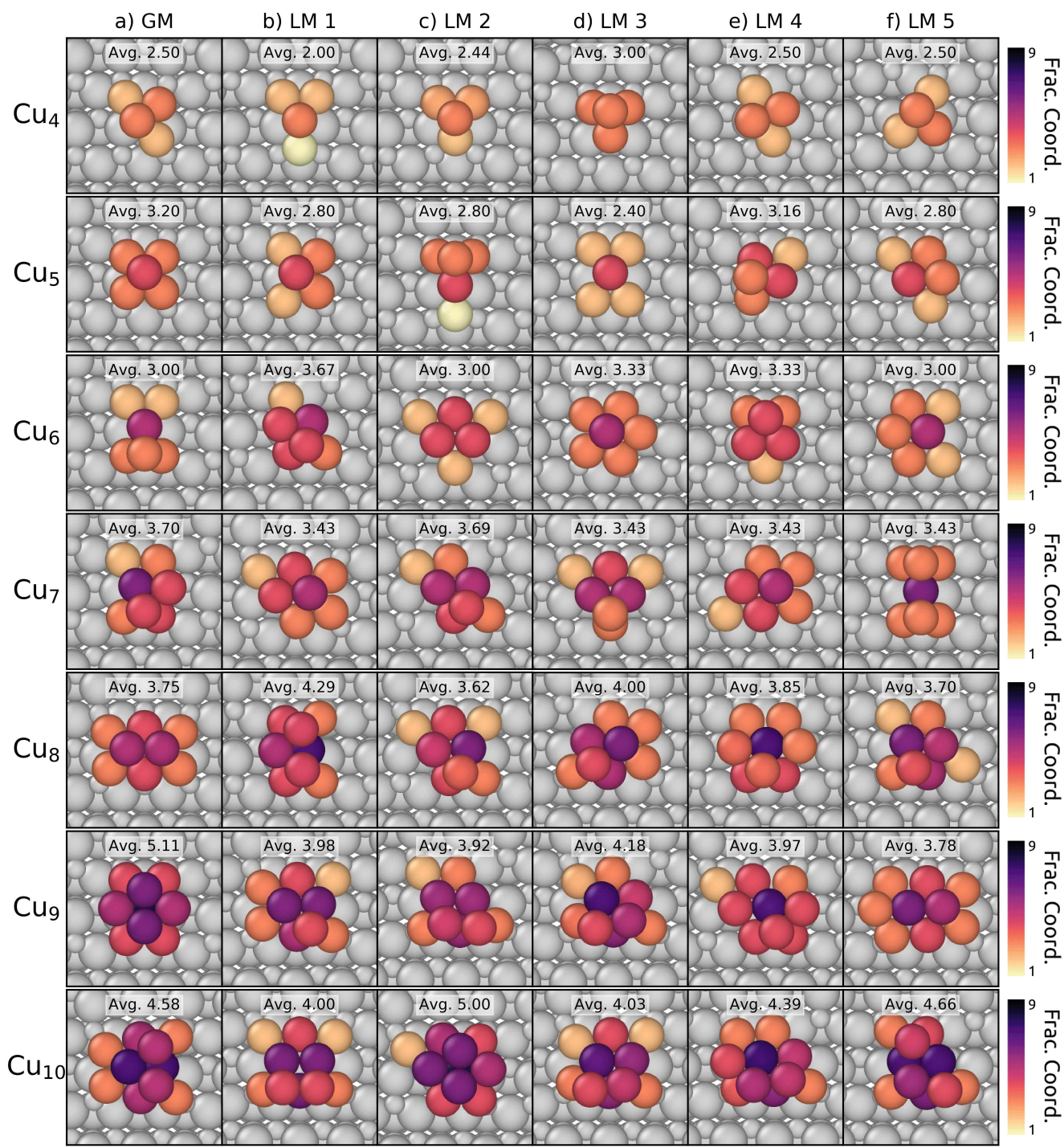}
    \caption{Minima from Fig.~\ref{fig:gaminima410} colored by their fractional coordination number. Average coordination numbers are given as insets.}
    \label{fig:gaminima410fraccoord}
\end{figure*}

\begin{figure*}
    \centering
    \includegraphics[width=\linewidth]{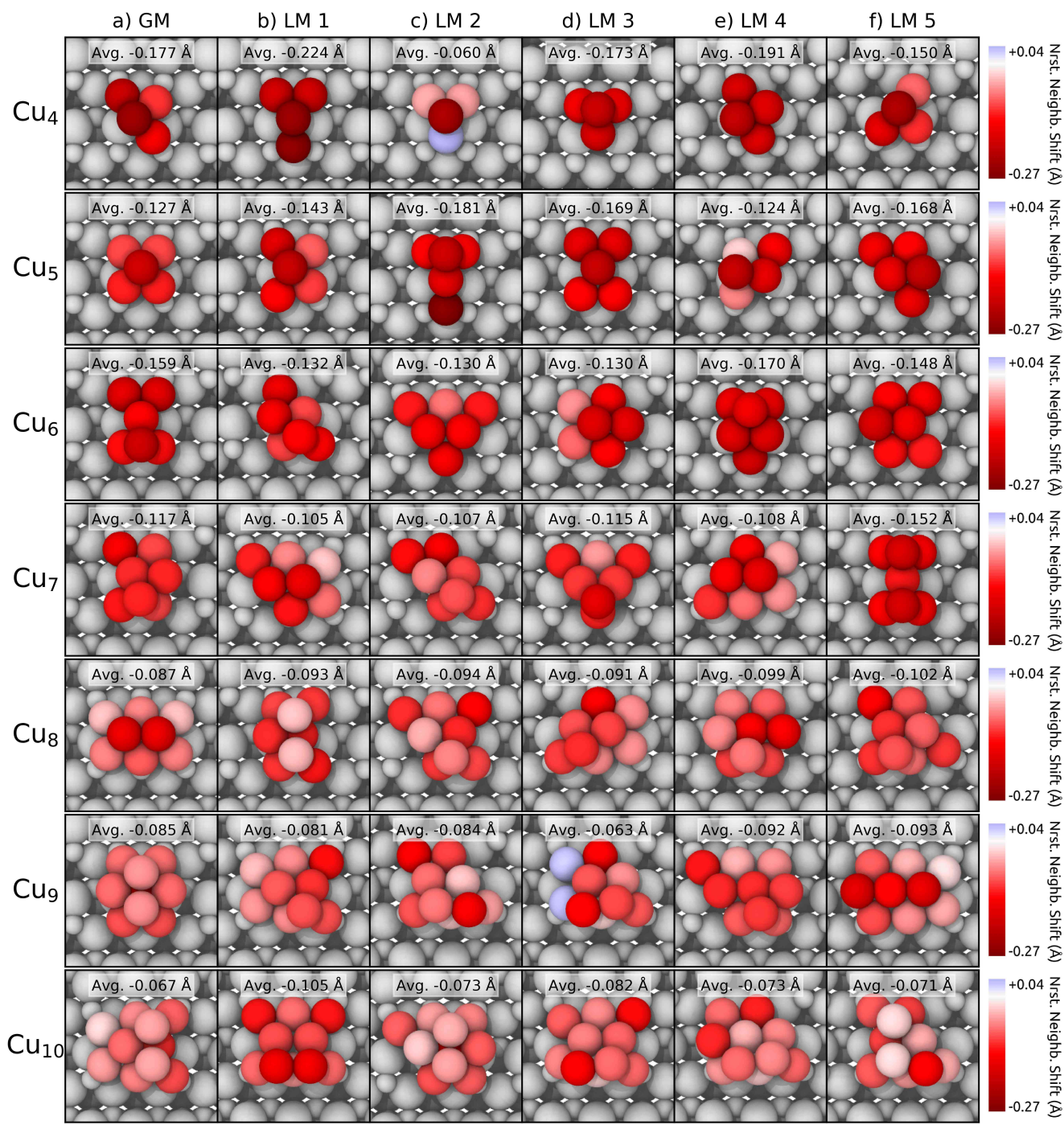}
    \caption{Minima from Fig.~\ref{fig:gaminima410} colored by their shift from the bulk Cu nearest neighbor distance (2.55~\r{A}), with average shift indicated in the insets. Red indicates atoms that have a negative shift from this distance, that is, atoms that are closer together than in the bulk. Blue (only visible in Cu$_4$ c and Cu$_9$ d) indicates the opposite. White indicates no shift.}
    \label{fig:gaminima410nnshift}
\end{figure*}

In the next step, we have analyzed a number of geometric properties of the obtained clusters to identify possible structural trends. Specifically, we have looked at: 

\begin{itemize}
    \item The fractional coordination number~\cite{iannuzzi_efficient_2003} of the atoms in the cluster. The fractional coordination number converts the usual integer-valued coordination number into a continuous value. For an atom within a certain distance of the central atom, like $r_0 = 2.55$~\r{A}, the nearest neighbor distance in bulk copper, the coordination number contribution is 1. For atoms beyond, smaller values which rapidly decrease with increasing distance from the coordinated atom are used. In effect atoms beyond the nearest neighbor distance are considered as partially coordinating. Here, this avoids rapidly changing coordination numbers due to cluster atoms being slightly beyond the nearest neighbor distance, which frequently occurs for small metal clusters.
    \item The height of the cluster, as given by the difference in $z$ coordinates of the top and bottom-most atoms in the cluster.
    \item The maximum distance of any atom to the center of mass (COM) of the cluster. This shows the maximum extent of the cluster.
    \item The average nearest neighbor distances, for each atom and for the whole cluster, with neighbors in the cluster up to 3.0~\r{A} away.
    \item The volume of the cluster as obtained from the QHull convex hull algorithm library~\cite{barber_quickhull_1996}. The convex hull algorithm returns the smallest convex polyhedron that contains a collection of points, i.e., the position of the cluster atoms. From this one can derive the  properties of a polyhedron such as volume and surface area, but also other properties, for instance if an arbitrary point is inside or outside the constructed polyhedron.
    \item The area of the atoms at the interface, also using QHull. Interface atoms are chosen following the definition in Sec.~\ref{sec:interface}, i.e., within 1.6~\r{A} vertical distance of the lowest cluster atom.
\end{itemize}

Some of these properties may have catalytic consequences, such as the coordination number of the cluster atoms, while others characterize the overall geometry of the clusters. We have plotted the results in Fig.~\ref{fig:cluster410properties}, showing the values for the GM, and the average value for the GM plus the first 5 LM, with the bars indicating the minimum and maximum values of the property within this group. 

In Fig.~\ref{fig:cluster410properties} a) we plot the fractional coordination number, and we can observe that this quantity is not yet saturated for clusters of size 10, since we are still in a rather small size regime that does not allow to approach the fcc bulk value of 12. The cluster height in b) does not show any clear trend for the different GMs, and there is quite a spread within each class, but the average shows a steadily increasing value in spite of the diversity of the cluster geometries. As expected, the average nearest neighbor distance in c) slowly approaches the equilibrium value (2.55~\r{A} in fcc Cu) from below, i.e., most cluster atoms are closer than in a bulk environment. This is related to the pattern imposed by the ZnO support as explained in the previous section but also a general property of small metal clusters, as the remaining bonds in undercoordinated atoms tend to be stronger and shorter.

Fig.~\ref{fig:cluster410properties} d) shows that the clusters slowly increase in radius, and is the only property where the GMs show a common behavior in that all of them are at the average or below indicating a more compact shape. The plot of interface area with the support in e) shows a rather slow growth past 6-7 atoms, which implies that the extra atoms coming into the clusters are not used to substantially increase the footprint of the cluster on the support. This might be related to the fact that the cluster needs enough atoms to bridge the sometimes significant gaps between oxygen adsorption sites on the surface. Finally, f) shows the interface area, i.e., footprint on the support, to volume ratio of the cluster, which supports the analysis of e). Note that most of the GM lie at the lower end of this range, that is, they tend to optimize their surface to volume ratio, but this trend may not necessarily extrapolate to larger clusters.

Unfortunately, it does not seem possible to easily differentiate the GMs by utilizing just the calculated properties, although some general trends can be discerned, such as GM usually being more compact (Fig.~\ref{fig:cluster410properties} d), and with a minimized interface surface to volume ratio (see Fig.~\ref{fig:cluster410properties} f). We ascribe this to the overall very small cluster size in this study.

Finally, we have also colored the clusters by those properties that can be expressed on a per-atom basis, as seen in Figs.~\ref{fig:gaminima410fraccoord} for the fractional coordination number and \ref{fig:gaminima410nnshift} for the shift from the fcc copper nearest neighbor distance. As expected from the property plots, we observe a gradual increase in the coordination numbers with cluster size. We observe the first fully surrounded/coordinated Cu atom with 8 other nearest neighbor Cu atoms at the center of the GM of Cu$_9$, (see Fig.~\ref{fig:gaminima410fraccoord} Cu$_9$ a). This is the first size that can present a full coordination sphere, and this happens to be the putative GM at this size. The only other fully coordinated cluster appears as a LM for Cu$_{10}$. Although it would be expected for fully coordinated structures to be energetically favored, this does not seem to be the case at these small cluster sizes.

%We only start observing Cu atoms that are completely surrounded by a layer of six first nearest neighbors starting with the GM of Cu$_9$ (see Fig.~\ref{fig:gaminima410fraccoord} Cu$_9$ a).

In Fig.~\ref{fig:gaminima410nnshift} we similarly observe a slow approach to the expected nearest neighbor distance of 2.55~\r{A}. Notice that only two clusters present positive deviations from this nearest neighbor distance (i.e., a value \textit{above} 2.55~\r{A}), LM 2 of Cu$_4$ and LM 3 of Cu$_9$.

%%%%%%%%%%%%%%%%%%%%%%%%%%%%%%%%%%%%%%%%%%%%%%%%%%%%%%%%%%%%%%%%%%%%%%%%%%%%
%%%%%%%%%%%%%%%%%%%%%%%%%%%%%%%%%%%%%%%%%%%%%%%%%%%%%%%%%%%%%%%%%%%%%%%%%%%%
\subsection{Frozen Surface Optimization} \label{sec:frozensurface}
%%%%%%%%%%%%%%%%%%%%%%%%%%%%%%%%%%%%%%%%%%%%%%%%%%%%%%%%%%%%%%%%%%%%%%%%%%%%
%%%%%%%%%%%%%%%%%%%%%%%%%%%%%%%%%%%%%%%%%%%%%%%%%%%%%%%%%%%%%%%%%%%%%%%%%%%%

\begin{figure*}
    \centering
    \includegraphics[width=\linewidth]{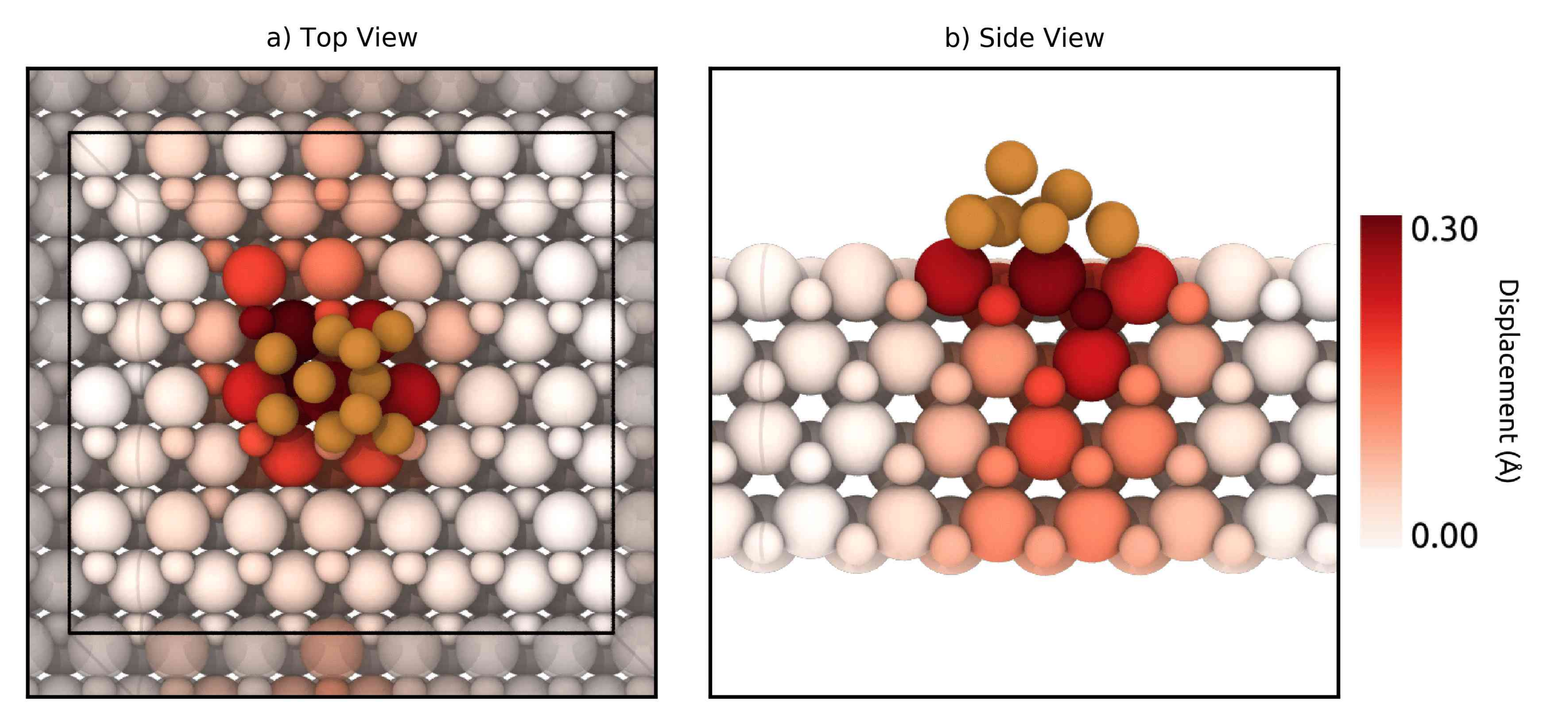}
    \caption{Structural changes in the ZnO support induced by the GM structure of Cu$_{10}$. a) Top view of the support atoms that are colored by their displacement compared to a clean ZnO slab of the same size. Transparent atoms correspond to periodic images outside the simulation box. The size of the Cu atoms has been reduced for visualization. b) Side view showing the cross-section of the slab %with a plane at y=11.15~\r{A}, 
    cutting through the middle of the cluster. }
    \label{fig:znodisturbances}
\end{figure*}

\begin{figure*}
    \centering
    \includegraphics[width=\linewidth]{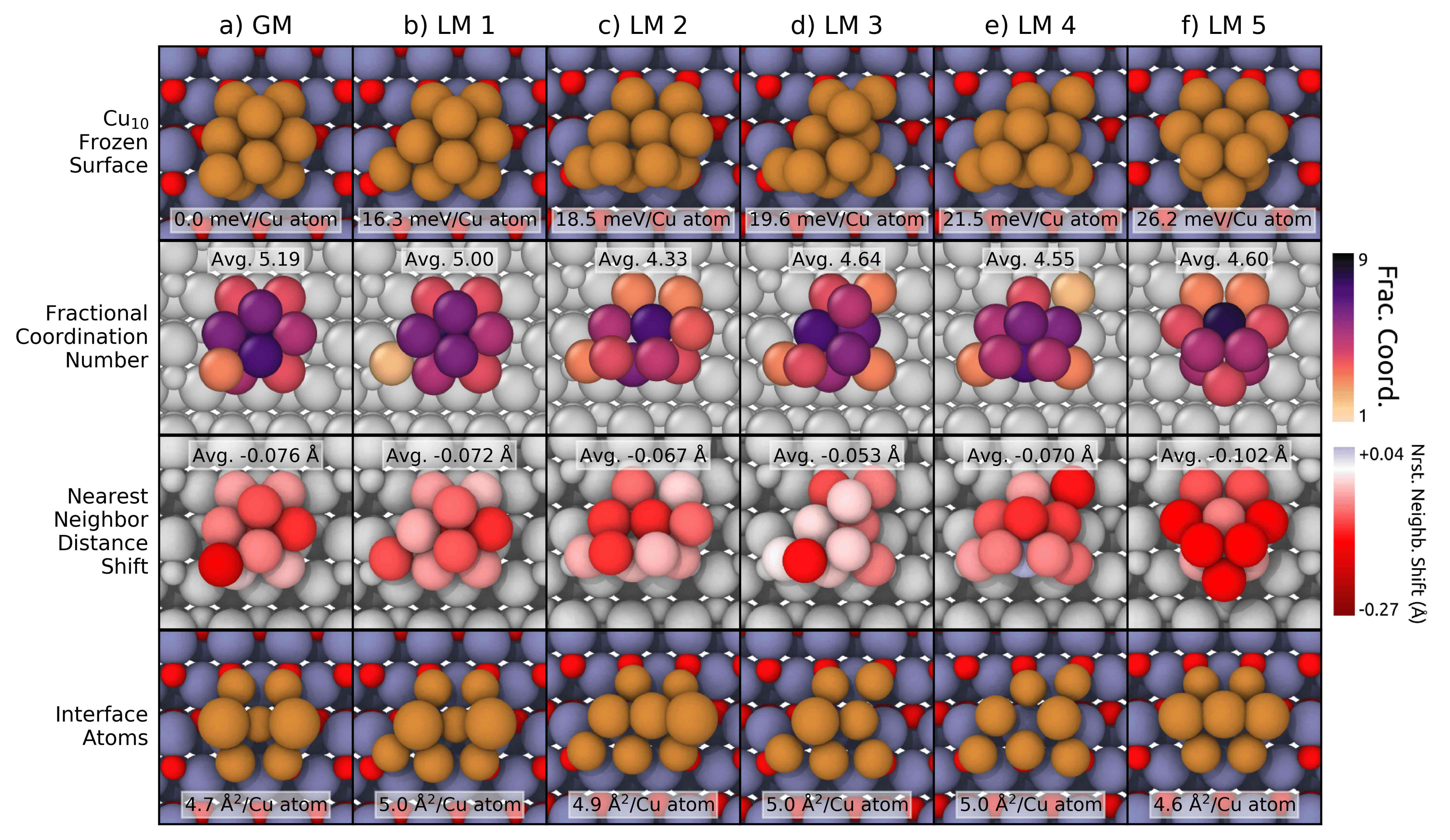}
    \caption{GM and first 5 LM for a GA search of Cu$_{10}$ on the frozen ZnO(10$\bar{1}$0) surface. Top: Normal view (flexible surface see Fig.~\ref{fig:gaminima410}). Middle Top: Fractional coordination number (flexible surface see Fig.~\ref{fig:gaminima410fraccoord}). Middle Bottom: Shift from the nearest neighbor distance of bulk copper (flexible surface see Fig.~\ref{fig:gaminima410nnshift}). Bottom: Interface atoms (flexible surface see Fig.~\ref{fig:gaminima410interface}).} 
    \label{fig:cluster10frozen}
\end{figure*}

An interesting question addresses the influence of the surface beyond offering merely a substrate for the cluster. In particular in \textit{ab-initio} based searches~\cite{davis_application_2016, ahussein_dft_2016} the surface is sometimes kept fixed to save on computational time during the multiple minimizations required. In other simulations the surface is kept frozen in first, coarse minimizations to save on resources and then the best clusters from these pre-optimizations are further optimized with more expensive methods and flexible surfaces. This could lead to biased results if the initial filtering employs a frozen support. But what happens, if we do not allow the surface to relax to the presence of the cluster?

To analyze this, we first investigate the effect of the cluster on the surface. As an example, Fig.~\ref{fig:znodisturbances} shows the distortions in the ZnO slab for the GM of Cu$_{10}$. In this case, some of the atoms below the cluster have been slightly displaced from their layer, being pulled towards the cluster. Fig.~\ref{fig:znodisturbances} a) shows in a top view the lateral distortion on the surface of the slab. We can see that very small lateral distortions reach all the way to the periodic boundary even for our very large surface cell, which shows another reason why utilizing large enough optimization cells is important. Smaller cells might inhibit or alter these distortions, and thus change the energetic ranking of the clusters. Further we can see that these distortions do not just follow the symmetry of the cluster but can exhibit pronounced directionalities close to the cluster due to the asymmetric orientation of the ZnO surface dimers. b) plots this effect in a cross section for the interior of the slab. In this case, we can see that the atoms are modified down to the bottom-most layer. This effect is also found for other clusters to varying degrees.
This shows the importance of utilizing multiple support layers, as some \textit{ab-initio} based GA searches reduce the size of the support down to even two layers only~\cite{davis_application_2016, ahussein_dft_2016, kolsbjerg_neural-network-enhanced_2018, bazhenov_globally_2019, engel_influence_2019} to save on computational cost, or freeze the bottom-most layers for the same purpose. These extra layers are not very expensive for the NNP, since its computation cost scales well with system size.

Overall, this effect might seem to be still rather small, but it clearly demonstrates that there is some notable interaction between the support and the cluster that requires the support to slightly deform. As this might have consequences for the identification of the minimum energy structures, 
we have repeated the GA search for the Cu$_{10}$ cluster to show how freezing the atoms of the surface affects the results of the search. The obtained GM and the first 5 LM for the frozen surface are shown in Fig.~\ref{fig:cluster10frozen}, in the same format as Fig.~\ref{fig:gaminima410} for the flexible surface. We note that in this case the energetic ordering of the clusters is different, and new cluster shapes appear that are not present in the flexible case. In fact, the GM of the flexible case was not observed among the energetically lowest 50 minima of the frozen surface search. In contrast, a structure similar to LM 2 of the original search (Fig.~\ref{fig:gaminima410} 10 c)) and to the GM at cluster size 9 (Fig.~\ref{fig:gaminima410} 9 a)) now become the GM and LM 1. We can thus see that the flexibility of the support can have a large impact on the relative stability of the studied clusters and on the identified LM, as overall the energy difference are quite subtle and not very large. 
Even though some of the correct GM and LM structures might reappear in a subsequent second optimization in which the frozen substrate constraint is lifted, at least if there are no sizeable barriers, usually only the best intermediate structures are further refined, and thus the best overall structures may be missed in such a second step.

%%%%%%%%%%%%%%%%%%%%%%%%%%%%%%%%%%%%%%%%%%%%%%%%%%%%%%%%%%%%%%%%%%%%%%%%%%%%
%%%%%%%%%%%%%%%%%%%%%%%%%%%%%%%%%%%%%%%%%%%%%%%%%%%%%%%%%%%%%%%%%%%%%%%%%%%%
\section{Conclusions} \label{sec:conclusion}
%%%%%%%%%%%%%%%%%%%%%%%%%%%%%%%%%%%%%%%%%%%%%%%%%%%%%%%%%%%%%%%%%%%%%%%%%%%%
%%%%%%%%%%%%%%%%%%%%%%%%%%%%%%%%%%%%%%%%%%%%%%%%%%%%%%%%%%%%%%%%%%%%%%%%%%%%

We have performed a genetic algorithm global optimization search on the most stable structures of copper clusters with 4 to 10 atoms on the (10$\bar{1}$0) surface of zinc oxide employing a high-dimensional neural network potential constructed for this purpose using DFT reference data. 
This search, which essentially provides results of first-principles quality at a fraction of the computational costs, can be performed in very large surface cells demonstrating the need to use a flexible support to obtain meaningful low-energy structures. For the obtained global and local minima we have extracted important structural trends that aid in the understanding of the interaction between both materials. We have shown that a number of structural families are consistently present among the low energy configurations, often exchanging their ranking as the cluster size changes. A detailed analysis of various geometric properties of these clusters has been performed. In general, the GM aim to minimize their size and their interface area to volume ratio. The energy differences within a given cluster size are often very small, and thus clusters are expected to interconvert at moderately elevated temperatures. In general, we find that the clusters prefer to interact with the substrate through the surface oxygen atoms, with the footprint of the cluster limited to a rather small surface area evidencing the competition between support-cluster and cluster-cluster interaction. 
With respect to the interfacial structure of the clusters and the support, we observe that at the interface the clusters arrange themselves in a continuum between Cu (111) and (110) structures, already starting at cluster size 5, and becoming more prevalent at larger cluster sizes. 

In summary, these results establish a complex picture of the interaction between the cluster and its support. The cluster needs to 1. interact with the support at specific sites, and within this constraint 2. minimize its energy. This kind of strong dependence between cluster morphology and support properties is already well known in the literature. Studies are available both for theoretical coarse models~\cite{henry_morphology_2005, eckhoff_structure_2017}, and more specific atomistic simulations and experimental results~\cite{tauster_strong_1987, pan_tuning_2017, figueiredo_understanding_2019}, and is known as strong metal-support interaction.

\section{Supplementary Material}

As described in the text, supplementary material is provided:

\begin{itemize}
   \item An example VASP input parameter file.
   \item Two tables with the parameters for the symmetry functions of the neural network potential.
   \item A plot of predicted vs. DFT energy for the structures in our dataset for the chosen potential.
\end{itemize}

\begin{acknowledgments}

We thank the Deutsche Forschungsgemeinschaft (DFG) for financial support (Be3264/10-1, project number 289217282 and INST186/1294-1 FUGG, project number 405832858). JB gratefully acknowledges a DFG Heisenberg professorship (Be3264/11-2, project number 329898176). We would also like to thank the North-German Supercomputing Alliance (HLRN) under project number nic00046 for computing time. 
\end{acknowledgments}

\section{Data Availability Statement}

The data that support the findings of this study are available from the corresponding author upon reasonable request.

\clearpage %Forcing a clearpage here otherwise the references get mixed up with all the extra images
\bibliography{bibliography.bib}

\begin{thebibliography}{118}%
\makeatletter
\providecommand \@ifxundefined [1]{%
 \@ifx{#1\undefined}
}%
\providecommand \@ifnum [1]{%
 \ifnum #1\expandafter \@firstoftwo
 \else \expandafter \@secondoftwo
 \fi
}%
\providecommand \@ifx [1]{%
 \ifx #1\expandafter \@firstoftwo
 \else \expandafter \@secondoftwo
 \fi
}%
\providecommand \natexlab [1]{#1}%
\providecommand \enquote  [1]{``#1''}%
\providecommand \bibnamefont  [1]{#1}%
\providecommand \bibfnamefont [1]{#1}%
\providecommand \citenamefont [1]{#1}%
\providecommand \href@noop [0]{\@secondoftwo}%
\providecommand \href [0]{\begingroup \@sanitize@url \@href}%
\providecommand \@href[1]{\@@startlink{#1}\@@href}%
\providecommand \@@href[1]{\endgroup#1\@@endlink}%
\providecommand \@sanitize@url [0]{\catcode `\\12\catcode `\$12\catcode
  `\&12\catcode `\#12\catcode `\^12\catcode `\_12\catcode `\%12\relax}%
\providecommand \@@startlink[1]{}%
\providecommand \@@endlink[0]{}%
\providecommand \url  [0]{\begingroup\@sanitize@url \@url }%
\providecommand \@url [1]{\endgroup\@href {#1}{\urlprefix }}%
\providecommand \urlprefix  [0]{URL }%
\providecommand \Eprint [0]{\href }%
\providecommand \doibase [0]{http://dx.doi.org/}%
\providecommand \selectlanguage [0]{\@gobble}%
\providecommand \bibinfo  [0]{\@secondoftwo}%
\providecommand \bibfield  [0]{\@secondoftwo}%
\providecommand \translation [1]{[#1]}%
\providecommand \BibitemOpen [0]{}%
\providecommand \bibitemStop [0]{}%
\providecommand \bibitemNoStop [0]{.\EOS\space}%
\providecommand \EOS [0]{\spacefactor3000\relax}%
\providecommand \BibitemShut  [1]{\csname bibitem#1\endcsname}%
\let\auto@bib@innerbib\@empty
%</preamble>
\bibitem [{\citenamefont {Gates}(1995)}]{gates_supported_1995}%
  \BibitemOpen
  \bibfield  {author} {\bibinfo {author} {\bibfnamefont {B.~C.}\ \bibnamefont
  {Gates}},\ }\bibfield  {title} {\enquote {\bibinfo {title} {Supported {Metal}
  {Clusters}: {Synthesis}, {Structure}, and {Catalysis}},}\ }\href {\doibase
  10.1021/cr00035a003} {\bibfield  {journal} {\bibinfo  {journal} {Chemical
  Reviews}\ }\textbf {\bibinfo {volume} {95}},\ \bibinfo {pages} {511--522}
  (\bibinfo {year} {1995})}\BibitemShut {NoStop}%
\bibitem [{\citenamefont {Santra}\ and\ \citenamefont
  {Goodman}(2002)}]{santra_oxide-supported_2002}%
  \BibitemOpen
  \bibfield  {author} {\bibinfo {author} {\bibfnamefont {A.~K.}\ \bibnamefont
  {Santra}}\ and\ \bibinfo {author} {\bibfnamefont {D.~W.}\ \bibnamefont
  {Goodman}},\ }\bibfield  {title} {\enquote {\bibinfo {title} {Oxide-supported
  metal clusters: models for heterogeneous catalysts},}\ }\href {\doibase
  10.1088/0953-8984/15/2/202} {\bibfield  {journal} {\bibinfo  {journal}
  {Journal of Physics: Condensed Matter}\ }\textbf {\bibinfo {volume} {15}},\
  \bibinfo {pages} {R31--R62} (\bibinfo {year} {2002})}\BibitemShut {NoStop}%
\bibitem [{\citenamefont {Wang}(2004)}]{wang_zinc_2004}%
  \BibitemOpen
  \bibfield  {author} {\bibinfo {author} {\bibfnamefont {Z.~L.}\ \bibnamefont
  {Wang}},\ }\bibfield  {title} {\enquote {\bibinfo {title} {Zinc oxide
  nanostructures: growth, properties and applications},}\ }\href {\doibase
  10.1088/0953-8984/16/25/R01} {\bibfield  {journal} {\bibinfo  {journal}
  {Journal of Physics: Condensed Matter}\ }\textbf {\bibinfo {volume} {16}},\
  \bibinfo {pages} {R829--R858} (\bibinfo {year} {2004})}\BibitemShut {NoStop}%
\bibitem [{\citenamefont {Heiz}\ and\ \citenamefont
  {Bullock}(2004)}]{heiz_fundamental_2004}%
  \BibitemOpen
  \bibfield  {author} {\bibinfo {author} {\bibfnamefont {U.}~\bibnamefont
  {Heiz}}\ and\ \bibinfo {author} {\bibfnamefont {E.~L.}\ \bibnamefont
  {Bullock}},\ }\bibfield  {title} {\enquote {\bibinfo {title} {Fundamental
  aspects of catalysis on supported metal clusters},}\ }\href {\doibase
  10.1039/B313560H} {\bibfield  {journal} {\bibinfo  {journal} {Journal of
  Materials Chemistry}\ }\textbf {\bibinfo {volume} {14}},\ \bibinfo {pages}
  {564--577} (\bibinfo {year} {2004})}\BibitemShut {NoStop}%
\bibitem [{\citenamefont {Yu}, \citenamefont {Porosoff},\ and\ \citenamefont
  {Chen}(2012)}]{yu_review_2012}%
  \BibitemOpen
  \bibfield  {author} {\bibinfo {author} {\bibfnamefont {W.}~\bibnamefont
  {Yu}}, \bibinfo {author} {\bibfnamefont {M.~D.}\ \bibnamefont {Porosoff}}, \
  and\ \bibinfo {author} {\bibfnamefont {J.~G.}\ \bibnamefont {Chen}},\
  }\bibfield  {title} {\enquote {\bibinfo {title} {Review of {Pt}-{Based}
  {Bimetallic} {Catalysis}: {From} {Model} {Surfaces} to {Supported}
  {Catalysts}},}\ }\href {\doibase 10.1021/cr300096b} {\bibfield  {journal}
  {\bibinfo  {journal} {Chemical Reviews}\ }\textbf {\bibinfo {volume} {112}},\
  \bibinfo {pages} {5780--5817} (\bibinfo {year} {2012})}\BibitemShut {NoStop}%
\bibitem [{\citenamefont {Cabria}, \citenamefont {López},\ and\ \citenamefont
  {Alonso}(2010)}]{cabria_theoretical_2010}%
  \BibitemOpen
  \bibfield  {author} {\bibinfo {author} {\bibfnamefont {I.}~\bibnamefont
  {Cabria}}, \bibinfo {author} {\bibfnamefont {M.~J.}\ \bibnamefont {López}},
  \ and\ \bibinfo {author} {\bibfnamefont {J.~A.}\ \bibnamefont {Alonso}},\
  }\bibfield  {title} {\enquote {\bibinfo {title} {Theoretical study of the
  transition from planar to three-dimensional structures of palladium clusters
  supported on graphene},}\ }\href {\doibase 10.1103/PhysRevB.81.035403}
  {\bibfield  {journal} {\bibinfo  {journal} {Physical Review B}\ }\textbf
  {\bibinfo {volume} {81}},\ \bibinfo {pages} {035403} (\bibinfo {year}
  {2010})}\BibitemShut {NoStop}%
\bibitem [{\citenamefont {Robles}\ and\ \citenamefont
  {Khanna}(2010)}]{robles_oxidation_2010}%
  \BibitemOpen
  \bibfield  {author} {\bibinfo {author} {\bibfnamefont {R.}~\bibnamefont
  {Robles}}\ and\ \bibinfo {author} {\bibfnamefont {S.~N.}\ \bibnamefont
  {Khanna}},\ }\bibfield  {title} {\enquote {\bibinfo {title} {Oxidation of
  {Pd$_n$} (n=1-7,~10) {Clusters} {Supported} on {Alumina}/{NiAl}(110)},}\
  }\href {\doibase 10.1103/PhysRevB.82.085428} {\bibfield  {journal} {\bibinfo
  {journal} {Physical Review B}\ }\textbf {\bibinfo {volume} {82}},\ \bibinfo
  {pages} {085428} (\bibinfo {year} {2010})}\BibitemShut {NoStop}%
\bibitem [{\citenamefont {Tang}, \citenamefont {Yang},\ and\ \citenamefont
  {Dai}(2012)}]{tang_theoretical_2012}%
  \BibitemOpen
  \bibfield  {author} {\bibinfo {author} {\bibfnamefont {Y.}~\bibnamefont
  {Tang}}, \bibinfo {author} {\bibfnamefont {Z.}~\bibnamefont {Yang}}, \ and\
  \bibinfo {author} {\bibfnamefont {X.}~\bibnamefont {Dai}},\ }\bibfield
  {title} {\enquote {\bibinfo {title} {A theoretical simulation on the
  catalytic oxidation of {CO} on {Pt}/graphene},}\ }\href {\doibase
  10.1039/C2CP41441D} {\bibfield  {journal} {\bibinfo  {journal} {Physical
  Chemistry Chemical Physics}\ }\textbf {\bibinfo {volume} {14}},\ \bibinfo
  {pages} {16566--16572} (\bibinfo {year} {2012})}\BibitemShut {NoStop}%
\bibitem [{\citenamefont {Heiz}\ \emph {et~al.}(1997)\citenamefont {Heiz},
  \citenamefont {Vanolli}, \citenamefont {Trento},\ and\ \citenamefont
  {Schneider}}]{heiz_chemical_1997}%
  \BibitemOpen
  \bibfield  {author} {\bibinfo {author} {\bibfnamefont {U.}~\bibnamefont
  {Heiz}}, \bibinfo {author} {\bibfnamefont {F.}~\bibnamefont {Vanolli}},
  \bibinfo {author} {\bibfnamefont {L.}~\bibnamefont {Trento}}, \ and\ \bibinfo
  {author} {\bibfnamefont {W.-D.}\ \bibnamefont {Schneider}},\ }\bibfield
  {title} {\enquote {\bibinfo {title} {Chemical reactivity of size-selected
  supported clusters: {An} experimental setup},}\ }\href {\doibase
  10.1063/1.1148113} {\bibfield  {journal} {\bibinfo  {journal} {Review of
  Scientific Instruments}\ }\textbf {\bibinfo {volume} {68}},\ \bibinfo {pages}
  {1986--1994} (\bibinfo {year} {1997})}\BibitemShut {NoStop}%
\bibitem [{\citenamefont {Haas}\ \emph {et~al.}(2000)\citenamefont {Haas},
  \citenamefont {Menck}, \citenamefont {Brune}, \citenamefont {Barth},
  \citenamefont {Venables},\ and\ \citenamefont {Kern}}]{haas_nucleation_2000}%
  \BibitemOpen
  \bibfield  {author} {\bibinfo {author} {\bibfnamefont {G.}~\bibnamefont
  {Haas}}, \bibinfo {author} {\bibfnamefont {A.}~\bibnamefont {Menck}},
  \bibinfo {author} {\bibfnamefont {H.}~\bibnamefont {Brune}}, \bibinfo
  {author} {\bibfnamefont {J.~V.}\ \bibnamefont {Barth}}, \bibinfo {author}
  {\bibfnamefont {J.~A.}\ \bibnamefont {Venables}}, \ and\ \bibinfo {author}
  {\bibfnamefont {K.}~\bibnamefont {Kern}},\ }\bibfield  {title} {\enquote
  {\bibinfo {title} {Nucleation and growth of supported clusters at defect
  sites: {Pd}/{MgO}(001)},}\ }\href {\doibase 10.1103/PhysRevB.61.11105}
  {\bibfield  {journal} {\bibinfo  {journal} {Physical Review B}\ }\textbf
  {\bibinfo {volume} {61}},\ \bibinfo {pages} {11105--11108} (\bibinfo {year}
  {2000})}\BibitemShut {NoStop}%
\bibitem [{\citenamefont {Yamaguchi}\ and\ \citenamefont
  {Iglesia}(2010)}]{yamaguchi_catalytic_2010}%
  \BibitemOpen
  \bibfield  {author} {\bibinfo {author} {\bibfnamefont {A.}~\bibnamefont
  {Yamaguchi}}\ and\ \bibinfo {author} {\bibfnamefont {E.}~\bibnamefont
  {Iglesia}},\ }\bibfield  {title} {\enquote {\bibinfo {title} {Catalytic
  activation and reforming of methane on supported palladium clusters},}\
  }\href {\doibase 10.1016/j.jcat.2010.06.001} {\bibfield  {journal} {\bibinfo
  {journal} {Journal of Catalysis}\ }\textbf {\bibinfo {volume} {274}},\
  \bibinfo {pages} {52--63} (\bibinfo {year} {2010})}\BibitemShut {NoStop}%
\bibitem [{\citenamefont {Peters}\ \emph {et~al.}(2013)\citenamefont {Peters},
  \citenamefont {Peredkov}, \citenamefont {Neeb}, \citenamefont {Eberhardt},\
  and\ \citenamefont {Al-Hada}}]{peters_size-dependent_2013}%
  \BibitemOpen
  \bibfield  {author} {\bibinfo {author} {\bibfnamefont {S.}~\bibnamefont
  {Peters}}, \bibinfo {author} {\bibfnamefont {S.}~\bibnamefont {Peredkov}},
  \bibinfo {author} {\bibfnamefont {M.}~\bibnamefont {Neeb}}, \bibinfo {author}
  {\bibfnamefont {W.}~\bibnamefont {Eberhardt}}, \ and\ \bibinfo {author}
  {\bibfnamefont {M.}~\bibnamefont {Al-Hada}},\ }\bibfield  {title} {\enquote
  {\bibinfo {title} {Size-dependent {XPS} spectra of small supported
  {Au}-clusters},}\ }\href {\doibase 10.1016/j.susc.2012.09.024} {\bibfield
  {journal} {\bibinfo  {journal} {Surface Science}\ }\textbf {\bibinfo {volume}
  {608}},\ \bibinfo {pages} {129--134} (\bibinfo {year} {2013})}\BibitemShut
  {NoStop}%
\bibitem [{\citenamefont {Bowker}\ \emph {et~al.}(1988)\citenamefont {Bowker},
  \citenamefont {Hadden}, \citenamefont {Houghton}, \citenamefont {Hyland},\
  and\ \citenamefont {Waugh}}]{bowker_mechanism_1988}%
  \BibitemOpen
  \bibfield  {author} {\bibinfo {author} {\bibfnamefont {M.}~\bibnamefont
  {Bowker}}, \bibinfo {author} {\bibfnamefont {R.~A.}\ \bibnamefont {Hadden}},
  \bibinfo {author} {\bibfnamefont {H.}~\bibnamefont {Houghton}}, \bibinfo
  {author} {\bibfnamefont {J.~N.~K.}\ \bibnamefont {Hyland}}, \ and\ \bibinfo
  {author} {\bibfnamefont {K.~C.}\ \bibnamefont {Waugh}},\ }\bibfield  {title}
  {\enquote {\bibinfo {title} {The mechanism of methanol synthesis on
  copper/zinc oxide/alumina catalysts},}\ }\href {\doibase
  10.1016/0021-9517(88)90209-6} {\bibfield  {journal} {\bibinfo  {journal}
  {Journal of Catalysis}\ }\textbf {\bibinfo {volume} {109}},\ \bibinfo {pages}
  {263--273} (\bibinfo {year} {1988})}\BibitemShut {NoStop}%
\bibitem [{\citenamefont {Saito}\ \emph {et~al.}(1996)\citenamefont {Saito},
  \citenamefont {Fujitani}, \citenamefont {Takeuchi},\ and\ \citenamefont
  {Watanabe}}]{saito_development_1996}%
  \BibitemOpen
  \bibfield  {author} {\bibinfo {author} {\bibfnamefont {M.}~\bibnamefont
  {Saito}}, \bibinfo {author} {\bibfnamefont {T.}~\bibnamefont {Fujitani}},
  \bibinfo {author} {\bibfnamefont {M.}~\bibnamefont {Takeuchi}}, \ and\
  \bibinfo {author} {\bibfnamefont {T.}~\bibnamefont {Watanabe}},\ }\bibfield
  {title} {\enquote {\bibinfo {title} {Development of copper/zinc oxide-based
  multicomponent catalysts for methanol synthesis from carbon dioxide and
  hydrogen},}\ }\href {\doibase 10.1016/0926-860X(95)00305-3} {\bibfield
  {journal} {\bibinfo  {journal} {Applied Catalysis A: General}\ }\bibinfo
  {series} {Chemical {Engineering} and {Catalysis}},\ \textbf {\bibinfo
  {volume} {138}},\ \bibinfo {pages} {311--318} (\bibinfo {year}
  {1996})}\BibitemShut {NoStop}%
\bibitem [{\citenamefont {Sá}\ \emph {et~al.}(2010)\citenamefont {Sá},
  \citenamefont {Silva}, \citenamefont {Brandão}, \citenamefont {Sousa},\ and\
  \citenamefont {Mendes}}]{sa_catalysts_2010}%
  \BibitemOpen
  \bibfield  {author} {\bibinfo {author} {\bibfnamefont {S.}~\bibnamefont
  {Sá}}, \bibinfo {author} {\bibfnamefont {H.}~\bibnamefont {Silva}}, \bibinfo
  {author} {\bibfnamefont {L.}~\bibnamefont {Brandão}}, \bibinfo {author}
  {\bibfnamefont {J.~M.}\ \bibnamefont {Sousa}}, \ and\ \bibinfo {author}
  {\bibfnamefont {A.}~\bibnamefont {Mendes}},\ }\bibfield  {title} {\enquote
  {\bibinfo {title} {Catalysts for methanol steam reforming—{A} review},}\
  }\href {\doibase 10.1016/j.apcatb.2010.06.015} {\bibfield  {journal}
  {\bibinfo  {journal} {Applied Catalysis B: Environmental}\ }\textbf {\bibinfo
  {volume} {99}},\ \bibinfo {pages} {43--57} (\bibinfo {year}
  {2010})}\BibitemShut {NoStop}%
\bibitem [{\citenamefont {French}\ \emph {et~al.}(2008)\citenamefont {French},
  \citenamefont {Sokol}, \citenamefont {Catlow},\ and\ \citenamefont
  {Sherwood}}]{french_growth_2008}%
  \BibitemOpen
  \bibfield  {author} {\bibinfo {author} {\bibfnamefont {S.~A.}\ \bibnamefont
  {French}}, \bibinfo {author} {\bibfnamefont {A.~A.}\ \bibnamefont {Sokol}},
  \bibinfo {author} {\bibfnamefont {C.~R.~A.}\ \bibnamefont {Catlow}}, \ and\
  \bibinfo {author} {\bibfnamefont {P.}~\bibnamefont {Sherwood}},\ }\bibfield
  {title} {\enquote {\bibinfo {title} {The {Growth} of {Copper} {Clusters} over
  {ZnO}: the {Competition} between {Planar} and {Polyhedral} {Clusters}},}\
  }\href {\doibase 10.1021/jp709821h} {\bibfield  {journal} {\bibinfo
  {journal} {The Journal of Physical Chemistry C}\ }\textbf {\bibinfo {volume}
  {112}},\ \bibinfo {pages} {7420--7430} (\bibinfo {year} {2008})}\BibitemShut
  {NoStop}%
\bibitem [{\citenamefont {Cheng}\ \emph {et~al.}(2014)\citenamefont {Cheng},
  \citenamefont {Liang}, \citenamefont {Nie}, \citenamefont {Choudhary},
  \citenamefont {Phillpot}, \citenamefont {Asthagiri},\ and\ \citenamefont
  {Sinnott}}]{cheng_cu_2014}%
  \BibitemOpen
  \bibfield  {author} {\bibinfo {author} {\bibfnamefont {Y.-T.}\ \bibnamefont
  {Cheng}}, \bibinfo {author} {\bibfnamefont {T.}~\bibnamefont {Liang}},
  \bibinfo {author} {\bibfnamefont {X.}~\bibnamefont {Nie}}, \bibinfo {author}
  {\bibfnamefont {K.}~\bibnamefont {Choudhary}}, \bibinfo {author}
  {\bibfnamefont {S.~R.}\ \bibnamefont {Phillpot}}, \bibinfo {author}
  {\bibfnamefont {A.}~\bibnamefont {Asthagiri}}, \ and\ \bibinfo {author}
  {\bibfnamefont {S.~B.}\ \bibnamefont {Sinnott}},\ }\bibfield  {title}
  {\enquote {\bibinfo {title} {Cu cluster deposition on {ZnO} (10$\bar{1}$0):
  {Morphology} and growth mode predicted from molecular dynamics
  simulations},}\ }\href {\doibase 10.1016/j.susc.2013.10.025} {\bibfield
  {journal} {\bibinfo  {journal} {Surface Science}\ }\textbf {\bibinfo {volume}
  {621}},\ \bibinfo {pages} {109--116} (\bibinfo {year} {2014})}\BibitemShut
  {NoStop}%
\bibitem [{\citenamefont {Artrith}, \citenamefont {Hiller},\ and\ \citenamefont
  {Behler}(2013)}]{artrith_neural_2013}%
  \BibitemOpen
  \bibfield  {author} {\bibinfo {author} {\bibfnamefont {N.}~\bibnamefont
  {Artrith}}, \bibinfo {author} {\bibfnamefont {B.}~\bibnamefont {Hiller}}, \
  and\ \bibinfo {author} {\bibfnamefont {J.}~\bibnamefont {Behler}},\
  }\bibfield  {title} {\enquote {\bibinfo {title} {Neural network potentials
  for metals and oxides – {First} applications to copper clusters at zinc
  oxide},}\ }\href {\doibase 10.1002/pssb.201248370} {\bibfield  {journal}
  {\bibinfo  {journal} {physica status solidi (b)}\ }\textbf {\bibinfo {volume}
  {250}},\ \bibinfo {pages} {1191--1203} (\bibinfo {year} {2013})}\BibitemShut
  {NoStop}%
\bibitem [{\citenamefont {Hellström}\ and\ \citenamefont
  {Behler}(2016)}]{hellstrom_structure_2016}%
  \BibitemOpen
  \bibfield  {author} {\bibinfo {author} {\bibfnamefont {M.}~\bibnamefont
  {Hellström}}\ and\ \bibinfo {author} {\bibfnamefont {J.}~\bibnamefont
  {Behler}},\ }\bibfield  {title} {\enquote {\bibinfo {title} {Structure of
  aqueous {NaOH} solutions: insights from neural-network-based molecular
  dynamics simulations},}\ }\href {\doibase 10.1039/C6CP06547C} {\bibfield
  {journal} {\bibinfo  {journal} {Physical Chemistry Chemical Physics}\
  }\textbf {\bibinfo {volume} {19}},\ \bibinfo {pages} {82--96} (\bibinfo
  {year} {2016})}\BibitemShut {NoStop}%
\bibitem [{\citenamefont {Mora-Fonz}\ \emph {et~al.}(2017)\citenamefont
  {Mora-Fonz}, \citenamefont {Lazauskas}, \citenamefont {Woodley},
  \citenamefont {Bromley}, \citenamefont {Catlow},\ and\ \citenamefont
  {Sokol}}]{mora-fonz_development_2017}%
  \BibitemOpen
  \bibfield  {author} {\bibinfo {author} {\bibfnamefont {D.}~\bibnamefont
  {Mora-Fonz}}, \bibinfo {author} {\bibfnamefont {T.}~\bibnamefont
  {Lazauskas}}, \bibinfo {author} {\bibfnamefont {S.~M.}\ \bibnamefont
  {Woodley}}, \bibinfo {author} {\bibfnamefont {S.~T.}\ \bibnamefont
  {Bromley}}, \bibinfo {author} {\bibfnamefont {C.~R.~A.}\ \bibnamefont
  {Catlow}}, \ and\ \bibinfo {author} {\bibfnamefont {A.~A.}\ \bibnamefont
  {Sokol}},\ }\bibfield  {title} {\enquote {\bibinfo {title} {Development of
  {Interatomic} {Potentials} for {Supported} {Nanoparticles}: {The} {Cu}/{ZnO}
  {Case}},}\ }\href {\doibase 10.1021/acs.jpcc.7b04502} {\bibfield  {journal}
  {\bibinfo  {journal} {The Journal of Physical Chemistry C}\ }\textbf
  {\bibinfo {volume} {121}},\ \bibinfo {pages} {16831--16844} (\bibinfo {year}
  {2017})}\BibitemShut {NoStop}%
\bibitem [{\citenamefont {Dulub}, \citenamefont {Boatner},\ and\ \citenamefont
  {Diebold}(2002)}]{dulub_stm_2002}%
  \BibitemOpen
  \bibfield  {author} {\bibinfo {author} {\bibfnamefont {O.}~\bibnamefont
  {Dulub}}, \bibinfo {author} {\bibfnamefont {L.~A.}\ \bibnamefont {Boatner}},
  \ and\ \bibinfo {author} {\bibfnamefont {U.}~\bibnamefont {Diebold}},\
  }\bibfield  {title} {\enquote {\bibinfo {title} {{STM} study of {Cu} growth
  on the {ZnO}(10$\bar{1}$0) surface},}\ }\href {\doibase
  10.1016/S0039-6028(02)01107-X} {\bibfield  {journal} {\bibinfo  {journal}
  {Surface Science}\ }\textbf {\bibinfo {volume} {504}},\ \bibinfo {pages}
  {271--281} (\bibinfo {year} {2002})}\BibitemShut {NoStop}%
\bibitem [{\citenamefont {Koplitz}, \citenamefont {Dulub},\ and\ \citenamefont
  {Diebold}(2003)}]{koplitz_stm_2003}%
  \BibitemOpen
  \bibfield  {author} {\bibinfo {author} {\bibfnamefont {L.~V.}\ \bibnamefont
  {Koplitz}}, \bibinfo {author} {\bibfnamefont {O.}~\bibnamefont {Dulub}}, \
  and\ \bibinfo {author} {\bibfnamefont {U.}~\bibnamefont {Diebold}},\
  }\bibfield  {title} {\enquote {\bibinfo {title} {{STM} {Study} of {Copper}
  {Growth} on {ZnO}(0001)-{Zn} and {ZnO}(000$\bar{1}$)-{O} {Surfaces}},}\
  }\href {\doibase 10.1021/jp0352175} {\bibfield  {journal} {\bibinfo
  {journal} {The Journal of Physical Chemistry B}\ }\textbf {\bibinfo {volume}
  {107}},\ \bibinfo {pages} {10583--10590} (\bibinfo {year}
  {2003})}\BibitemShut {NoStop}%
\bibitem [{\citenamefont {Kroll}\ and\ \citenamefont
  {Köhler}(2007)}]{kroll_small_2007}%
  \BibitemOpen
  \bibfield  {author} {\bibinfo {author} {\bibfnamefont {M.}~\bibnamefont
  {Kroll}}\ and\ \bibinfo {author} {\bibfnamefont {U.}~\bibnamefont
  {Köhler}},\ }\bibfield  {title} {\enquote {\bibinfo {title} {Small
  {Cu}-clusters on {ZnO}(0001)–{Zn}: {Nucleation} and annealing behavior},}\
  }\href {\doibase 10.1016/j.susc.2007.03.013} {\bibfield  {journal} {\bibinfo
  {journal} {Surface Science}\ }\textbf {\bibinfo {volume} {601}},\ \bibinfo
  {pages} {2182--2188} (\bibinfo {year} {2007})}\BibitemShut {NoStop}%
\bibitem [{\citenamefont {Behrens}\ \emph {et~al.}(2012)\citenamefont
  {Behrens}, \citenamefont {Studt}, \citenamefont {Kasatkin}, \citenamefont
  {Kühl}, \citenamefont {Hävecker}, \citenamefont {Abild-Pedersen},
  \citenamefont {Zander}, \citenamefont {Girgsdies}, \citenamefont {Kurr},
  \citenamefont {Kniep}, \citenamefont {Tovar}, \citenamefont {Fischer},
  \citenamefont {Nørskov},\ and\ \citenamefont
  {Schlögl}}]{behrens_active_2012}%
  \BibitemOpen
  \bibfield  {author} {\bibinfo {author} {\bibfnamefont {M.}~\bibnamefont
  {Behrens}}, \bibinfo {author} {\bibfnamefont {F.}~\bibnamefont {Studt}},
  \bibinfo {author} {\bibfnamefont {I.}~\bibnamefont {Kasatkin}}, \bibinfo
  {author} {\bibfnamefont {S.}~\bibnamefont {Kühl}}, \bibinfo {author}
  {\bibfnamefont {M.}~\bibnamefont {Hävecker}}, \bibinfo {author}
  {\bibfnamefont {F.}~\bibnamefont {Abild-Pedersen}}, \bibinfo {author}
  {\bibfnamefont {S.}~\bibnamefont {Zander}}, \bibinfo {author} {\bibfnamefont
  {F.}~\bibnamefont {Girgsdies}}, \bibinfo {author} {\bibfnamefont
  {P.}~\bibnamefont {Kurr}}, \bibinfo {author} {\bibfnamefont {B.-L.}\
  \bibnamefont {Kniep}}, \bibinfo {author} {\bibfnamefont {M.}~\bibnamefont
  {Tovar}}, \bibinfo {author} {\bibfnamefont {R.~W.}\ \bibnamefont {Fischer}},
  \bibinfo {author} {\bibfnamefont {J.~K.}\ \bibnamefont {Nørskov}}, \ and\
  \bibinfo {author} {\bibfnamefont {R.}~\bibnamefont {Schlögl}},\ }\bibfield
  {title} {\enquote {\bibinfo {title} {The {Active} {Site} of {Methanol}
  {Synthesis} over {Cu}/{ZnO}/{Al$_2$O$_3$} {Industrial} {Catalysts}},}\ }\href
  {\doibase 10.1126/science.1219831} {\bibfield  {journal} {\bibinfo  {journal}
  {Science}\ }\textbf {\bibinfo {volume} {336}},\ \bibinfo {pages} {893--897}
  (\bibinfo {year} {2012})}\BibitemShut {NoStop}%
\bibitem [{\citenamefont {Kuld}\ \emph {et~al.}(2016)\citenamefont {Kuld},
  \citenamefont {Thorhauge}, \citenamefont {Falsig}, \citenamefont {Elkjær},
  \citenamefont {Helveg}, \citenamefont {Chorkendorff},\ and\ \citenamefont
  {Sehested}}]{kuld_quantifying_2016}%
  \BibitemOpen
  \bibfield  {author} {\bibinfo {author} {\bibfnamefont {S.}~\bibnamefont
  {Kuld}}, \bibinfo {author} {\bibfnamefont {M.}~\bibnamefont {Thorhauge}},
  \bibinfo {author} {\bibfnamefont {H.}~\bibnamefont {Falsig}}, \bibinfo
  {author} {\bibfnamefont {C.~F.}\ \bibnamefont {Elkjær}}, \bibinfo {author}
  {\bibfnamefont {S.}~\bibnamefont {Helveg}}, \bibinfo {author} {\bibfnamefont
  {I.}~\bibnamefont {Chorkendorff}}, \ and\ \bibinfo {author} {\bibfnamefont
  {J.}~\bibnamefont {Sehested}},\ }\bibfield  {title} {\enquote {\bibinfo
  {title} {Quantifying the promotion of {Cu} catalysts by {ZnO} for methanol
  synthesis},}\ }\href {\doibase 10.1126/science.aaf0718} {\bibfield  {journal}
  {\bibinfo  {journal} {Science}\ }\textbf {\bibinfo {volume} {352}},\ \bibinfo
  {pages} {969--974} (\bibinfo {year} {2016})}\BibitemShut {NoStop}%
\bibitem [{\citenamefont {Kattel}\ \emph {et~al.}(2017)\citenamefont {Kattel},
  \citenamefont {Ramírez}, \citenamefont {Chen}, \citenamefont {Rodriguez},\
  and\ \citenamefont {Liu}}]{kattel_active_2017}%
  \BibitemOpen
  \bibfield  {author} {\bibinfo {author} {\bibfnamefont {S.}~\bibnamefont
  {Kattel}}, \bibinfo {author} {\bibfnamefont {P.~J.}\ \bibnamefont
  {Ramírez}}, \bibinfo {author} {\bibfnamefont {J.~G.}\ \bibnamefont {Chen}},
  \bibinfo {author} {\bibfnamefont {J.~A.}\ \bibnamefont {Rodriguez}}, \ and\
  \bibinfo {author} {\bibfnamefont {P.}~\bibnamefont {Liu}},\ }\bibfield
  {title} {\enquote {\bibinfo {title} {Active sites for {CO$_2$} hydrogenation
  to methanol on {Cu}/{ZnO} catalysts},}\ }\href {\doibase
  10.1126/science.aal3573} {\bibfield  {journal} {\bibinfo  {journal}
  {Science}\ }\textbf {\bibinfo {volume} {355}},\ \bibinfo {pages} {1296--1299}
  (\bibinfo {year} {2017})}\BibitemShut {NoStop}%
\bibitem [{\citenamefont {Zhou}\ \emph {et~al.}(2004)\citenamefont {Zhou},
  \citenamefont {Wadley}, \citenamefont {Filhol},\ and\ \citenamefont
  {Neurock}}]{zhou_modified_2004}%
  \BibitemOpen
  \bibfield  {author} {\bibinfo {author} {\bibfnamefont {X.~W.}\ \bibnamefont
  {Zhou}}, \bibinfo {author} {\bibfnamefont {H.~N.~G.}\ \bibnamefont {Wadley}},
  \bibinfo {author} {\bibfnamefont {J.-S.}\ \bibnamefont {Filhol}}, \ and\
  \bibinfo {author} {\bibfnamefont {M.~N.}\ \bibnamefont {Neurock}},\
  }\bibfield  {title} {\enquote {\bibinfo {title} {Modified charge
  transfer--embedded atom method potential for metal/metal oxide systems},}\
  }\href {\doibase 10.1103/PhysRevB.69.035402} {\bibfield  {journal} {\bibinfo
  {journal} {Physical Review B}\ }\textbf {\bibinfo {volume} {69}},\ \bibinfo
  {pages} {035402} (\bibinfo {year} {2004})}\BibitemShut {NoStop}%
\bibitem [{\citenamefont {van Duin}\ \emph {et~al.}(2010)\citenamefont {van
  Duin}, \citenamefont {Bryantsev}, \citenamefont {Diallo}, \citenamefont
  {Goddard}, \citenamefont {Rahaman}, \citenamefont {Doren}, \citenamefont
  {Raymand},\ and\ \citenamefont {Hermansson}}]{P2884}%
  \BibitemOpen
  \bibfield  {author} {\bibinfo {author} {\bibfnamefont {A.~C.~T.}\
  \bibnamefont {van Duin}}, \bibinfo {author} {\bibfnamefont {V.~S.}\
  \bibnamefont {Bryantsev}}, \bibinfo {author} {\bibfnamefont {M.~S.}\
  \bibnamefont {Diallo}}, \bibinfo {author} {\bibfnamefont {W.~A.}\
  \bibnamefont {Goddard}, \bibfnamefont {III}}, \bibinfo {author}
  {\bibfnamefont {O.}~\bibnamefont {Rahaman}}, \bibinfo {author} {\bibfnamefont
  {D.~J.}\ \bibnamefont {Doren}}, \bibinfo {author} {\bibfnamefont
  {D.}~\bibnamefont {Raymand}}, \ and\ \bibinfo {author} {\bibfnamefont
  {K.}~\bibnamefont {Hermansson}},\ }\bibfield  {title} {\enquote {\bibinfo
  {title} {Development and validation of a reaxff reactive force field for {Cu}
  {Cation}/{Water} interactions and {Copper} {Metal}/{Metal} {Oxide}/{Metal}
  {Hydroxide} {Condensed} {Phases}},}\ }\href@noop {} {\bibfield  {journal}
  {\bibinfo  {journal} {J. Phys. Chem. A}\ }\textbf {\bibinfo {volume} {114}},\
  \bibinfo {pages} {9507} (\bibinfo {year} {2010})}\BibitemShut {NoStop}%
\bibitem [{\citenamefont {Handley}\ and\ \citenamefont
  {Popelier}(2010)}]{P2559}%
  \BibitemOpen
  \bibfield  {author} {\bibinfo {author} {\bibfnamefont {C.~M.}\ \bibnamefont
  {Handley}}\ and\ \bibinfo {author} {\bibfnamefont {P.~L.~A.}\ \bibnamefont
  {Popelier}},\ }\bibfield  {title} {\enquote {\bibinfo {title} {Potential
  energy surfaces fitted by artificial neural networks},}\ }\href@noop {}
  {\bibfield  {journal} {\bibinfo  {journal} {J. Phys. Chem. A}\ }\textbf
  {\bibinfo {volume} {114}},\ \bibinfo {pages} {3371--3383} (\bibinfo {year}
  {2010})}\BibitemShut {NoStop}%
\bibitem [{\citenamefont {Behler}(2016)}]{behler_perspective_2016}%
  \BibitemOpen
  \bibfield  {author} {\bibinfo {author} {\bibfnamefont {J.}~\bibnamefont
  {Behler}},\ }\bibfield  {title} {\enquote {\bibinfo {title} {Perspective:
  {Machine} learning potentials for atomistic simulations},}\ }\href {\doibase
  10.1063/1.4966192} {\bibfield  {journal} {\bibinfo  {journal} {The Journal of
  Chemical Physics}\ }\textbf {\bibinfo {volume} {145}},\ \bibinfo {pages}
  {170901} (\bibinfo {year} {2016})}\BibitemShut {NoStop}%
\bibitem [{\citenamefont {Botu}\ \emph {et~al.}(2017)\citenamefont {Botu},
  \citenamefont {Batra}, \citenamefont {Chapman},\ and\ \citenamefont
  {Ramprasad}}]{P4938}%
  \BibitemOpen
  \bibfield  {author} {\bibinfo {author} {\bibfnamefont {V.}~\bibnamefont
  {Botu}}, \bibinfo {author} {\bibfnamefont {R.}~\bibnamefont {Batra}},
  \bibinfo {author} {\bibfnamefont {J.}~\bibnamefont {Chapman}}, \ and\
  \bibinfo {author} {\bibfnamefont {R.}~\bibnamefont {Ramprasad}},\ }\bibfield
  {title} {\enquote {\bibinfo {title} {Machine learning force fields:
  Construction, validation, and outlook},}\ }\href@noop {} {\bibfield
  {journal} {\bibinfo  {journal} {J. Phys. Chem. C}\ }\textbf {\bibinfo
  {volume} {121}},\ \bibinfo {pages} {511--522} (\bibinfo {year}
  {2017})}\BibitemShut {NoStop}%
\bibitem [{\citenamefont {Blank}\ \emph {et~al.}(1995)\citenamefont {Blank},
  \citenamefont {Brown}, \citenamefont {Calhoun},\ and\ \citenamefont
  {Doren}}]{P0316}%
  \BibitemOpen
  \bibfield  {author} {\bibinfo {author} {\bibfnamefont {T.~B.}\ \bibnamefont
  {Blank}}, \bibinfo {author} {\bibfnamefont {S.~D.}\ \bibnamefont {Brown}},
  \bibinfo {author} {\bibfnamefont {A.~W.}\ \bibnamefont {Calhoun}}, \ and\
  \bibinfo {author} {\bibfnamefont {D.~J.}\ \bibnamefont {Doren}},\ }\bibfield
  {title} {\enquote {\bibinfo {title} {Neural network models of potential
  energy surfaces},}\ }\href@noop {} {\bibfield  {journal} {\bibinfo  {journal}
  {J. Chem. Phys.}\ }\textbf {\bibinfo {volume} {103}},\ \bibinfo {pages}
  {4129--4137} (\bibinfo {year} {1995})}\BibitemShut {NoStop}%
\bibitem [{\citenamefont {Lorenz}, \citenamefont {Gro\ss},\ and\ \citenamefont
  {Scheffler}(2004)}]{P0421}%
  \BibitemOpen
  \bibfield  {author} {\bibinfo {author} {\bibfnamefont {S.}~\bibnamefont
  {Lorenz}}, \bibinfo {author} {\bibfnamefont {A.}~\bibnamefont {Gro\ss}}, \
  and\ \bibinfo {author} {\bibfnamefont {M.}~\bibnamefont {Scheffler}},\
  }\bibfield  {title} {\enquote {\bibinfo {title} {Representing
  high-dimensional potential-energy surfaces for reactions at surfaces by
  neural networks},}\ }\href@noop {} {\bibfield  {journal} {\bibinfo  {journal}
  {Chem. Phys. Lett.}\ }\textbf {\bibinfo {volume} {395}},\ \bibinfo {pages}
  {210--215} (\bibinfo {year} {2004})}\BibitemShut {NoStop}%
\bibitem [{\citenamefont {Behler}, \citenamefont {Lorenz},\ and\ \citenamefont
  {Reuter}(2007)}]{P1388}%
  \BibitemOpen
  \bibfield  {author} {\bibinfo {author} {\bibfnamefont {J.}~\bibnamefont
  {Behler}}, \bibinfo {author} {\bibfnamefont {S.}~\bibnamefont {Lorenz}}, \
  and\ \bibinfo {author} {\bibfnamefont {K.}~\bibnamefont {Reuter}},\
  }\bibfield  {title} {\enquote {\bibinfo {title} {Representing
  molecule-surface interactions with symmetry-adapted neural networks},}\
  }\href@noop {} {\bibfield  {journal} {\bibinfo  {journal} {J. Chem. Phys.}\
  }\textbf {\bibinfo {volume} {127}},\ \bibinfo {pages} {014705} (\bibinfo
  {year} {2007})}\BibitemShut {NoStop}%
\bibitem [{\citenamefont {Sch{\"u}tt}\ \emph {et~al.}(2018)\citenamefont
  {Sch{\"u}tt}, \citenamefont {Sauceda}, \citenamefont {Kindermans},
  \citenamefont {Tkatchenko},\ and\ \citenamefont {M{\"u}ller}}]{P5366}%
  \BibitemOpen
  \bibfield  {author} {\bibinfo {author} {\bibfnamefont {K.~T.}\ \bibnamefont
  {Sch{\"u}tt}}, \bibinfo {author} {\bibfnamefont {H.~E.}\ \bibnamefont
  {Sauceda}}, \bibinfo {author} {\bibfnamefont {P.-J.}\ \bibnamefont
  {Kindermans}}, \bibinfo {author} {\bibfnamefont {A.}~\bibnamefont
  {Tkatchenko}}, \ and\ \bibinfo {author} {\bibfnamefont {K.-R.}\ \bibnamefont
  {M{\"u}ller}},\ }\bibfield  {title} {\enquote {\bibinfo {title} {Schnet - a
  deep learning architecture for molecules and materials},}\ }\href@noop {}
  {\bibfield  {journal} {\bibinfo  {journal} {J. Chem. Phys.}\ }\textbf
  {\bibinfo {volume} {148}},\ \bibinfo {pages} {241722} (\bibinfo {year}
  {2018})}\BibitemShut {NoStop}%
\bibitem [{\citenamefont {Unke}\ and\ \citenamefont {Meuwly}(2019)}]{P5577}%
  \BibitemOpen
  \bibfield  {author} {\bibinfo {author} {\bibfnamefont {O.~T.}\ \bibnamefont
  {Unke}}\ and\ \bibinfo {author} {\bibfnamefont {M.}~\bibnamefont {Meuwly}},\
  }\bibfield  {title} {\enquote {\bibinfo {title} {Physnet: A neural network
  for predicting energies, forces, dipole moments, and partial charges},}\
  }\href@noop {} {\bibfield  {journal} {\bibinfo  {journal} {J. Chem. Theory
  Comput.}\ }\textbf {\bibinfo {volume} {15}},\ \bibinfo {pages} {3678--3693}
  (\bibinfo {year} {2019})}\BibitemShut {NoStop}%
\bibitem [{\citenamefont {Jiang}\ and\ \citenamefont
  {Guo}(2013)}]{jiang_permutation_2013}%
  \BibitemOpen
  \bibfield  {author} {\bibinfo {author} {\bibfnamefont {B.}~\bibnamefont
  {Jiang}}\ and\ \bibinfo {author} {\bibfnamefont {H.}~\bibnamefont {Guo}},\
  }\bibfield  {title} {\enquote {\bibinfo {title} {Permutation invariant
  polynomial neural network approach to fitting potential energy surfaces},}\
  }\href {\doibase 10.1063/1.4817187} {\bibfield  {journal} {\bibinfo
  {journal} {The Journal of Chemical Physics}\ }\textbf {\bibinfo {volume}
  {139}},\ \bibinfo {pages} {054112} (\bibinfo {year} {2013})}\BibitemShut
  {NoStop}%
\bibitem [{\citenamefont {Smith}, \citenamefont {Isayev},\ and\ \citenamefont
  {Roitberg}(2017)}]{P4945}%
  \BibitemOpen
  \bibfield  {author} {\bibinfo {author} {\bibfnamefont {J.~S.}\ \bibnamefont
  {Smith}}, \bibinfo {author} {\bibfnamefont {O.}~\bibnamefont {Isayev}}, \
  and\ \bibinfo {author} {\bibfnamefont {A.~E.}\ \bibnamefont {Roitberg}},\
  }\bibfield  {title} {\enquote {\bibinfo {title} {Ani-1: An extensible neural
  network potential with dft accuracy at force field computational cost},}\
  }\href@noop {} {\bibfield  {journal} {\bibinfo  {journal} {Chem. Sci.}\
  }\textbf {\bibinfo {volume} {8}},\ \bibinfo {pages} {3192--3203} (\bibinfo
  {year} {2017})}\BibitemShut {NoStop}%
\bibitem [{\citenamefont {Ghasemi}\ \emph {et~al.}(2015)\citenamefont
  {Ghasemi}, \citenamefont {Hofstetter}, \citenamefont {Saha},\ and\
  \citenamefont {Goedecker}}]{P4419}%
  \BibitemOpen
  \bibfield  {author} {\bibinfo {author} {\bibfnamefont {S.~A.}\ \bibnamefont
  {Ghasemi}}, \bibinfo {author} {\bibfnamefont {A.}~\bibnamefont {Hofstetter}},
  \bibinfo {author} {\bibfnamefont {S.}~\bibnamefont {Saha}}, \ and\ \bibinfo
  {author} {\bibfnamefont {S.}~\bibnamefont {Goedecker}},\ }\bibfield  {title}
  {\enquote {\bibinfo {title} {Interatomic potentials for ionic systems with
  density functional accuracy based on charge densities obtained by a neural
  network},}\ }\href@noop {} {\bibfield  {journal} {\bibinfo  {journal} {Phys.
  Rev. B}\ }\textbf {\bibinfo {volume} {92}},\ \bibinfo {pages} {045131}
  (\bibinfo {year} {2015})}\BibitemShut {NoStop}%
\bibitem [{\citenamefont {Bartók}\ \emph {et~al.}(2010)\citenamefont
  {Bartók}, \citenamefont {Payne}, \citenamefont {Kondor},\ and\ \citenamefont
  {Csányi}}]{bartok_gaussian_2010}%
  \BibitemOpen
  \bibfield  {author} {\bibinfo {author} {\bibfnamefont {A.~P.}\ \bibnamefont
  {Bartók}}, \bibinfo {author} {\bibfnamefont {M.~C.}\ \bibnamefont {Payne}},
  \bibinfo {author} {\bibfnamefont {R.}~\bibnamefont {Kondor}}, \ and\ \bibinfo
  {author} {\bibfnamefont {G.}~\bibnamefont {Csányi}},\ }\bibfield  {title}
  {\enquote {\bibinfo {title} {Gaussian {Approximation} {Potentials}: {The}
  {Accuracy} of {Quantum} {Mechanics}, without the {Electrons}},}\ }\href
  {\doibase 10.1103/PhysRevLett.104.136403} {\bibfield  {journal} {\bibinfo
  {journal} {Physical Review Letters}\ }\textbf {\bibinfo {volume} {104}},\
  \bibinfo {pages} {136403} (\bibinfo {year} {2010})}\BibitemShut {NoStop}%
\bibitem [{\citenamefont {Shapeev}(2016)}]{shapeev_moment_2016}%
  \BibitemOpen
  \bibfield  {author} {\bibinfo {author} {\bibfnamefont {A.~V.}\ \bibnamefont
  {Shapeev}},\ }\bibfield  {title} {\enquote {\bibinfo {title} {Moment {Tensor}
  {Potentials}: {A} {Class} of {Systematically} {Improvable} {Interatomic}
  {Potentials}},}\ }\href {\doibase 10.1137/15M1054183} {\bibfield  {journal}
  {\bibinfo  {journal} {Multiscale Modeling \& Simulation}\ }\textbf {\bibinfo
  {volume} {14}},\ \bibinfo {pages} {1153--1173} (\bibinfo {year}
  {2016})}\BibitemShut {NoStop}%
\bibitem [{\citenamefont {Thompson}\ \emph {et~al.}(2015)\citenamefont
  {Thompson}, \citenamefont {Swiler}, \citenamefont {Trott}, \citenamefont
  {Foiles},\ and\ \citenamefont {Tucker}}]{thompson_spectral_2015}%
  \BibitemOpen
  \bibfield  {author} {\bibinfo {author} {\bibfnamefont {A.~P.}\ \bibnamefont
  {Thompson}}, \bibinfo {author} {\bibfnamefont {L.~P.}\ \bibnamefont
  {Swiler}}, \bibinfo {author} {\bibfnamefont {C.~R.}\ \bibnamefont {Trott}},
  \bibinfo {author} {\bibfnamefont {S.~M.}\ \bibnamefont {Foiles}}, \ and\
  \bibinfo {author} {\bibfnamefont {G.~J.}\ \bibnamefont {Tucker}},\ }\bibfield
   {title} {\enquote {\bibinfo {title} {Spectral neighbor analysis method for
  automated generation of quantum-accurate interatomic potentials},}\ }\href
  {\doibase 10.1016/j.jcp.2014.12.018} {\bibfield  {journal} {\bibinfo
  {journal} {Journal of Computational Physics}\ }\textbf {\bibinfo {volume}
  {285}},\ \bibinfo {pages} {316--330} (\bibinfo {year} {2015})}\BibitemShut
  {NoStop}%
\bibitem [{\citenamefont {Behler}\ and\ \citenamefont
  {Parrinello}(2007)}]{behler_generalized_2007}%
  \BibitemOpen
  \bibfield  {author} {\bibinfo {author} {\bibfnamefont {J.}~\bibnamefont
  {Behler}}\ and\ \bibinfo {author} {\bibfnamefont {M.}~\bibnamefont
  {Parrinello}},\ }\bibfield  {title} {\enquote {\bibinfo {title} {Generalized
  {Neural}-{Network} {Representation} of {High}-{Dimensional}
  {Potential}-{Energy} {Surfaces}},}\ }\href {\doibase
  10.1103/PhysRevLett.98.146401} {\bibfield  {journal} {\bibinfo  {journal}
  {Physical Review Letters}\ }\textbf {\bibinfo {volume} {98}},\ \bibinfo
  {pages} {146401} (\bibinfo {year} {2007})}\BibitemShut {NoStop}%
\bibitem [{\citenamefont {Behler}(2015)}]{behler_constructing_2015}%
  \BibitemOpen
  \bibfield  {author} {\bibinfo {author} {\bibfnamefont {J.}~\bibnamefont
  {Behler}},\ }\bibfield  {title} {\enquote {\bibinfo {title} {Constructing
  high-dimensional neural network potentials: {A} tutorial review},}\ }\href
  {\doibase 10.1002/qua.24890} {\bibfield  {journal} {\bibinfo  {journal}
  {International Journal of Quantum Chemistry}\ }\textbf {\bibinfo {volume}
  {115}},\ \bibinfo {pages} {1032--1050} (\bibinfo {year} {2015})}\BibitemShut
  {NoStop}%
\bibitem [{\citenamefont {Behler}(2017)}]{behler_first_2017}%
  \BibitemOpen
  \bibfield  {author} {\bibinfo {author} {\bibfnamefont {J.}~\bibnamefont
  {Behler}},\ }\bibfield  {title} {\enquote {\bibinfo {title} {First
  {Principles} {Neural} {Network} {Potentials} for {Reactive} {Simulations} of
  {Large} {Molecular} and {Condensed} {Systems}},}\ }\href {\doibase
  10.1002/anie.201703114} {\bibfield  {journal} {\bibinfo  {journal}
  {Angewandte Chemie International Edition}\ }\textbf {\bibinfo {volume}
  {56}},\ \bibinfo {pages} {12828--12840} (\bibinfo {year} {2017})}\BibitemShut
  {NoStop}%
\bibitem [{\citenamefont {Artrith}\ and\ \citenamefont {Behler}(2012)}]{P3114}%
  \BibitemOpen
  \bibfield  {author} {\bibinfo {author} {\bibfnamefont {N.}~\bibnamefont
  {Artrith}}\ and\ \bibinfo {author} {\bibfnamefont {J.}~\bibnamefont
  {Behler}},\ }\bibfield  {title} {\enquote {\bibinfo {title} {High-dimensional
  neural network potentials for metal surfaces: A prototype study for
  copper},}\ }\href@noop {} {\bibfield  {journal} {\bibinfo  {journal} {Phys.
  Rev. B}\ }\textbf {\bibinfo {volume} {85}},\ \bibinfo {pages} {045439}
  (\bibinfo {year} {2012})}\BibitemShut {NoStop}%
\bibitem [{\citenamefont {Weinreich}\ \emph {et~al.}(2020)\citenamefont
  {Weinreich}, \citenamefont {Römer}, \citenamefont {Paleico},\ and\
  \citenamefont {Behler}}]{weinreich_properties_2020}%
  \BibitemOpen
  \bibfield  {author} {\bibinfo {author} {\bibfnamefont {J.}~\bibnamefont
  {Weinreich}}, \bibinfo {author} {\bibfnamefont {A.}~\bibnamefont {Römer}},
  \bibinfo {author} {\bibfnamefont {M.~L.}\ \bibnamefont {Paleico}}, \ and\
  \bibinfo {author} {\bibfnamefont {J.}~\bibnamefont {Behler}},\ }\bibfield
  {title} {\enquote {\bibinfo {title} {Properties of alpha-{Brass}
  {Nanoparticles}. 1. {Neural} {Network} {Potential} {Energy} {Surface}},}\
  }\href {\doibase 10.1021/acs.jpcc.0c00559} {\bibfield  {journal} {\bibinfo
  {journal} {The Journal of Physical Chemistry C}\ }\textbf {\bibinfo {volume}
  {124}},\ \bibinfo {pages} {12682--12695} (\bibinfo {year} {2020})},\ \bibinfo
  {note} {publisher: American Chemical Society}\BibitemShut {NoStop}%
\bibitem [{\citenamefont {Artrith}, \citenamefont {Morawietz},\ and\
  \citenamefont {Behler}(2011)}]{artrith_high-dimensional_2011}%
  \BibitemOpen
  \bibfield  {author} {\bibinfo {author} {\bibfnamefont {N.}~\bibnamefont
  {Artrith}}, \bibinfo {author} {\bibfnamefont {T.}~\bibnamefont {Morawietz}},
  \ and\ \bibinfo {author} {\bibfnamefont {J.}~\bibnamefont {Behler}},\
  }\bibfield  {title} {\enquote {\bibinfo {title} {High-dimensional
  neural-network potentials for multicomponent systems: {Applications} to zinc
  oxide},}\ }\href {\doibase 10.1103/PhysRevB.83.153101} {\bibfield  {journal}
  {\bibinfo  {journal} {Physical Review B}\ }\textbf {\bibinfo {volume} {83}},\
  \bibinfo {pages} {153101} (\bibinfo {year} {2011})}\BibitemShut {NoStop}%
\bibitem [{\citenamefont {Kondati~Natarajan}\ and\ \citenamefont
  {Behler}(2017{\natexlab{a}})}]{P4886}%
  \BibitemOpen
  \bibfield  {author} {\bibinfo {author} {\bibfnamefont {S.}~\bibnamefont
  {Kondati~Natarajan}}\ and\ \bibinfo {author} {\bibfnamefont {J.}~\bibnamefont
  {Behler}},\ }\bibfield  {title} {\enquote {\bibinfo {title} {Self-diffusion
  of surface defects at copper-water interfaces},}\ }\href@noop {} {\bibfield
  {journal} {\bibinfo  {journal} {J. Phys. Chem. C}\ }\textbf {\bibinfo
  {volume} {121}},\ \bibinfo {pages} {4368} (\bibinfo {year}
  {2017}{\natexlab{a}})}\BibitemShut {NoStop}%
\bibitem [{\citenamefont {Quaranta}, \citenamefont {Hellstrom},\ and\
  \citenamefont {Behler}(2017)}]{P4988}%
  \BibitemOpen
  \bibfield  {author} {\bibinfo {author} {\bibfnamefont {V.}~\bibnamefont
  {Quaranta}}, \bibinfo {author} {\bibfnamefont {M.}~\bibnamefont {Hellstrom}},
  \ and\ \bibinfo {author} {\bibfnamefont {J.}~\bibnamefont {Behler}},\
  }\bibfield  {title} {\enquote {\bibinfo {title} {Proton transfer mechanisms
  at the {Water}-{ZnO} interface: The role of presolvation},}\ }\href@noop {}
  {\bibfield  {journal} {\bibinfo  {journal} {J. Phys. Chem. Lett.}\ }\textbf
  {\bibinfo {volume} {8}},\ \bibinfo {pages} {1476} (\bibinfo {year}
  {2017})}\BibitemShut {NoStop}%
\bibitem [{\citenamefont {Michalewicz}\ and\ \citenamefont
  {Janikow}(1991)}]{michalewicz_genetic_1991}%
  \BibitemOpen
  \bibfield  {author} {\bibinfo {author} {\bibfnamefont {Z.}~\bibnamefont
  {Michalewicz}}\ and\ \bibinfo {author} {\bibfnamefont {C.~Z.}\ \bibnamefont
  {Janikow}},\ }\bibfield  {title} {\enquote {\bibinfo {title} {Genetic
  algorithms for numerical optimization},}\ }\href {\doibase
  10.1007/BF01889983} {\bibfield  {journal} {\bibinfo  {journal} {Statistics
  and Computing}\ }\textbf {\bibinfo {volume} {1}},\ \bibinfo {pages} {75--91}
  (\bibinfo {year} {1991})}\BibitemShut {NoStop}%
\bibitem [{\citenamefont {Deaven}\ and\ \citenamefont
  {Ho}(1995)}]{deaven_molecular_1995}%
  \BibitemOpen
  \bibfield  {author} {\bibinfo {author} {\bibfnamefont {D.~M.}\ \bibnamefont
  {Deaven}}\ and\ \bibinfo {author} {\bibfnamefont {K.~M.}\ \bibnamefont
  {Ho}},\ }\bibfield  {title} {\enquote {\bibinfo {title} {Molecular {Geometry}
  {Optimization} with a {Genetic} {Algorithm}},}\ }\href {\doibase
  10.1103/PhysRevLett.75.288} {\bibfield  {journal} {\bibinfo  {journal}
  {Physical Review Letters}\ }\textbf {\bibinfo {volume} {75}},\ \bibinfo
  {pages} {288--291} (\bibinfo {year} {1995})}\BibitemShut {NoStop}%
\bibitem [{\citenamefont {Kolsbjerg}, \citenamefont {Peterson},\ and\
  \citenamefont {Hammer}(2018)}]{kolsbjerg_neural-network-enhanced_2018}%
  \BibitemOpen
  \bibfield  {author} {\bibinfo {author} {\bibfnamefont {E.~L.}\ \bibnamefont
  {Kolsbjerg}}, \bibinfo {author} {\bibfnamefont {A.~A.}\ \bibnamefont
  {Peterson}}, \ and\ \bibinfo {author} {\bibfnamefont {B.}~\bibnamefont
  {Hammer}},\ }\bibfield  {title} {\enquote {\bibinfo {title}
  {Neural-network-enhanced evolutionary algorithm applied to supported metal
  nanoparticles},}\ }\href {\doibase 10.1103/PhysRevB.97.195424} {\bibfield
  {journal} {\bibinfo  {journal} {Physical Review B}\ }\textbf {\bibinfo
  {volume} {97}},\ \bibinfo {pages} {195424} (\bibinfo {year}
  {2018})}\BibitemShut {NoStop}%
\bibitem [{\citenamefont {Jennings}\ \emph {et~al.}(2019)\citenamefont
  {Jennings}, \citenamefont {Lysgaard}, \citenamefont {Hummelshøj},
  \citenamefont {Vegge},\ and\ \citenamefont
  {Bligaard}}]{jennings_genetic_2019}%
  \BibitemOpen
  \bibfield  {author} {\bibinfo {author} {\bibfnamefont {P.~C.}\ \bibnamefont
  {Jennings}}, \bibinfo {author} {\bibfnamefont {S.}~\bibnamefont {Lysgaard}},
  \bibinfo {author} {\bibfnamefont {J.~S.}\ \bibnamefont {Hummelshøj}},
  \bibinfo {author} {\bibfnamefont {T.}~\bibnamefont {Vegge}}, \ and\ \bibinfo
  {author} {\bibfnamefont {T.}~\bibnamefont {Bligaard}},\ }\bibfield  {title}
  {\enquote {\bibinfo {title} {Genetic algorithms for computational materials
  discovery accelerated by machine learning},}\ }\href {\doibase
  10.1038/s41524-019-0181-4} {\bibfield  {journal} {\bibinfo  {journal} {npj
  Computational Materials}\ }\textbf {\bibinfo {volume} {5}},\ \bibinfo {pages}
  {1--6} (\bibinfo {year} {2019})}\BibitemShut {NoStop}%
\bibitem [{\citenamefont {Wan}\ \emph {et~al.}(2018)\citenamefont {Wan},
  \citenamefont {Wei}, \citenamefont {Wang}, \citenamefont {Wang},
  \citenamefont {Zhou}, \citenamefont {Lin}, \citenamefont {Xie},\ and\
  \citenamefont {Guo}}]{wan_single_2018}%
  \BibitemOpen
  \bibfield  {author} {\bibinfo {author} {\bibfnamefont {Q.}~\bibnamefont
  {Wan}}, \bibinfo {author} {\bibfnamefont {F.}~\bibnamefont {Wei}}, \bibinfo
  {author} {\bibfnamefont {Y.}~\bibnamefont {Wang}}, \bibinfo {author}
  {\bibfnamefont {F.}~\bibnamefont {Wang}}, \bibinfo {author} {\bibfnamefont
  {L.}~\bibnamefont {Zhou}}, \bibinfo {author} {\bibfnamefont {S.}~\bibnamefont
  {Lin}}, \bibinfo {author} {\bibfnamefont {D.}~\bibnamefont {Xie}}, \ and\
  \bibinfo {author} {\bibfnamefont {H.}~\bibnamefont {Guo}},\ }\bibfield
  {title} {\enquote {\bibinfo {title} {Single atom detachment from {Cu}
  clusters, and diffusion and trapping on {CeO$_2$}(111): implications in
  {Ostwald} ripening and atomic redispersion},}\ }\href {\doibase
  10.1039/C8NR06232C} {\bibfield  {journal} {\bibinfo  {journal} {Nanoscale}\
  }\textbf {\bibinfo {volume} {10}},\ \bibinfo {pages} {17893--17901} (\bibinfo
  {year} {2018})}\BibitemShut {NoStop}%
\bibitem [{\citenamefont {Hornik}(1991)}]{hornik_approximation_1991}%
  \BibitemOpen
  \bibfield  {author} {\bibinfo {author} {\bibfnamefont {K.}~\bibnamefont
  {Hornik}},\ }\bibfield  {title} {\enquote {\bibinfo {title} {Approximation
  capabilities of multilayer feedforward networks},}\ }\href {\doibase
  10.1016/0893-6080(91)90009-T} {\bibfield  {journal} {\bibinfo  {journal}
  {Neural Networks}\ }\textbf {\bibinfo {volume} {4}},\ \bibinfo {pages}
  {251--257} (\bibinfo {year} {1991})}\BibitemShut {NoStop}%
\bibitem [{\citenamefont {Behler}(2011)}]{P2882}%
  \BibitemOpen
  \bibfield  {author} {\bibinfo {author} {\bibfnamefont {J.}~\bibnamefont
  {Behler}},\ }\bibfield  {title} {\enquote {\bibinfo {title} {Atom-centered
  symmetry functions for constructing high-dimensional neural network
  potentials},}\ }\href@noop {} {\bibfield  {journal} {\bibinfo  {journal} {J.
  Chem. Phys.}\ }\textbf {\bibinfo {volume} {134}},\ \bibinfo {pages} {074106}
  (\bibinfo {year} {2011})}\BibitemShut {NoStop}%
\bibitem [{\citenamefont {Behler}(2014)}]{P4106}%
  \BibitemOpen
  \bibfield  {author} {\bibinfo {author} {\bibfnamefont {J.}~\bibnamefont
  {Behler}},\ }\bibfield  {title} {\enquote {\bibinfo {title} {Representing
  potential energy surfaces by high-dimensional neural network potentials},}\
  }\href@noop {} {\bibfield  {journal} {\bibinfo  {journal} {J. Phys.: Condens.
  Matter}\ }\textbf {\bibinfo {volume} {26}},\ \bibinfo {pages} {183001}
  (\bibinfo {year} {2014})}\BibitemShut {NoStop}%
\bibitem [{\citenamefont {Jensen}\ \emph {et~al.}(2014)\citenamefont {Jensen},
  \citenamefont {Lysgaard}, \citenamefont {Quaade},\ and\ \citenamefont
  {Vegge}}]{jensen_designing_2014}%
  \BibitemOpen
  \bibfield  {author} {\bibinfo {author} {\bibfnamefont {P.~B.}\ \bibnamefont
  {Jensen}}, \bibinfo {author} {\bibfnamefont {S.}~\bibnamefont {Lysgaard}},
  \bibinfo {author} {\bibfnamefont {U.~J.}\ \bibnamefont {Quaade}}, \ and\
  \bibinfo {author} {\bibfnamefont {T.}~\bibnamefont {Vegge}},\ }\bibfield
  {title} {\enquote {\bibinfo {title} {Designing mixed metal halide ammines for
  ammonia storage using density functional theory and genetic algorithms},}\
  }\href {\doibase 10.1039/C4CP03133D} {\bibfield  {journal} {\bibinfo
  {journal} {Physical Chemistry Chemical Physics}\ }\textbf {\bibinfo {volume}
  {16}},\ \bibinfo {pages} {19732--19740} (\bibinfo {year} {2014})}\BibitemShut
  {NoStop}%
\bibitem [{\citenamefont {Bozkurt}\ \emph {et~al.}(2018)\citenamefont
  {Bozkurt}, \citenamefont {Perez}, \citenamefont {Hovius}, \citenamefont
  {Browning},\ and\ \citenamefont {Rothlisberger}}]{bozkurt_genetic_2018}%
  \BibitemOpen
  \bibfield  {author} {\bibinfo {author} {\bibfnamefont {E.}~\bibnamefont
  {Bozkurt}}, \bibinfo {author} {\bibfnamefont {M.~A.~S.}\ \bibnamefont
  {Perez}}, \bibinfo {author} {\bibfnamefont {R.}~\bibnamefont {Hovius}},
  \bibinfo {author} {\bibfnamefont {N.~J.}\ \bibnamefont {Browning}}, \ and\
  \bibinfo {author} {\bibfnamefont {U.}~\bibnamefont {Rothlisberger}},\
  }\bibfield  {title} {\enquote {\bibinfo {title} {Genetic {Algorithm} {Based}
  {Design} and {Experimental} {Characterization} of a {Highly} {Thermostable}
  {Metalloprotein}},}\ }\href {\doibase 10.1021/jacs.7b10660} {\bibfield
  {journal} {\bibinfo  {journal} {Journal of the American Chemical Society}\
  }\textbf {\bibinfo {volume} {140}},\ \bibinfo {pages} {4517--4521} (\bibinfo
  {year} {2018})}\BibitemShut {NoStop}%
\bibitem [{\citenamefont {Vilhelmsen}\ and\ \citenamefont
  {Hammer}(2012)}]{vilhelmsen_systematic_2012}%
  \BibitemOpen
  \bibfield  {author} {\bibinfo {author} {\bibfnamefont {L.~B.}\ \bibnamefont
  {Vilhelmsen}}\ and\ \bibinfo {author} {\bibfnamefont {B.}~\bibnamefont
  {Hammer}},\ }\bibfield  {title} {\enquote {\bibinfo {title} {Systematic
  {Study} of {Au$_6$} to {Au$_{12}$} {Gold} {Clusters} on {MgO}(100) {F}
  {Centers} {Using} {Density}-{Functional} {Theory}},}\ }\href {\doibase
  10.1103/PhysRevLett.108.126101} {\bibfield  {journal} {\bibinfo  {journal}
  {Physical Review Letters}\ }\textbf {\bibinfo {volume} {108}},\ \bibinfo
  {pages} {126101} (\bibinfo {year} {2012})}\BibitemShut {NoStop}%
\bibitem [{\citenamefont {Vilhelmsen}\ and\ \citenamefont
  {Hammer}(2014{\natexlab{a}})}]{vilhelmsen_identification_2014}%
  \BibitemOpen
  \bibfield  {author} {\bibinfo {author} {\bibfnamefont {L.~B.}\ \bibnamefont
  {Vilhelmsen}}\ and\ \bibinfo {author} {\bibfnamefont {B.}~\bibnamefont
  {Hammer}},\ }\bibfield  {title} {\enquote {\bibinfo {title} {Identification
  of the {Catalytic} {Site} at the {Interface} {Perimeter} of {Au} {Clusters}
  on {Rutile} {TiO$_2$}(110)},}\ }\href {\doibase 10.1021/cs500202f} {\bibfield
   {journal} {\bibinfo  {journal} {ACS Catalysis}\ }\textbf {\bibinfo {volume}
  {4}},\ \bibinfo {pages} {1626--1631} (\bibinfo {year}
  {2014}{\natexlab{a}})}\BibitemShut {NoStop}%
\bibitem [{\citenamefont {Huang}\ \emph {et~al.}(2019)\citenamefont {Huang},
  \citenamefont {Jiang}, \citenamefont {Liang}, \citenamefont {Wu},
  \citenamefont {Li},\ and\ \citenamefont {Hou}}]{huang_structural_2019}%
  \BibitemOpen
  \bibfield  {author} {\bibinfo {author} {\bibfnamefont {P.}~\bibnamefont
  {Huang}}, \bibinfo {author} {\bibfnamefont {Y.}~\bibnamefont {Jiang}},
  \bibinfo {author} {\bibfnamefont {T.}~\bibnamefont {Liang}}, \bibinfo
  {author} {\bibfnamefont {E.}~\bibnamefont {Wu}}, \bibinfo {author}
  {\bibfnamefont {J.}~\bibnamefont {Li}}, \ and\ \bibinfo {author}
  {\bibfnamefont {J.}~\bibnamefont {Hou}},\ }\bibfield  {title} {\enquote
  {\bibinfo {title} {Structural exploration of {Au$_x$M}- ({M} = {Si}, {Ge},
  {Sn}; x = 9–12) clusters with a revised genetic algorithm},}\ }\href
  {\doibase 10.1039/C9RA01019J} {\bibfield  {journal} {\bibinfo  {journal} {RSC
  Advances}\ }\textbf {\bibinfo {volume} {9}},\ \bibinfo {pages} {7432--7439}
  (\bibinfo {year} {2019})}\BibitemShut {NoStop}%
\bibitem [{\citenamefont {Buendía}\ \emph {et~al.}(2017)\citenamefont
  {Buendía}, \citenamefont {Vargas}, \citenamefont {Johnston},\ and\
  \citenamefont {Beltrán}}]{buendia_study_2017}%
  \BibitemOpen
  \bibfield  {author} {\bibinfo {author} {\bibfnamefont {F.}~\bibnamefont
  {Buendía}}, \bibinfo {author} {\bibfnamefont {J.~A.}\ \bibnamefont
  {Vargas}}, \bibinfo {author} {\bibfnamefont {R.~L.}\ \bibnamefont
  {Johnston}}, \ and\ \bibinfo {author} {\bibfnamefont {M.~R.}\ \bibnamefont
  {Beltrán}},\ }\bibfield  {title} {\enquote {\bibinfo {title} {Study of the
  stability of small {AuRh} clusters found by a {Genetic} {Algorithm}
  methodology},}\ }\href {\doibase 10.1016/j.comptc.2017.09.008} {\bibfield
  {journal} {\bibinfo  {journal} {Computational and Theoretical Chemistry}\
  }\textbf {\bibinfo {volume} {1119}},\ \bibinfo {pages} {51--58} (\bibinfo
  {year} {2017})}\BibitemShut {NoStop}%
\bibitem [{\citenamefont {Heydariyan}\ \emph {et~al.}(2018)\citenamefont
  {Heydariyan}, \citenamefont {Nouri}, \citenamefont {Alaei}, \citenamefont
  {Allahyari},\ and\ \citenamefont {Niehaus}}]{heydariyan_new_2018}%
  \BibitemOpen
  \bibfield  {author} {\bibinfo {author} {\bibfnamefont {S.}~\bibnamefont
  {Heydariyan}}, \bibinfo {author} {\bibfnamefont {M.~R.}\ \bibnamefont
  {Nouri}}, \bibinfo {author} {\bibfnamefont {M.}~\bibnamefont {Alaei}},
  \bibinfo {author} {\bibfnamefont {Z.}~\bibnamefont {Allahyari}}, \ and\
  \bibinfo {author} {\bibfnamefont {T.~A.}\ \bibnamefont {Niehaus}},\
  }\bibfield  {title} {\enquote {\bibinfo {title} {New candidates for the
  global minimum of medium-sized silicon clusters: {A} hybrid {DFTB}/{DFT}
  genetic algorithm applied to {Si$_n$}, n = 8-80},}\ }\href {\doibase
  10.1063/1.5037159} {\bibfield  {journal} {\bibinfo  {journal} {The Journal of
  Chemical Physics}\ }\textbf {\bibinfo {volume} {149}},\ \bibinfo {pages}
  {074313} (\bibinfo {year} {2018})}\BibitemShut {NoStop}%
\bibitem [{\citenamefont {Wales}\ and\ \citenamefont
  {Doye}(1997)}]{wales_global_1997}%
  \BibitemOpen
  \bibfield  {author} {\bibinfo {author} {\bibfnamefont {D.~J.}\ \bibnamefont
  {Wales}}\ and\ \bibinfo {author} {\bibfnamefont {J.~P.~K.}\ \bibnamefont
  {Doye}},\ }\bibfield  {title} {\enquote {\bibinfo {title} {Global
  {Optimization} by {Basin}-{Hopping} and the {Lowest} {Energy} {Structures} of
  {Lennard}-{Jones} {Clusters} {Containing} up to 110 {Atoms}},}\ }\href
  {\doibase 10.1021/jp970984n} {\bibfield  {journal} {\bibinfo  {journal} {The
  Journal of Physical Chemistry A}\ }\textbf {\bibinfo {volume} {101}},\
  \bibinfo {pages} {5111--5116} (\bibinfo {year} {1997})}\BibitemShut {NoStop}%
\bibitem [{\citenamefont {Vilhelmsen}\ and\ \citenamefont
  {Hammer}(2014{\natexlab{b}})}]{vilhelmsen_genetic_2014}%
  \BibitemOpen
  \bibfield  {author} {\bibinfo {author} {\bibfnamefont {L.~B.}\ \bibnamefont
  {Vilhelmsen}}\ and\ \bibinfo {author} {\bibfnamefont {B.}~\bibnamefont
  {Hammer}},\ }\bibfield  {title} {\enquote {\bibinfo {title} {A genetic
  algorithm for first principles global structure optimization of supported
  nano structures},}\ }\href {\doibase 10.1063/1.4886337} {\bibfield  {journal}
  {\bibinfo  {journal} {The Journal of Chemical Physics}\ }\textbf {\bibinfo
  {volume} {141}},\ \bibinfo {pages} {044711} (\bibinfo {year}
  {2014}{\natexlab{b}})}\BibitemShut {NoStop}%
\bibitem [{\citenamefont {Rondina}\ and\ \citenamefont
  {Da~Silva}(2013)}]{rondina_revised_2013}%
  \BibitemOpen
  \bibfield  {author} {\bibinfo {author} {\bibfnamefont {G.~G.}\ \bibnamefont
  {Rondina}}\ and\ \bibinfo {author} {\bibfnamefont {J.~L.~F.}\ \bibnamefont
  {Da~Silva}},\ }\bibfield  {title} {\enquote {\bibinfo {title} {Revised
  {Basin}-{Hopping} {Monte} {Carlo} {Algorithm} for {Structure} {Optimization}
  of {Clusters} and {Nanoparticles}},}\ }\href {\doibase 10.1021/ci400224z}
  {\bibfield  {journal} {\bibinfo  {journal} {Journal of Chemical Information
  and Modeling}\ }\textbf {\bibinfo {volume} {53}},\ \bibinfo {pages}
  {2282--2298} (\bibinfo {year} {2013})}\BibitemShut {NoStop}%
\bibitem [{\citenamefont {Kirkpatrick}, \citenamefont {Gelatt},\ and\
  \citenamefont {Vecchi}(1983)}]{kirkpatrick_optimization_1983}%
  \BibitemOpen
  \bibfield  {author} {\bibinfo {author} {\bibfnamefont {S.}~\bibnamefont
  {Kirkpatrick}}, \bibinfo {author} {\bibfnamefont {C.~D.}\ \bibnamefont
  {Gelatt}}, \ and\ \bibinfo {author} {\bibfnamefont {M.~P.}\ \bibnamefont
  {Vecchi}},\ }\bibfield  {title} {\enquote {\bibinfo {title} {Optimization by
  {Simulated} {Annealing}},}\ }\href {\doibase 10.1126/science.220.4598.671}
  {\bibfield  {journal} {\bibinfo  {journal} {Science}\ }\textbf {\bibinfo
  {volume} {220}},\ \bibinfo {pages} {671--680} (\bibinfo {year}
  {1983})}\BibitemShut {NoStop}%
\bibitem [{\citenamefont {Goedecker}(2004)}]{goedecker_minima_2004}%
  \BibitemOpen
  \bibfield  {author} {\bibinfo {author} {\bibfnamefont {S.}~\bibnamefont
  {Goedecker}},\ }\bibfield  {title} {\enquote {\bibinfo {title} {Minima
  hopping: {An} efficient search method for the global minimum of the potential
  energy surface of complex molecular systems},}\ }\href {\doibase
  10.1063/1.1724816} {\bibfield  {journal} {\bibinfo  {journal} {The Journal of
  Chemical Physics}\ }\textbf {\bibinfo {volume} {120}},\ \bibinfo {pages}
  {9911--9917} (\bibinfo {year} {2004})}\BibitemShut {NoStop}%
\bibitem [{\citenamefont {Vilhelmsen}, \citenamefont {Walton},\ and\
  \citenamefont {Sholl}(2012)}]{vilhelmsen_structure_2012}%
  \BibitemOpen
  \bibfield  {author} {\bibinfo {author} {\bibfnamefont {L.~B.}\ \bibnamefont
  {Vilhelmsen}}, \bibinfo {author} {\bibfnamefont {K.~S.}\ \bibnamefont
  {Walton}}, \ and\ \bibinfo {author} {\bibfnamefont {D.~S.}\ \bibnamefont
  {Sholl}},\ }\bibfield  {title} {\enquote {\bibinfo {title} {Structure and
  {Mobility} of {Metal} {Clusters} in {MOFs}: {Au}, {Pd}, and {AuPd} {Clusters}
  in {MOF}-74},}\ }\href {\doibase 10.1021/ja305004a} {\bibfield  {journal}
  {\bibinfo  {journal} {Journal of the American Chemical Society}\ }\textbf
  {\bibinfo {volume} {134}},\ \bibinfo {pages} {12807--12816} (\bibinfo {year}
  {2012})}\BibitemShut {NoStop}%
\bibitem [{\citenamefont {Johnston}(2003)}]{johnston_evolving_2003}%
  \BibitemOpen
  \bibfield  {author} {\bibinfo {author} {\bibfnamefont {R.~L.}\ \bibnamefont
  {Johnston}},\ }\bibfield  {title} {\enquote {\bibinfo {title} {Evolving
  better nanoparticles: {Genetic} algorithms for optimising cluster
  geometries},}\ }\href {\doibase 10.1039/B305686D} {\bibfield  {journal}
  {\bibinfo  {journal} {Dalton Transactions}\ ,\ \bibinfo {pages} {4193--4207}}
  (\bibinfo {year} {2003})}\BibitemShut {NoStop}%
\bibitem [{\citenamefont {Rahm}\ and\ \citenamefont
  {Erhart}(2017)}]{rahm_beyond_2017}%
  \BibitemOpen
  \bibfield  {author} {\bibinfo {author} {\bibfnamefont {J.~M.}\ \bibnamefont
  {Rahm}}\ and\ \bibinfo {author} {\bibfnamefont {P.}~\bibnamefont {Erhart}},\
  }\bibfield  {title} {\enquote {\bibinfo {title} {Beyond {Magic} {Numbers}:
  {Atomic} {Scale} {Equilibrium} {Nanoparticle} {Shapes} for {Any} {Size}},}\
  }\href {\doibase 10.1021/acs.nanolett.7b02761} {\bibfield  {journal}
  {\bibinfo  {journal} {Nano Letters}\ }\textbf {\bibinfo {volume} {17}},\
  \bibinfo {pages} {5775--5781} (\bibinfo {year} {2017})}\BibitemShut {NoStop}%
\bibitem [{\citenamefont {Paleico}\ and\ \citenamefont
  {Behler}(2020)}]{paleico_flexible_2020}%
  \BibitemOpen
  \bibfield  {author} {\bibinfo {author} {\bibfnamefont {M.~L.}\ \bibnamefont
  {Paleico}}\ and\ \bibinfo {author} {\bibfnamefont {J.}~\bibnamefont
  {Behler}},\ }\bibfield  {title} {\enquote {\bibinfo {title} {A flexible and
  adaptive grid algorithm for global optimization utilizing basin hopping
  {Monte} {Carlo}},}\ }\href {\doibase 10.1063/1.5142363} {\bibfield  {journal}
  {\bibinfo  {journal} {The Journal of Chemical Physics}\ }\textbf {\bibinfo
  {volume} {152}},\ \bibinfo {pages} {094109} (\bibinfo {year}
  {2020})}\BibitemShut {NoStop}%
\bibitem [{\citenamefont {Hohenberg}\ and\ \citenamefont
  {Kohn}(1964)}]{hohenberg_inhomogeneous_1964}%
  \BibitemOpen
  \bibfield  {author} {\bibinfo {author} {\bibfnamefont {P.}~\bibnamefont
  {Hohenberg}}\ and\ \bibinfo {author} {\bibfnamefont {W.}~\bibnamefont
  {Kohn}},\ }\bibfield  {title} {\enquote {\bibinfo {title} {Inhomogeneous
  {Electron} {Gas}},}\ }\href {\doibase 10.1103/PhysRev.136.B864} {\bibfield
  {journal} {\bibinfo  {journal} {Physical Review}\ }\textbf {\bibinfo {volume}
  {136}},\ \bibinfo {pages} {B864--B871} (\bibinfo {year} {1964})}\BibitemShut
  {NoStop}%
\bibitem [{\citenamefont {Kohn}\ and\ \citenamefont
  {Sham}(1965)}]{kohn_self-consistent_1965}%
  \BibitemOpen
  \bibfield  {author} {\bibinfo {author} {\bibfnamefont {W.}~\bibnamefont
  {Kohn}}\ and\ \bibinfo {author} {\bibfnamefont {L.~J.}\ \bibnamefont
  {Sham}},\ }\bibfield  {title} {\enquote {\bibinfo {title} {Self-{Consistent}
  {Equations} {Including} {Exchange} and {Correlation} {Effects}},}\ }\href
  {\doibase 10.1103/PhysRev.140.A1133} {\bibfield  {journal} {\bibinfo
  {journal} {Physical Review}\ }\textbf {\bibinfo {volume} {140}},\ \bibinfo
  {pages} {A1133--A1138} (\bibinfo {year} {1965})}\BibitemShut {NoStop}%
\bibitem [{\citenamefont {Kresse}\ and\ \citenamefont
  {Furthmüller}(1996)}]{kresse_efficient_1996}%
  \BibitemOpen
  \bibfield  {author} {\bibinfo {author} {\bibfnamefont {G.}~\bibnamefont
  {Kresse}}\ and\ \bibinfo {author} {\bibfnamefont {J.}~\bibnamefont
  {Furthmüller}},\ }\bibfield  {title} {\enquote {\bibinfo {title} {Efficient
  iterative schemes for ab initio total-energy calculations using a plane-wave
  basis set},}\ }\href {\doibase 10.1103/PhysRevB.54.11169} {\bibfield
  {journal} {\bibinfo  {journal} {Physical Review B}\ }\textbf {\bibinfo
  {volume} {54}},\ \bibinfo {pages} {11169--11186} (\bibinfo {year}
  {1996})}\BibitemShut {NoStop}%
\bibitem [{\citenamefont {Kresse}\ and\ \citenamefont
  {Joubert}(1999)}]{kresse_ultrasoft_1999}%
  \BibitemOpen
  \bibfield  {author} {\bibinfo {author} {\bibfnamefont {G.}~\bibnamefont
  {Kresse}}\ and\ \bibinfo {author} {\bibfnamefont {D.}~\bibnamefont
  {Joubert}},\ }\bibfield  {title} {\enquote {\bibinfo {title} {From ultrasoft
  pseudopotentials to the projector augmented-wave method},}\ }\href {\doibase
  10.1103/PhysRevB.59.1758} {\bibfield  {journal} {\bibinfo  {journal}
  {Physical Review B}\ }\textbf {\bibinfo {volume} {59}},\ \bibinfo {pages}
  {1758--1775} (\bibinfo {year} {1999})}\BibitemShut {NoStop}%
\bibitem [{\citenamefont {Blöchl}(1994)}]{blochl_projector_1994}%
  \BibitemOpen
  \bibfield  {author} {\bibinfo {author} {\bibfnamefont {P.~E.}\ \bibnamefont
  {Blöchl}},\ }\bibfield  {title} {\enquote {\bibinfo {title} {Projector
  augmented-wave method},}\ }\href {\doibase 10.1103/PhysRevB.50.17953}
  {\bibfield  {journal} {\bibinfo  {journal} {Physical Review B}\ }\textbf
  {\bibinfo {volume} {50}},\ \bibinfo {pages} {17953--17979} (\bibinfo {year}
  {1994})}\BibitemShut {NoStop}%
\bibitem [{\citenamefont {Kresse}\ and\ \citenamefont
  {Hafner}(1994)}]{kresse_norm-conserving_1994}%
  \BibitemOpen
  \bibfield  {author} {\bibinfo {author} {\bibfnamefont {G.}~\bibnamefont
  {Kresse}}\ and\ \bibinfo {author} {\bibfnamefont {J.}~\bibnamefont
  {Hafner}},\ }\bibfield  {title} {\enquote {\bibinfo {title} {Norm-conserving
  and ultrasoft pseudopotentials for first-row and transition elements},}\
  }\href {\doibase 10.1088/0953-8984/6/40/015} {\bibfield  {journal} {\bibinfo
  {journal} {Journal of Physics: Condensed Matter}\ }\textbf {\bibinfo {volume}
  {6}},\ \bibinfo {pages} {8245--8257} (\bibinfo {year} {1994})}\BibitemShut
  {NoStop}%
\bibitem [{\citenamefont {Perdew}, \citenamefont {Burke},\ and\ \citenamefont
  {Ernzerhof}(1996)}]{perdew_generalized_1996}%
  \BibitemOpen
  \bibfield  {author} {\bibinfo {author} {\bibfnamefont {J.~P.}\ \bibnamefont
  {Perdew}}, \bibinfo {author} {\bibfnamefont {K.}~\bibnamefont {Burke}}, \
  and\ \bibinfo {author} {\bibfnamefont {M.}~\bibnamefont {Ernzerhof}},\
  }\bibfield  {title} {\enquote {\bibinfo {title} {Generalized {Gradient}
  {Approximation} {Made} {Simple}},}\ }\href {\doibase
  10.1103/PhysRevLett.77.3865} {\bibfield  {journal} {\bibinfo  {journal}
  {Physical Review Letters}\ }\textbf {\bibinfo {volume} {77}},\ \bibinfo
  {pages} {3865--3868} (\bibinfo {year} {1996})}\BibitemShut {NoStop}%
\bibitem [{\citenamefont {Kondati~Natarajan}\ and\ \citenamefont
  {Behler}(2017{\natexlab{b}})}]{kondati_natarajan_self-diffusion_2017}%
  \BibitemOpen
  \bibfield  {author} {\bibinfo {author} {\bibfnamefont {S.}~\bibnamefont
  {Kondati~Natarajan}}\ and\ \bibinfo {author} {\bibfnamefont {J.}~\bibnamefont
  {Behler}},\ }\bibfield  {title} {\enquote {\bibinfo {title} {Self-{Diffusion}
  of {Surface} {Defects} at {Copper}–{Water} {Interfaces}},}\ }\href
  {\doibase 10.1021/acs.jpcc.6b12657} {\bibfield  {journal} {\bibinfo
  {journal} {The Journal of Physical Chemistry C}\ }\textbf {\bibinfo {volume}
  {121}},\ \bibinfo {pages} {4368--4383} (\bibinfo {year}
  {2017}{\natexlab{b}})}\BibitemShut {NoStop}%
\bibitem [{\citenamefont {Quaranta}, \citenamefont {Behler},\ and\
  \citenamefont {Hellström}(2019)}]{quaranta_structure_2019}%
  \BibitemOpen
  \bibfield  {author} {\bibinfo {author} {\bibfnamefont {V.}~\bibnamefont
  {Quaranta}}, \bibinfo {author} {\bibfnamefont {J.}~\bibnamefont {Behler}}, \
  and\ \bibinfo {author} {\bibfnamefont {M.}~\bibnamefont {Hellström}},\
  }\bibfield  {title} {\enquote {\bibinfo {title} {Structure and {Dynamics} of
  the {Liquid}–{Water}/{Zinc}-{Oxide} {Interface} from {Machine} {Learning}
  {Potential} {Simulations}},}\ }\href {\doibase 10.1021/acs.jpcc.8b10781}
  {\bibfield  {journal} {\bibinfo  {journal} {The Journal of Physical Chemistry
  C}\ }\textbf {\bibinfo {volume} {123}},\ \bibinfo {pages} {1293--1304}
  (\bibinfo {year} {2019})}\BibitemShut {NoStop}%
\bibitem [{\citenamefont {Morawietz}\ \emph {et~al.}(2016)\citenamefont
  {Morawietz}, \citenamefont {Singraber}, \citenamefont {Dellago},\ and\
  \citenamefont {Behler}}]{P4556}%
  \BibitemOpen
  \bibfield  {author} {\bibinfo {author} {\bibfnamefont {T.}~\bibnamefont
  {Morawietz}}, \bibinfo {author} {\bibfnamefont {A.}~\bibnamefont
  {Singraber}}, \bibinfo {author} {\bibfnamefont {C.}~\bibnamefont {Dellago}},
  \ and\ \bibinfo {author} {\bibfnamefont {J.}~\bibnamefont {Behler}},\
  }\bibfield  {title} {\enquote {\bibinfo {title} {How van der waals
  interactions determine the unique properties of water},}\ }\href@noop {}
  {\bibfield  {journal} {\bibinfo  {journal} {PNAS}\ }\textbf {\bibinfo
  {volume} {113}},\ \bibinfo {pages} {8368} (\bibinfo {year}
  {2016})}\BibitemShut {NoStop}%
\bibitem [{\citenamefont {Grimme}\ \emph {et~al.}(2010)\citenamefont {Grimme},
  \citenamefont {Antony}, \citenamefont {Ehrlich},\ and\ \citenamefont
  {Krieg}}]{P3112}%
  \BibitemOpen
  \bibfield  {author} {\bibinfo {author} {\bibfnamefont {S.}~\bibnamefont
  {Grimme}}, \bibinfo {author} {\bibfnamefont {J.}~\bibnamefont {Antony}},
  \bibinfo {author} {\bibfnamefont {S.}~\bibnamefont {Ehrlich}}, \ and\
  \bibinfo {author} {\bibfnamefont {H.}~\bibnamefont {Krieg}},\ }\bibfield
  {title} {\enquote {\bibinfo {title} {A consistent and accurate ab initio
  parametrization of density functional dispersion correction (dft-d) for the
  94 elements h-pu},}\ }\href@noop {} {\bibfield  {journal} {\bibinfo
  {journal} {J. Chem. Phys.}\ }\textbf {\bibinfo {volume} {132}},\ \bibinfo
  {pages} {154104} (\bibinfo {year} {2010})}\BibitemShut {NoStop}%
\bibitem [{\citenamefont {Tkatchenko}\ \emph {et~al.}(2012)\citenamefont
  {Tkatchenko}, \citenamefont {Robert A.~DiStasio}, \citenamefont {Car},\ and\
  \citenamefont {Scheffler}}]{P3296}%
  \BibitemOpen
  \bibfield  {author} {\bibinfo {author} {\bibfnamefont {A.}~\bibnamefont
  {Tkatchenko}}, \bibinfo {author} {\bibfnamefont {J.}~\bibnamefont {Robert
  A.~DiStasio}}, \bibinfo {author} {\bibfnamefont {R.}~\bibnamefont {Car}}, \
  and\ \bibinfo {author} {\bibfnamefont {M.}~\bibnamefont {Scheffler}},\
  }\bibfield  {title} {\enquote {\bibinfo {title} {Accurate and efficient
  method for many-body van der waals interactions},}\ }\href@noop {} {\bibfield
   {journal} {\bibinfo  {journal} {Phys. Rev. Lett.}\ }\textbf {\bibinfo
  {volume} {108}},\ \bibinfo {pages} {236402} (\bibinfo {year}
  {2012})}\BibitemShut {NoStop}%
\bibitem [{\citenamefont {Bučko}\ \emph {et~al.}(2014)\citenamefont {Bučko},
  \citenamefont {Lebègue}, \citenamefont {Ángyán},\ and\ \citenamefont
  {Hafner}}]{bucko_extending_2014}%
  \BibitemOpen
  \bibfield  {author} {\bibinfo {author} {\bibfnamefont {T.}~\bibnamefont
  {Bučko}}, \bibinfo {author} {\bibfnamefont {S.}~\bibnamefont {Lebègue}},
  \bibinfo {author} {\bibfnamefont {J.~G.}\ \bibnamefont {Ángyán}}, \ and\
  \bibinfo {author} {\bibfnamefont {J.}~\bibnamefont {Hafner}},\ }\bibfield
  {title} {\enquote {\bibinfo {title} {Extending the applicability of the
  {Tkatchenko}-{Scheffler} dispersion correction via iterative {Hirshfeld}
  partitioning},}\ }\href {\doibase 10.1063/1.4890003} {\bibfield  {journal}
  {\bibinfo  {journal} {The Journal of Chemical Physics}\ }\textbf {\bibinfo
  {volume} {141}},\ \bibinfo {pages} {034114} (\bibinfo {year}
  {2014})}\BibitemShut {NoStop}%
\bibitem [{\citenamefont {Antony}\ and\ \citenamefont
  {Grimme}(2006)}]{antony_density_2006}%
  \BibitemOpen
  \bibfield  {author} {\bibinfo {author} {\bibfnamefont {J.}~\bibnamefont
  {Antony}}\ and\ \bibinfo {author} {\bibfnamefont {S.}~\bibnamefont
  {Grimme}},\ }\bibfield  {title} {\enquote {\bibinfo {title} {Density
  functional theory including dispersion corrections for intermolecular
  interactions in a large benchmark set of biologically relevant molecules},}\
  }\href {\doibase 10.1039/B612585A} {\bibfield  {journal} {\bibinfo  {journal}
  {Physical Chemistry Chemical Physics}\ }\textbf {\bibinfo {volume} {8}},\
  \bibinfo {pages} {5287--5293} (\bibinfo {year} {2006})}\BibitemShut {NoStop}%
\bibitem [{\citenamefont {Kofke}\ and\ \citenamefont
  {Glandt}(1988)}]{kofke_monte_1988}%
  \BibitemOpen
  \bibfield  {author} {\bibinfo {author} {\bibfnamefont {D.~A.}\ \bibnamefont
  {Kofke}}\ and\ \bibinfo {author} {\bibfnamefont {E.~D.}\ \bibnamefont
  {Glandt}},\ }\bibfield  {title} {\enquote {\bibinfo {title} {Monte {Carlo}
  simulation of multicomponent equilibria in a semigrand canonical ensemble},}\
  }\href {\doibase 10.1080/00268978800100743} {\bibfield  {journal} {\bibinfo
  {journal} {Molecular Physics}\ }\textbf {\bibinfo {volume} {64}},\ \bibinfo
  {pages} {1105--1131} (\bibinfo {year} {1988})}\BibitemShut {NoStop}%
\bibitem [{\citenamefont {Imbalzano}\ \emph {et~al.}(2018)\citenamefont
  {Imbalzano}, \citenamefont {Anelli}, \citenamefont {Giofre}, \citenamefont
  {Klees}, \citenamefont {Behler},\ and\ \citenamefont {Ceriotti}}]{P5398}%
  \BibitemOpen
  \bibfield  {author} {\bibinfo {author} {\bibfnamefont {G.}~\bibnamefont
  {Imbalzano}}, \bibinfo {author} {\bibfnamefont {A.}~\bibnamefont {Anelli}},
  \bibinfo {author} {\bibfnamefont {D.}~\bibnamefont {Giofre}}, \bibinfo
  {author} {\bibfnamefont {S.}~\bibnamefont {Klees}}, \bibinfo {author}
  {\bibfnamefont {J.}~\bibnamefont {Behler}}, \ and\ \bibinfo {author}
  {\bibfnamefont {M.}~\bibnamefont {Ceriotti}},\ }\bibfield  {title} {\enquote
  {\bibinfo {title} {Automatic selection of atomic fingerprints and reference
  configurations for machine-learning potentials},}\ }\href@noop {} {\bibfield
  {journal} {\bibinfo  {journal} {J. Chem. Phys.}\ }\textbf {\bibinfo {volume}
  {148}},\ \bibinfo {pages} {241730} (\bibinfo {year} {2018})}\BibitemShut
  {NoStop}%
\bibitem [{\citenamefont {Eckhoff}\ and\ \citenamefont
  {Behler}(2019)}]{eckhoff_molecular_2019}%
  \BibitemOpen
  \bibfield  {author} {\bibinfo {author} {\bibfnamefont {M.}~\bibnamefont
  {Eckhoff}}\ and\ \bibinfo {author} {\bibfnamefont {J.}~\bibnamefont
  {Behler}},\ }\bibfield  {title} {\enquote {\bibinfo {title} {From {Molecular}
  {Fragments} to the {Bulk}: {Development} of a {Neural} {Network} {Potential}
  for {MOF}-5},}\ }\href {\doibase 10.1021/acs.jctc.8b01288} {\bibfield
  {journal} {\bibinfo  {journal} {Journal of Chemical Theory and Computation}\
  }\textbf {\bibinfo {volume} {15}},\ \bibinfo {pages} {3793--3809} (\bibinfo
  {year} {2019})}\BibitemShut {NoStop}%
\bibitem [{\citenamefont {Liu}\ and\ \citenamefont
  {Nocedal}(1989)}]{liu_limited_1989}%
  \BibitemOpen
  \bibfield  {author} {\bibinfo {author} {\bibfnamefont {D.~C.}\ \bibnamefont
  {Liu}}\ and\ \bibinfo {author} {\bibfnamefont {J.}~\bibnamefont {Nocedal}},\
  }\bibfield  {title} {\enquote {\bibinfo {title} {On the limited memory {BFGS}
  method for large scale optimization},}\ }\href {\doibase 10.1007/BF01589116}
  {\bibfield  {journal} {\bibinfo  {journal} {Mathematical Programming}\
  }\textbf {\bibinfo {volume} {45}},\ \bibinfo {pages} {503--528} (\bibinfo
  {year} {1989})}\BibitemShut {NoStop}%
\bibitem [{\citenamefont {Larsen}\ \emph {et~al.}(2017)\citenamefont {Larsen},
  \citenamefont {Mortensen}, \citenamefont {Blomqvist}, \citenamefont
  {Castelli}, \citenamefont {Christensen}, \citenamefont {Dulak}, \citenamefont
  {Friis}, \citenamefont {Groves}, \citenamefont {Hammer}, \citenamefont
  {Hargus}, \citenamefont {Hermes}, \citenamefont {Jennings}, \citenamefont
  {Jensen}, \citenamefont {Kermode}, \citenamefont {Kitchin}, \citenamefont
  {Kolsbjerg}, \citenamefont {Kubal}, \citenamefont {Kaasbjerg}, \citenamefont
  {Lysgaard}, \citenamefont {Maronsson}, \citenamefont {Maxson}, \citenamefont
  {Olsen}, \citenamefont {Pastewka}, \citenamefont {Peterson}, \citenamefont
  {Rostgaard}, \citenamefont {Schiøtz}, \citenamefont {Schütt}, \citenamefont
  {Strange}, \citenamefont {Thygesen}, \citenamefont {Vegge}, \citenamefont
  {Vilhelmsen}, \citenamefont {Walter}, \citenamefont {Zeng},\ and\
  \citenamefont {Jacobsen}}]{larsen_atomic_2017}%
  \BibitemOpen
  \bibfield  {author} {\bibinfo {author} {\bibfnamefont {A.~H.}\ \bibnamefont
  {Larsen}}, \bibinfo {author} {\bibfnamefont {J.~J.}\ \bibnamefont
  {Mortensen}}, \bibinfo {author} {\bibfnamefont {J.}~\bibnamefont
  {Blomqvist}}, \bibinfo {author} {\bibfnamefont {I.~E.}\ \bibnamefont
  {Castelli}}, \bibinfo {author} {\bibfnamefont {R.}~\bibnamefont
  {Christensen}}, \bibinfo {author} {\bibfnamefont {M.}~\bibnamefont {Dulak}},
  \bibinfo {author} {\bibfnamefont {J.}~\bibnamefont {Friis}}, \bibinfo
  {author} {\bibfnamefont {M.~N.}\ \bibnamefont {Groves}}, \bibinfo {author}
  {\bibfnamefont {B.}~\bibnamefont {Hammer}}, \bibinfo {author} {\bibfnamefont
  {C.}~\bibnamefont {Hargus}}, \bibinfo {author} {\bibfnamefont {E.~D.}\
  \bibnamefont {Hermes}}, \bibinfo {author} {\bibfnamefont {P.~C.}\
  \bibnamefont {Jennings}}, \bibinfo {author} {\bibfnamefont {P.~B.}\
  \bibnamefont {Jensen}}, \bibinfo {author} {\bibfnamefont {J.}~\bibnamefont
  {Kermode}}, \bibinfo {author} {\bibfnamefont {J.~R.}\ \bibnamefont
  {Kitchin}}, \bibinfo {author} {\bibfnamefont {E.~L.}\ \bibnamefont
  {Kolsbjerg}}, \bibinfo {author} {\bibfnamefont {J.}~\bibnamefont {Kubal}},
  \bibinfo {author} {\bibfnamefont {K.}~\bibnamefont {Kaasbjerg}}, \bibinfo
  {author} {\bibfnamefont {S.}~\bibnamefont {Lysgaard}}, \bibinfo {author}
  {\bibfnamefont {J.~B.}\ \bibnamefont {Maronsson}}, \bibinfo {author}
  {\bibfnamefont {T.}~\bibnamefont {Maxson}}, \bibinfo {author} {\bibfnamefont
  {T.}~\bibnamefont {Olsen}}, \bibinfo {author} {\bibfnamefont
  {L.}~\bibnamefont {Pastewka}}, \bibinfo {author} {\bibfnamefont
  {A.}~\bibnamefont {Peterson}}, \bibinfo {author} {\bibfnamefont
  {C.}~\bibnamefont {Rostgaard}}, \bibinfo {author} {\bibfnamefont
  {J.}~\bibnamefont {Schiøtz}}, \bibinfo {author} {\bibfnamefont
  {O.}~\bibnamefont {Schütt}}, \bibinfo {author} {\bibfnamefont
  {M.}~\bibnamefont {Strange}}, \bibinfo {author} {\bibfnamefont {K.~S.}\
  \bibnamefont {Thygesen}}, \bibinfo {author} {\bibfnamefont {T.}~\bibnamefont
  {Vegge}}, \bibinfo {author} {\bibfnamefont {L.}~\bibnamefont {Vilhelmsen}},
  \bibinfo {author} {\bibfnamefont {M.}~\bibnamefont {Walter}}, \bibinfo
  {author} {\bibfnamefont {Z.}~\bibnamefont {Zeng}}, \ and\ \bibinfo {author}
  {\bibfnamefont {K.~W.}\ \bibnamefont {Jacobsen}},\ }\bibfield  {title}
  {\enquote {\bibinfo {title} {The atomic simulation environment—a {Python}
  library for working with atoms},}\ }\href {\doibase 10.1088/1361-648X/aa680e}
  {\bibfield  {journal} {\bibinfo  {journal} {Journal of Physics: Condensed
  Matter}\ }\textbf {\bibinfo {volume} {29}},\ \bibinfo {pages} {273002}
  (\bibinfo {year} {2017})}\BibitemShut {NoStop}%
\bibitem [{\citenamefont {Cui}\ \emph {et~al.}(2017)\citenamefont {Cui},
  \citenamefont {Lu}, \citenamefont {Jiang}, \citenamefont {Cao},\ and\
  \citenamefont {Meng}}]{cui_phase_2017}%
  \BibitemOpen
  \bibfield  {author} {\bibinfo {author} {\bibfnamefont {M.}~\bibnamefont
  {Cui}}, \bibinfo {author} {\bibfnamefont {H.}~\bibnamefont {Lu}}, \bibinfo
  {author} {\bibfnamefont {H.}~\bibnamefont {Jiang}}, \bibinfo {author}
  {\bibfnamefont {Z.}~\bibnamefont {Cao}}, \ and\ \bibinfo {author}
  {\bibfnamefont {X.}~\bibnamefont {Meng}},\ }\bibfield  {title} {\enquote
  {\bibinfo {title} {Phase {Diagram} of {Continuous} {Binary} {Nanoalloys}:
  {Size}, {Shape}, and {Segregation} {Effects}},}\ }\href {\doibase
  10.1038/srep41990} {\bibfield  {journal} {\bibinfo  {journal} {Scientific
  Reports}\ }\textbf {\bibinfo {volume} {7}},\ \bibinfo {pages} {1--10}
  (\bibinfo {year} {2017})}\BibitemShut {NoStop}%
\bibitem [{\citenamefont {Singraber}, \citenamefont {Behler},\ and\
  \citenamefont {Dellago}(2019)}]{singraber_library-based_2019}%
  \BibitemOpen
  \bibfield  {author} {\bibinfo {author} {\bibfnamefont {A.}~\bibnamefont
  {Singraber}}, \bibinfo {author} {\bibfnamefont {J.}~\bibnamefont {Behler}}, \
  and\ \bibinfo {author} {\bibfnamefont {C.}~\bibnamefont {Dellago}},\
  }\bibfield  {title} {\enquote {\bibinfo {title} {Library-{Based} {LAMMPS}
  {Implementation} of {High}-{Dimensional} {Neural} {Network} {Potentials}},}\
  }\href {\doibase 10.1021/acs.jctc.8b00770} {\bibfield  {journal} {\bibinfo
  {journal} {Journal of Chemical Theory and Computation}\ }\textbf {\bibinfo
  {volume} {15}},\ \bibinfo {pages} {1827--1840} (\bibinfo {year}
  {2019})}\BibitemShut {NoStop}%
\bibitem [{\citenamefont {Plimpton}(1995)}]{plimpton_fast_1995}%
  \BibitemOpen
  \bibfield  {author} {\bibinfo {author} {\bibfnamefont {S.}~\bibnamefont
  {Plimpton}},\ }\bibfield  {title} {\enquote {\bibinfo {title} {Fast
  {Parallel} {Algorithms} for {Short}-{Range} {Molecular} {Dynamics}},}\ }\href
  {\doibase 10.1006/jcph.1995.1039} {\bibfield  {journal} {\bibinfo  {journal}
  {Journal of Computational Physics}\ }\textbf {\bibinfo {volume} {117}},\
  \bibinfo {pages} {1--19} (\bibinfo {year} {1995})}\BibitemShut {NoStop}%
\bibitem [{\citenamefont {Lide}(2009)}]{lide_david_r_handbook_2009}%
  \BibitemOpen
  \bibfield  {author} {\bibinfo {author} {\bibfnamefont {D.~R.}\ \bibnamefont
  {Lide}},\ }\href@noop {} {\emph {\bibinfo {title} {Handbook of {Chemistry}
  and {Physics}}}},\ \bibinfo {edition} {89th}\ ed.\ (\bibinfo  {publisher}
  {CRC Press},\ \bibinfo {year} {2009})\BibitemShut {NoStop}%
\bibitem [{\citenamefont {Meyer}\ and\ \citenamefont
  {Marx}(2003)}]{meyer_density-functional_2003}%
  \BibitemOpen
  \bibfield  {author} {\bibinfo {author} {\bibfnamefont {B.}~\bibnamefont
  {Meyer}}\ and\ \bibinfo {author} {\bibfnamefont {D.}~\bibnamefont {Marx}},\
  }\bibfield  {title} {\enquote {\bibinfo {title} {Density-functional study of
  the structure and stability of {ZnO} surfaces},}\ }\href {\doibase
  10.1103/PhysRevB.67.035403} {\bibfield  {journal} {\bibinfo  {journal}
  {Physical Review B}\ }\textbf {\bibinfo {volume} {67}},\ \bibinfo {pages}
  {035403} (\bibinfo {year} {2003})}\BibitemShut {NoStop}%
\bibitem [{\citenamefont {Henry}(2005)}]{henry_morphology_2005}%
  \BibitemOpen
  \bibfield  {author} {\bibinfo {author} {\bibfnamefont {C.~R.}\ \bibnamefont
  {Henry}},\ }\bibfield  {title} {\enquote {\bibinfo {title} {Morphology of
  supported nanoparticles},}\ }\href {\doibase 10.1016/j.progsurf.2005.09.004}
  {\bibfield  {journal} {\bibinfo  {journal} {Progress in Surface Science}\
  }\textbf {\bibinfo {volume} {80}},\ \bibinfo {pages} {92--116} (\bibinfo
  {year} {2005})}\BibitemShut {NoStop}%
\bibitem [{\citenamefont {Henkelman}, \citenamefont {Uberuaga},\ and\
  \citenamefont {Jonsson}(2000)}]{P0951}%
  \BibitemOpen
  \bibfield  {author} {\bibinfo {author} {\bibfnamefont {G.}~\bibnamefont
  {Henkelman}}, \bibinfo {author} {\bibfnamefont {B.~P.}\ \bibnamefont
  {Uberuaga}}, \ and\ \bibinfo {author} {\bibfnamefont {H.}~\bibnamefont
  {Jonsson}},\ }\bibfield  {title} {\enquote {\bibinfo {title} {A climbing
  image nudged elastic band method for finding saddle points and minimum energy
  paths},}\ }\href@noop {} {\bibfield  {journal} {\bibinfo  {journal} {J. Chem.
  Phys.}\ }\textbf {\bibinfo {volume} {113}},\ \bibinfo {pages} {9901}
  (\bibinfo {year} {2000})}\BibitemShut {NoStop}%
\bibitem [{\citenamefont {Eckhoff}, \citenamefont {Schebarchov},\ and\
  \citenamefont {Wales}(2017)}]{eckhoff_structure_2017}%
  \BibitemOpen
  \bibfield  {author} {\bibinfo {author} {\bibfnamefont {M.}~\bibnamefont
  {Eckhoff}}, \bibinfo {author} {\bibfnamefont {D.}~\bibnamefont
  {Schebarchov}}, \ and\ \bibinfo {author} {\bibfnamefont {D.~J.}\ \bibnamefont
  {Wales}},\ }\bibfield  {title} {\enquote {\bibinfo {title} {Structure and
  {Thermodynamics} of {Metal} {Clusters} on {Atomically} {Smooth}
  {Substrates}},}\ }\href {\doibase 10.1021/acs.jpclett.7b02543} {\bibfield
  {journal} {\bibinfo  {journal} {The Journal of Physical Chemistry Letters}\
  }\textbf {\bibinfo {volume} {8}},\ \bibinfo {pages} {5402--5407} (\bibinfo
  {year} {2017})}\BibitemShut {NoStop}%
\bibitem [{\citenamefont {Wulff}(1901)}]{wulff_zur_1901}%
  \BibitemOpen
  \bibfield  {author} {\bibinfo {author} {\bibfnamefont {G.}~\bibnamefont
  {Wulff}},\ }\bibfield  {title} {\enquote {\bibinfo {title} {Zur {Frage} der
  {Geschwindigkeit} des {Wachsthums} und der {Auflösung} der
  {Krystallflächen}},}\ }\href {\doibase 10.1524/zkri.1901.34.1.449}
  {\bibfield  {journal} {\bibinfo  {journal} {Zeitschrift für Kristallographie
  - Crystalline Materials}\ }\textbf {\bibinfo {volume} {34}},\ \bibinfo
  {pages} {449--530} (\bibinfo {year} {1901})}\BibitemShut {NoStop}%
\bibitem [{\citenamefont {Ringe}, \citenamefont {Van~Duyne},\ and\
  \citenamefont {Marks}(2011)}]{ringe_wulff_2011}%
  \BibitemOpen
  \bibfield  {author} {\bibinfo {author} {\bibfnamefont {E.}~\bibnamefont
  {Ringe}}, \bibinfo {author} {\bibfnamefont {R.~P.}\ \bibnamefont
  {Van~Duyne}}, \ and\ \bibinfo {author} {\bibfnamefont {L.~D.}\ \bibnamefont
  {Marks}},\ }\bibfield  {title} {\enquote {\bibinfo {title} {Wulff
  {Construction} for {Alloy} {Nanoparticles}},}\ }\href {\doibase
  10.1021/nl2018146} {\bibfield  {journal} {\bibinfo  {journal} {Nano Letters}\
  }\textbf {\bibinfo {volume} {11}},\ \bibinfo {pages} {3399--3403} (\bibinfo
  {year} {2011})}\BibitemShut {NoStop}%
\bibitem [{\citenamefont {Tauster}(1987)}]{tauster_strong_1987}%
  \BibitemOpen
  \bibfield  {author} {\bibinfo {author} {\bibfnamefont {S.~J.}\ \bibnamefont
  {Tauster}},\ }\bibfield  {title} {\enquote {\bibinfo {title} {Strong
  metal-support interactions},}\ }\href {\doibase 10.1021/ar00143a001}
  {\bibfield  {journal} {\bibinfo  {journal} {Accounts of Chemical Research}\
  }\textbf {\bibinfo {volume} {20}},\ \bibinfo {pages} {389--394} (\bibinfo
  {year} {1987})}\BibitemShut {NoStop}%
\bibitem [{\citenamefont {Pan}\ \emph {et~al.}(2017)\citenamefont {Pan},
  \citenamefont {Tsai}, \citenamefont {Su}, \citenamefont {Rick}, \citenamefont
  {Akalework}, \citenamefont {Agegnehu}, \citenamefont {Cheng},\ and\
  \citenamefont {Hwang}}]{pan_tuning_2017}%
  \BibitemOpen
  \bibfield  {author} {\bibinfo {author} {\bibfnamefont {C.-J.}\ \bibnamefont
  {Pan}}, \bibinfo {author} {\bibfnamefont {M.-C.}\ \bibnamefont {Tsai}},
  \bibinfo {author} {\bibfnamefont {W.-N.}\ \bibnamefont {Su}}, \bibinfo
  {author} {\bibfnamefont {J.}~\bibnamefont {Rick}}, \bibinfo {author}
  {\bibfnamefont {N.~G.}\ \bibnamefont {Akalework}}, \bibinfo {author}
  {\bibfnamefont {A.~K.}\ \bibnamefont {Agegnehu}}, \bibinfo {author}
  {\bibfnamefont {S.-Y.}\ \bibnamefont {Cheng}}, \ and\ \bibinfo {author}
  {\bibfnamefont {B.-J.}\ \bibnamefont {Hwang}},\ }\bibfield  {title} {\enquote
  {\bibinfo {title} {Tuning and exploiting {Strong} {Metal}-{Support}
  {Interaction} ({SMSI}) in {Heterogeneous} {Catalysis}},}\ }\href {\doibase
  10.1016/j.jtice.2017.02.012} {\bibfield  {journal} {\bibinfo  {journal}
  {Journal of the Taiwan Institute of Chemical Engineers}\ }\textbf {\bibinfo
  {volume} {74}},\ \bibinfo {pages} {154--186} (\bibinfo {year}
  {2017})}\BibitemShut {NoStop}%
\bibitem [{\citenamefont {Figueiredo}\ \emph {et~al.}(2019)\citenamefont
  {Figueiredo}, \citenamefont {Della~Mea}, \citenamefont {Segala},
  \citenamefont {Baptista}, \citenamefont {Escudero}, \citenamefont
  {Pérez-Dieste},\ and\ \citenamefont
  {Bernardi}}]{figueiredo_understanding_2019}%
  \BibitemOpen
  \bibfield  {author} {\bibinfo {author} {\bibfnamefont {W.~T.}\ \bibnamefont
  {Figueiredo}}, \bibinfo {author} {\bibfnamefont {G.~B.}\ \bibnamefont
  {Della~Mea}}, \bibinfo {author} {\bibfnamefont {M.}~\bibnamefont {Segala}},
  \bibinfo {author} {\bibfnamefont {D.~L.}\ \bibnamefont {Baptista}}, \bibinfo
  {author} {\bibfnamefont {C.}~\bibnamefont {Escudero}}, \bibinfo {author}
  {\bibfnamefont {V.}~\bibnamefont {Pérez-Dieste}}, \ and\ \bibinfo {author}
  {\bibfnamefont {F.}~\bibnamefont {Bernardi}},\ }\bibfield  {title} {\enquote
  {\bibinfo {title} {Understanding the {Strong} {Metal}–{Support}
  {Interaction} ({SMSI}) {Effect} in {Cu$_x$Ni$_{1-x}$}/{CeO$_2$} (0
  {\textless} x {\textless} 1) {Nanoparticles} for {Enhanced} {Catalysis}},}\
  }\href {\doibase 10.1021/acsanm.9b00569} {\bibfield  {journal} {\bibinfo
  {journal} {ACS Applied Nano Materials}\ }\textbf {\bibinfo {volume} {2}},\
  \bibinfo {pages} {2559--2573} (\bibinfo {year} {2019})}\BibitemShut {NoStop}%
\bibitem [{\citenamefont {Koda}\ \emph {et~al.}(2016)\citenamefont {Koda},
  \citenamefont {Bechstedt}, \citenamefont {Marques},\ and\ \citenamefont
  {Teles}}]{koda_coincidence_2016}%
  \BibitemOpen
  \bibfield  {author} {\bibinfo {author} {\bibfnamefont {D.~S.}\ \bibnamefont
  {Koda}}, \bibinfo {author} {\bibfnamefont {F.}~\bibnamefont {Bechstedt}},
  \bibinfo {author} {\bibfnamefont {M.}~\bibnamefont {Marques}}, \ and\
  \bibinfo {author} {\bibfnamefont {L.~K.}\ \bibnamefont {Teles}},\ }\bibfield
  {title} {\enquote {\bibinfo {title} {Coincidence {Lattices} of {2D}
  {Crystals}: {Heterostructure} {Predictions} and {Applications}},}\ }\href
  {\doibase 10.1021/acs.jpcc.6b01496} {\bibfield  {journal} {\bibinfo
  {journal} {The Journal of Physical Chemistry C}\ }\textbf {\bibinfo {volume}
  {120}},\ \bibinfo {pages} {10895--10908} (\bibinfo {year}
  {2016})}\BibitemShut {NoStop}%
\bibitem [{\citenamefont {Mavrikakis}, \citenamefont {Hammer},\ and\
  \citenamefont {Nørskov}(1998)}]{mavrikakis_effect_1998}%
  \BibitemOpen
  \bibfield  {author} {\bibinfo {author} {\bibfnamefont {M.}~\bibnamefont
  {Mavrikakis}}, \bibinfo {author} {\bibfnamefont {B.}~\bibnamefont {Hammer}},
  \ and\ \bibinfo {author} {\bibfnamefont {J.~K.}\ \bibnamefont {Nørskov}},\
  }\bibfield  {title} {\enquote {\bibinfo {title} {Effect of {Strain} on the
  {Reactivity} of {Metal} {Surfaces}},}\ }\href {\doibase
  10.1103/PhysRevLett.81.2819} {\bibfield  {journal} {\bibinfo  {journal}
  {Physical Review Letters}\ }\textbf {\bibinfo {volume} {81}},\ \bibinfo
  {pages} {2819--2822} (\bibinfo {year} {1998})}\BibitemShut {NoStop}%
\bibitem [{\citenamefont {Amakawa}\ \emph {et~al.}(2013)\citenamefont
  {Amakawa}, \citenamefont {Sun}, \citenamefont {Guo}, \citenamefont
  {Hävecker}, \citenamefont {Kube}, \citenamefont {Wachs}, \citenamefont
  {Lwin}, \citenamefont {Frenkel}, \citenamefont {Patlolla}, \citenamefont
  {Hermann}, \citenamefont {Schlögl},\ and\ \citenamefont
  {Trunschke}}]{amakawa_how_2013}%
  \BibitemOpen
  \bibfield  {author} {\bibinfo {author} {\bibfnamefont {K.}~\bibnamefont
  {Amakawa}}, \bibinfo {author} {\bibfnamefont {L.}~\bibnamefont {Sun}},
  \bibinfo {author} {\bibfnamefont {C.}~\bibnamefont {Guo}}, \bibinfo {author}
  {\bibfnamefont {M.}~\bibnamefont {Hävecker}}, \bibinfo {author}
  {\bibfnamefont {P.}~\bibnamefont {Kube}}, \bibinfo {author} {\bibfnamefont
  {I.~E.}\ \bibnamefont {Wachs}}, \bibinfo {author} {\bibfnamefont
  {S.}~\bibnamefont {Lwin}}, \bibinfo {author} {\bibfnamefont {A.~I.}\
  \bibnamefont {Frenkel}}, \bibinfo {author} {\bibfnamefont {A.}~\bibnamefont
  {Patlolla}}, \bibinfo {author} {\bibfnamefont {K.}~\bibnamefont {Hermann}},
  \bibinfo {author} {\bibfnamefont {R.}~\bibnamefont {Schlögl}}, \ and\
  \bibinfo {author} {\bibfnamefont {A.}~\bibnamefont {Trunschke}},\ }\bibfield
  {title} {\enquote {\bibinfo {title} {How {Strain} {Affects} the {Reactivity}
  of {Surface} {Metal} {Oxide} {Catalysts}},}\ }\href {\doibase
  10.1002/anie.201306620} {\bibfield  {journal} {\bibinfo  {journal}
  {Angewandte Chemie International Edition}\ }\textbf {\bibinfo {volume}
  {52}},\ \bibinfo {pages} {13553--13557} (\bibinfo {year} {2013})}\BibitemShut
  {NoStop}%
\bibitem [{\citenamefont {Bissett}\ \emph {et~al.}(2013)\citenamefont
  {Bissett}, \citenamefont {Konabe}, \citenamefont {Okada}, \citenamefont
  {Tsuji},\ and\ \citenamefont {Ago}}]{bissett_enhanced_2013}%
  \BibitemOpen
  \bibfield  {author} {\bibinfo {author} {\bibfnamefont {M.~A.}\ \bibnamefont
  {Bissett}}, \bibinfo {author} {\bibfnamefont {S.}~\bibnamefont {Konabe}},
  \bibinfo {author} {\bibfnamefont {S.}~\bibnamefont {Okada}}, \bibinfo
  {author} {\bibfnamefont {M.}~\bibnamefont {Tsuji}}, \ and\ \bibinfo {author}
  {\bibfnamefont {H.}~\bibnamefont {Ago}},\ }\bibfield  {title} {\enquote
  {\bibinfo {title} {Enhanced {Chemical} {Reactivity} of {Graphene} {Induced}
  by {Mechanical} {Strain}},}\ }\href {\doibase 10.1021/nn404746h} {\bibfield
  {journal} {\bibinfo  {journal} {ACS Nano}\ }\textbf {\bibinfo {volume} {7}},\
  \bibinfo {pages} {10335--10343} (\bibinfo {year} {2013})}\BibitemShut
  {NoStop}%
\bibitem [{\citenamefont {Tyson}\ and\ \citenamefont
  {Miller}(1977)}]{tyson_surface_1977}%
  \BibitemOpen
  \bibfield  {author} {\bibinfo {author} {\bibfnamefont {W.~R.}\ \bibnamefont
  {Tyson}}\ and\ \bibinfo {author} {\bibfnamefont {W.~A.}\ \bibnamefont
  {Miller}},\ }\bibfield  {title} {\enquote {\bibinfo {title} {Surface free
  energies of solid metals: {Estimation} from liquid surface tension
  measurements},}\ }\href {\doibase 10.1016/0039-6028(77)90442-3} {\bibfield
  {journal} {\bibinfo  {journal} {Surface Science}\ }\textbf {\bibinfo {volume}
  {62}},\ \bibinfo {pages} {267--276} (\bibinfo {year} {1977})}\BibitemShut
  {NoStop}%
\bibitem [{\citenamefont {Tran}\ \emph {et~al.}(2016)\citenamefont {Tran},
  \citenamefont {Xu}, \citenamefont {Radhakrishnan}, \citenamefont {Winston},
  \citenamefont {Sun}, \citenamefont {Persson},\ and\ \citenamefont
  {Ong}}]{tran_surface_2016}%
  \BibitemOpen
  \bibfield  {author} {\bibinfo {author} {\bibfnamefont {R.}~\bibnamefont
  {Tran}}, \bibinfo {author} {\bibfnamefont {Z.}~\bibnamefont {Xu}}, \bibinfo
  {author} {\bibfnamefont {B.}~\bibnamefont {Radhakrishnan}}, \bibinfo {author}
  {\bibfnamefont {D.}~\bibnamefont {Winston}}, \bibinfo {author} {\bibfnamefont
  {W.}~\bibnamefont {Sun}}, \bibinfo {author} {\bibfnamefont {K.~A.}\
  \bibnamefont {Persson}}, \ and\ \bibinfo {author} {\bibfnamefont {S.~P.}\
  \bibnamefont {Ong}},\ }\bibfield  {title} {\enquote {\bibinfo {title}
  {Surface energies of elemental crystals},}\ }\href {\doibase
  10.1038/sdata.2016.80} {\bibfield  {journal} {\bibinfo  {journal} {Scientific
  Data}\ }\textbf {\bibinfo {volume} {3}},\ \bibinfo {pages} {1--13} (\bibinfo
  {year} {2016})}\BibitemShut {NoStop}%
\bibitem [{\citenamefont {Iannuzzi}, \citenamefont {Laio},\ and\ \citenamefont
  {Parrinello}(2003)}]{iannuzzi_efficient_2003}%
  \BibitemOpen
  \bibfield  {author} {\bibinfo {author} {\bibfnamefont {M.}~\bibnamefont
  {Iannuzzi}}, \bibinfo {author} {\bibfnamefont {A.}~\bibnamefont {Laio}}, \
  and\ \bibinfo {author} {\bibfnamefont {M.}~\bibnamefont {Parrinello}},\
  }\bibfield  {title} {\enquote {\bibinfo {title} {Efficient {Exploration} of
  {Reactive} {Potential} {Energy} {Surfaces} {Using} {Car}-{Parrinello}
  {Molecular} {Dynamics}},}\ }\href {\doibase 10.1103/PhysRevLett.90.238302}
  {\bibfield  {journal} {\bibinfo  {journal} {Physical Review Letters}\
  }\textbf {\bibinfo {volume} {90}},\ \bibinfo {pages} {238302} (\bibinfo
  {year} {2003})}\BibitemShut {NoStop}%
\bibitem [{\citenamefont {Barber}, \citenamefont {Dobkin},\ and\ \citenamefont
  {Huhdanpaa}(1996)}]{barber_quickhull_1996}%
  \BibitemOpen
  \bibfield  {author} {\bibinfo {author} {\bibfnamefont {C.~B.}\ \bibnamefont
  {Barber}}, \bibinfo {author} {\bibfnamefont {D.~P.}\ \bibnamefont {Dobkin}},
  \ and\ \bibinfo {author} {\bibfnamefont {H.}~\bibnamefont {Huhdanpaa}},\
  }\bibfield  {title} {\enquote {\bibinfo {title} {The quickhull algorithm for
  convex hulls},}\ }\href {\doibase 10.1145/235815.235821} {\bibfield
  {journal} {\bibinfo  {journal} {ACM Transactions on Mathematical Software}\
  }\textbf {\bibinfo {volume} {22}},\ \bibinfo {pages} {469--483} (\bibinfo
  {year} {1996})}\BibitemShut {NoStop}%
\bibitem [{\citenamefont {Davis}, \citenamefont {Horswell},\ and\ \citenamefont
  {Johnston}(2016)}]{davis_application_2016}%
  \BibitemOpen
  \bibfield  {author} {\bibinfo {author} {\bibfnamefont {J.~B.~A.}\
  \bibnamefont {Davis}}, \bibinfo {author} {\bibfnamefont {S.~L.}\ \bibnamefont
  {Horswell}}, \ and\ \bibinfo {author} {\bibfnamefont {R.~L.}\ \bibnamefont
  {Johnston}},\ }\bibfield  {title} {\enquote {\bibinfo {title} {Application of
  a {Parallel} {Genetic} {Algorithm} to the {Global} {Optimization} of
  {Gas}-{Phase} and {Supported} {Gold}–{Iridium} {Sub}-{Nanoalloys}},}\
  }\href {\doibase 10.1021/acs.jpcc.5b10226} {\bibfield  {journal} {\bibinfo
  {journal} {The Journal of Physical Chemistry C}\ }\textbf {\bibinfo {volume}
  {120}},\ \bibinfo {pages} {3759--3765} (\bibinfo {year} {2016})}\BibitemShut
  {NoStop}%
\bibitem [{\citenamefont {A. Hussein}, \citenamefont {A. Davis},\ and\
  \citenamefont {L. Johnston}(2016)}]{ahussein_dft_2016}%
  \BibitemOpen
  \bibfield  {author} {\bibinfo {author} {\bibfnamefont {H.}~\bibnamefont
  {A. Hussein}}, \bibinfo {author} {\bibfnamefont {J.~B.}\ \bibnamefont
  {A. Davis}}, \ and\ \bibinfo {author} {\bibfnamefont {R.}~\bibnamefont
  {L. Johnston}},\ }\bibfield  {title} {\enquote {\bibinfo {title} {{DFT}
  global optimisation of gas-phase and {MgO}-supported sub-nanometre {AuPd}
  clusters},}\ }\href {\doibase 10.1039/C6CP03958H} {\bibfield  {journal}
  {\bibinfo  {journal} {Physical Chemistry Chemical Physics}\ }\textbf
  {\bibinfo {volume} {18}},\ \bibinfo {pages} {26133--26143} (\bibinfo {year}
  {2016})}\BibitemShut {NoStop}%
\bibitem [{\citenamefont {Bazhenov}\ and\ \citenamefont
  {Honkala}(2019)}]{bazhenov_globally_2019}%
  \BibitemOpen
  \bibfield  {author} {\bibinfo {author} {\bibfnamefont {A.~S.}\ \bibnamefont
  {Bazhenov}}\ and\ \bibinfo {author} {\bibfnamefont {K.}~\bibnamefont
  {Honkala}},\ }\bibfield  {title} {\enquote {\bibinfo {title} {Globally
  {Optimized} {Equilibrium} {Shapes} of {Zirconia}-{Supported} {Rh} and {Pt}
  {Nanoclusters}: {Insights} into {Site} {Assembly} and {Reactivity}},}\ }\href
  {\doibase 10.1021/acs.jpcc.9b00272} {\bibfield  {journal} {\bibinfo
  {journal} {The Journal of Physical Chemistry C}\ }\textbf {\bibinfo {volume}
  {123}},\ \bibinfo {pages} {7209--7216} (\bibinfo {year} {2019})}\BibitemShut
  {NoStop}%
\bibitem [{\citenamefont {Engel}, \citenamefont {Francis},\ and\ \citenamefont
  {Roldan}(2019)}]{engel_influence_2019}%
  \BibitemOpen
  \bibfield  {author} {\bibinfo {author} {\bibfnamefont {J.}~\bibnamefont
  {Engel}}, \bibinfo {author} {\bibfnamefont {S.}~\bibnamefont {Francis}}, \
  and\ \bibinfo {author} {\bibfnamefont {A.}~\bibnamefont {Roldan}},\
  }\bibfield  {title} {\enquote {\bibinfo {title} {The influence of support
  materials on the structural and electronic properties of gold nanoparticles
  – a {DFT} study},}\ }\href {\doibase 10.1039/C9CP03066B} {\bibfield
  {journal} {\bibinfo  {journal} {Physical Chemistry Chemical Physics}\
  }\textbf {\bibinfo {volume} {21}},\ \bibinfo {pages} {19011--19025} (\bibinfo
  {year} {2019})}\BibitemShut {NoStop}%
\end{thebibliography}%

\end{document}

\title[]{Supporting Information}
\author{Mart\'in Leandro Paleico}
\affiliation{Universit\"{a}t G\"{o}ttingen, Institut f\"{u}r Physikalische Chemie, Theoretische Chemie, Tammannstra\ss{}e 6, 37077 G\"{o}ttingen, Germany}
\email{martin.paleico@uni-goettingen.de}
\author{J\"org Behler}
\email{joerg.behler@uni-goettingen.de}
\affiliation{Universit\"{a}t G\"{o}ttingen, Institut f\"{u}r Physikalische Chemie, Theoretische Chemie, Tammannstra\ss{}e 6, 37077 G\"{o}ttingen, Germany}
\affiliation{International Center for Advanced Studies of Energy Conversion (ICASEC), Universit\"at G\"ottingen, Tammannstra\ss{}e 6, 37077 G\"ottingen, Germany}
\date{\today}

\maketitle

\section{Example VASP input file}

\begin{lstlisting}
SYSTEM=VASP CALCULATION

#INITIALIZATION
ISTART = 0     # restart from scratch
ICHARG = 2     # initial charge: from atomic charge densities
INIWAV = 1     # random initialization for wf
NELM   = 100    # maximum of NELM electronic steps
NELMIN =  2    # minimum of NELMIN convergence steps
NELMDL = -5    # no update/self consistency of charge for NELMDL steps, negative for only at the beggining
EDIFF  = 1.00E-06  # accuracy for electronic minimization
PREC = Accurate
GGA = PE #x-c potential
ALGO = Fast
#ADDGRID = .TRUE.

ENCUT  = 500   #energy cutoff of planewaves
ISMEAR = 0; SIGMA = 0.1;
#Gaussian(0)

#MISC
LREAL = A #projection in real or reciprocal space
NSIM = 8
LASPH = .TRUE. #aspherical contributions
\end{lstlisting}

\section{Symmetry Function Set}

\begin{table}[h]
    \centering
    \begin{tabular}{|c|c|c|c|}
    \hline
    Number & $\eta$ (eta) (1/\textrm{Bohr}$^2$) & R$_{\rm shift}$ (\textrm{Bohr}) & Cutoff (\textrm{Bohr}) \\
    \hline
    1 & 0.001 & 0.0 & 12.0 \\ 
    2 & 0.010 & 0.0 & 12.0 \\ 
    3 & 0.020 & 0.0 & 12.0 \\ 
    4 & 0.050 & 0.0 & 12.0 \\ 
    5 & 0.100 & 0.0 & 12.0 \\ 
    6 & 0.200 & 0.0 & 12.0 \\ 
    7 & 0.050 & 3.0 & 12.0 \\ 
    8 & 0.100 & 3.0 & 12.0 \\ 
    9 & 0.200 & 3.0 & 12.0 \\ 
    10 & 0.500 & 3.0 & 12.0 \\ 
    11 & 0.900 & 3.0 & 12.0 \\ 
    \hline
    \end{tabular}
    \caption{Parameters for radial symmetry functions for all element combinations.}
    \label{tab:rsf}
\end{table}

\begin{table}[h]
    \centering
    \begin{tabular}{|c|c|c|c|c|}
    \hline
    Number & $\eta$ (eta) (1/\textrm{Bohr}$^2$) & $\zeta$ (zeta) &  $\lambda$ (lambda) & Cutoff (\textrm{Bohr}) \\
    \hline
    1 & 0.001 & 1.0 & 1.0 & 12.0 \\ 
    2 & 0.001 & 2.0 & 1.0 & 12.0 \\ 
    3 & 0.001 & 4.0 & 1.0 & 12.0 \\ 
    4 & 0.001 & 16.0 & 1.0 & 12.0 \\ 
    5 & 0.001 & 1.0 & -1.0 & 12.0 \\ 
    6 & 0.001 & 2.0 & -1.0 & 12.0 \\ 
    7 & 0.001 & 4.0 & -1.0 & 12.0 \\ 
    8 & 0.001 & 16.0 & -1.0 & 12.0 \\ 
    9 & 0.003 & 1.0 & 1.0 & 12.0 \\ 
    10 & 0.003 & 4.0 & 1.0 & 12.0 \\ 
    11 & 0.003 & 1.0 & -1.0 & 12.0 \\ 
    12 & 0.003 & 4.0 & -1.0 & 12.0 \\ 
    \hline
    \end{tabular}
    \caption{Parameters for angular symmetry functions for all element combinations.}
    \label{tab:asf}
\end{table}

\section{Energy plot}

\begin{figure}
    \centering
    \includegraphics[width=\linewidth]{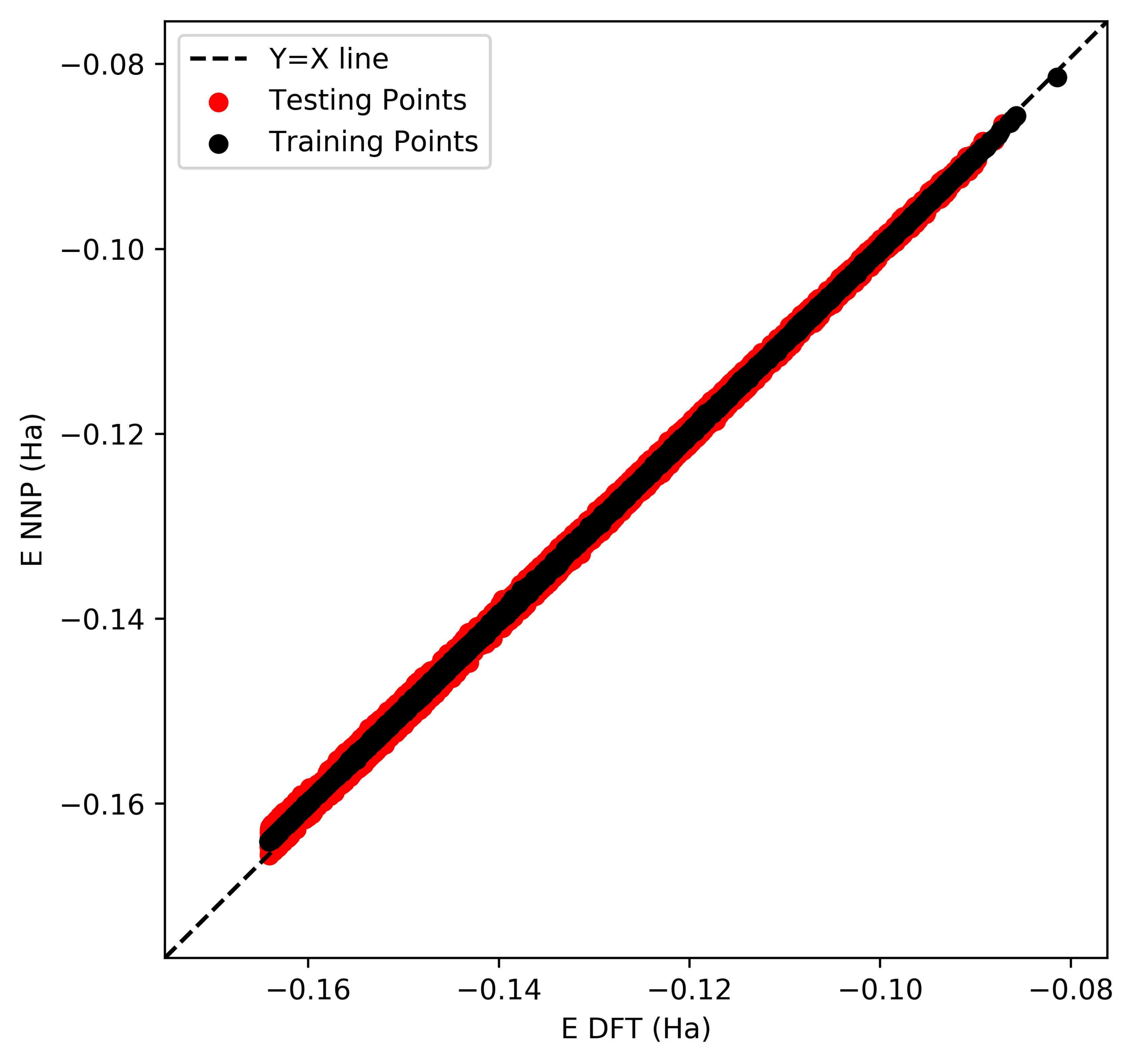}
    \caption{Predicted vs. DFT energy for structures in our dataset, for the chosen NNP architecture and fit.}
    \label{fig:my_label}
\end{figure}